%% file: sAdSgerbe_v2.tex
\documentclass{amsarte}

\usepackage{latexsym,pifont}
\usepackage{epsfig}
\usepackage{rotating}

\usepackage{ifpdf}
\ifpdf
\usepackage{epstopdf}
\usepackage{hyperref}
\else
\usepackage[hypertex]{hyperref}
\fi

\usepackage{extarrows}
\usepackage{amsfonts,amsmath,amssymb,mathrsfs,extarrows,MnSymbol}
\usepackage[all]{xy}

\usepackage{tikz}

\usepackage{cancel}

\usetikzlibrary{matrix,arrows}

\include{1armagedef}

\newtheorem{Thm}{Theorem}
\newtheorem{Prop}[Thm]{Proposition}
\newtheorem{Lem}[Thm]{Lemma}
\newtheorem{Cor}[Thm]{Corollary}
\theoremstyle{definition}
\newtheorem{Rem}[Thm]{Remark}
\newtheorem{Def}[Thm]{Definition}
\newtheorem{Eg}[Thm]{Example}
\newtheorem{Conv}[Thm]{Convention}
\newtheorem{Fact}[Thm]{Fact}
\newtheorem{Quest}[Thm]{Question}
\newtheorem{Prob}[Thm]{Problem}

\numberwithin{equation}{section} \numberwithin{Thm}{section}

\newcount\hour\newcount\minute
        \hour=\time \divide\hour by60 \minute=\time
        {\multiply\hour by60 \global\advance\minute by-\hour}
        \edef\militarytime{\number\hour:\ifnum\minute<10 0\fi\number\minute}

\begin{document}

\title{Equivariant Cartan--Eilenberg supergerbes\\
for the Green--Schwarz superbranes\\[2pt] III. The wrapping anomaly and the super-${\rm AdS}_5\x\bS^5\,$ background}

\author{Rafa\l ~R. ~Suszek}
\address{R.R.S.:\ Katedra Metod Matematycznych Fizyki, Wydzia\l ~Fizyki
Uniwersytetu Warszawskiego, ul.\ Pasteura 5, PL-02-093 Warszawa,
Poland} \email{suszek@fuw.edu.pl}

\begin{abstract}
This is a continuation of a programme, initiated in the work arXiv:1706.05682 [hep-th], of supersymmetry-equivariant geometrisation of the Green--Schwarz super-$(p+2)$-cocycles coupling to the topological charges carried by super-$p$-branes of the superstring theory on reductive homogeneous spaces of supersymmetry groups. In the present part, the ideas and geometro-algebraic tools developed previously are substantially enhanced, adapted to and applied in the physically significant curved backround of Metsaev and Tseytlin, determining the propagation of the critical superstring in the super-${\rm AdS}_5\x\bS^5\,$ geometry. The analysis brings to the fore the r\^ole, in the geometrisation scheme proposed, of the wrapping anomaly of the Poisson algebra of the Noether charges of the rigid symmetries of the relevant super-$\si$-model that lift the geometric symmetries of the supertarget. In particular, the significance of the charges quantifying the monodromy of the Gra\ss mann-odd coordinates in the Kosteleck\'y--Rabin-type quotient of the supertarget is emphasised. A trivial super-1-gerbe is associated with the Metsaev--Tseytlin super-3-cocycle over the super-${\rm AdS}_5\x\bS^5\,$ target. The issue of compatibility of the geometrisation with the \.In\"on\"u--Wigner contraction of the supersymmetry algebra to its flat-superspace counterpart is investigated at some length, revealing the rigidity of the relevant Cartan--Eilenberg cohomology and signalling an attractive potential alternative to the \emph{non-contractible} trivial super-1-gerbe constructed.
\end{abstract}

\void{\date{\today, \militarytime\,(GMT+1)}}

\maketitle

\tableofcontents

\section{Introduction}

A rigorous definition of the lagrangean field theory, termed \textbf{the non-linear $\si$-model}, describing simple geometric dynamics of topologically charged material points, loops and higher ($p-$)dimensional extended objects ($p$-branes) in an ambient smooth metric manifold $\,(\xcM,\txg)\,$ (termed \textbf{the target space}), modelled by smooth\footnote{In fact, in the most general scenario, $\,\bX\,$ is only patchwise smooth over $\,\Om_{p+1}\,$ and maps the latter into a disjoint sum of manifolds. This is the setting of a $\si$-model with defects, {\it cp} \Rcite{Runkel:2008gr}.} embeddings
\qq\nn
\bX\ :\ \Om_{p+1}\too\xcM
\qqq
of the $(p+1)$-dimensional closed \textbf{worldvolume} $\,\Om_{p+1},\ \p\Om_{p+1}=\emptyset\,$ of the extended object in the target space and affected by external fields coupling to their mass (the metric $\,\txg\,$ on $\,\xcM$) and charge (a closed $(p+2)$-form gauge field $\,\underset{\tx{\ciut{(p+2)}}}{\chi},\ \sfd\underset{\tx{\ciut{(p+2)}}}{\chi}=0$), has long been known to call for a geometrisation of the de Rham $(p+2)$-cocycle $\,\underset{\tx{\ciut{(p+2)}}}{\chi}\,$ whose effect on the dynamics is captured by the so-called topological Wess--Zumino (WZ) term
\qq\nn
S^{(p+1)}_{\rm WZ}[\bX]=\int_{\Om_{p+1}}\,\bX^*\sfd^{-1}\underset{\tx{\ciut{(p+2)}}}{\chi}\,.
\qqq
Here, $\,\sfd^{-1}\underset{\tx{\ciut{(p+2)}}}{\chi}\,$ is locally (that is for $\,\bX\,$ that factors through a contractible open subset $\,\cO\subset\xcM$) representable by a primitive of $\,\underset{\tx{\ciut{(p+2)}}}{\chi}$.\ The geometrisation takes the form of a principal $\bC^\x$-bundle, an abelian bundle gerbe, or a $p$-gerbe with connection $\,\underset{\tx{\ciut{(p)}}}{\cG}$,\ respectively. These are conveniently described by classes in the real Deligne--Beilinson cohomology of the target space in degree $p+2$ whose representatives are local differential-form data of a trivialisation of $\,\underset{\tx{\ciut{(p+2)}}}{\chi}\,$ over some ({\it e.g.}, good) open cover of $\,\xcM$.\ Accordingly, (the topological term of) \textbf{the Dirac--Feynman amplitude}
\qq\nn
\cA_{\rm DF(WZ)}[\bX]:=\ee^{\sfi\,S^{(p+1)}_{\rm WZ}[\bX]}
\qqq
acquires the precise interpretation of \textbf{the $(p+1)$-surface holonomy} of the $p$-gerbe $\,\underset{\tx{\ciut{(p)}}}{\cG}\,$ along $\,\bX(\Om_{p+1})$,
\qq\nn
\cA_{\rm DF(WZ)}\equiv{\rm Hol}_{\underset{\tx{\ciut{(p)}}}{\cG}}(\cdot)\ :\ C^\infty(\Om_{p+1},\xcM)\too\uj\,,
\qqq
assigning to $\,\underset{\tx{\ciut{(p)}}}{\cG}\,$ the image of the class
\qq\nn
[\bX^*\underset{\tx{\ciut{(p)}}}{\cG}]\in\vH^{p+1}\bigl(\Om_{p+1},\unl\uj\bigr)
\qqq
of its pullback along $\,\bX\,$ in the \Cv ech cohomology group $\,\vH^{p+1}\bigl(\Om_{p+1},\unl\uj\bigr)\,$ with values in the sheaf $\,\unl\uj\,$ of locally constant maps $\,\Om_{p+1}\too\uj\,$ under the isomorphism
\qq\nn
\vH^{p+1}\bigl(\Om_{p+1},\unl\uj\bigr)\cong\uj\,.
\qqq
The relevance of the geometric object thus associated with $\,\underset{\tx{\ciut{(p+2)}}}{\chi}\,$ in the classical field theory hinges upon the fact that a $p$-gerbe canonically determines the prequantum bundle of the $\si$-model through the so-called cohomological transgression. Upon polarisation, square-integrable sections of that bundle become wave functionals of the field theory under consideration. This and other ramifications and merits of the rigorous higher-geometric and -cohomological formulation of the classical field theory have been established in a long sequence of works \cite{Gawedzki:1987ak,Gawedzki:1999bq,Gawedzki:2002se,Gawedzki:2003pm,Gawedzki:2004tu,Schreiber:2005mi,Recknagel:2006hp,Gawedzki:2007uz,Fuchs:2007fw,Runkel:2008gr,Gawedzki:2008um,Gawedzki:2010rn,Suszek:2011hg,Suszek:2012ddg,Gawedzki:2012fu,Suszek:2013,Gawedzki:2010G} and recalled, together with the relevant technicalities, in \Rcite{Suszek:2017xlw}, to be referred to as Part I henceforth (the reference being inherited by section, proposition and theorem labels).

Incorporation of supersymmetry into the $\si$-model picture, with a sound theoretical motivation (such as, {\it e.g.}, cancellation of the tachyonic mode of the bosonic string), leads to the emergence of novel cohomological and geometric phenomena. In the formulation in which supersymmetries form a Lie supergroup $\,\txG\,$ acting transitively by automorphisms on the supermanifold $\,\xcM\,$ into which the previously introduced worldvolume $\,\Om_{p+1}\,$ is mapped by the lagrangean field $\,\bX\,$ of the ensuing \textbf{super-$\si$-model}, the cohomological novelty is due, in particular, to the discrepancy between the standard de Rham cohomology $\,H^\bullet(\xcM)\,$ of the \textbf{supertarget} $\,\xcM$,\ identical with the de Rham cohomology of its body $\,|\xcM|\,$ in virtue of the Kostant Theorem of \Rcite{Kostant:1975},\ and its supersymmetry-invariant refinement, $\,H^\bullet(\xcM)^\txG$.\ A glaring example of the discrepancy is provided by the super-Minkowski space $\,{\rm sMink}^{d,1\,\vert\,D_{d,1}}\,$ (defined precisely as a supermanifold in Sec.\,I.4.1) with the trivial de Rham cohomology (of the $d+1$ dimensional body $\,{\rm Mink}^{d,1}\equiv\bR^{d,1}$) and a non-trivial supersymmetry-invariant de Rham cohomology, containing, in particular, the Green--Schwarz (GS) super-$(p+2)$-cocycles $\,\underset{\tx{\ciut{(p+2)}}}{\chi}^{\rm GS}\,$ that define the standard (super-$p$-brane) super-$\si$-models with the supertarget $\,{\rm sMink}^{d,1\,\vert\,D_{d,1}}\,$ for distinguished values of $\,d\,$ and $\,p$.

The topological content of the supersymmetric refinement $\,H^\bullet(\xcM)^\txG\,$ of the de Rham cohomology discovered for $\,\xcM={\rm sMink}^{d,1\,\vert\,D_{d,1}}\,$ by Rabin and Crane in Refs.\,\cite{Rabin:1984rm,Rabin:1985tv} justifies a search for a (super)geometrisation of the physically distinguished GS super-$(p+2)$-cocycles, representing classes in $\,H^{p+2}(\xcM)^\txG$,\ fully analogous to the construction of $p$-gerbes in the standard (non-super-)geometry. Indeed, the nontrivial classes in $\,H^\bullet({\rm sMink}^{d,1\,\vert\,D_{d,1}})^{\bR^{d,1\,\vert\,D_{d,1}}}\,$ that trivialise in $\,H^\bullet({\rm sMink}^{d,1\,\vert\,D_{d,1}})\,$ are to be regarded as duals of certain non-trivial cycles in the orbifold $\,{\rm sMink}^{d,1\,\vert\,D_{d,1}}/\G_{\rm KR}\,$ of the super-Minkowski space by the action of the Kosteleck\'y--Rabin discrete supersymmetry group of \Rcite{Kostelecky:1983qu} engendered from trivial cycles in $\,{\rm sMink}^{d,1\,\vert\,D_{d,1}}\,$ through the orbifolding, and so we should actually think of the super-$\si$-model as a field theory on a supermanifold of the same type as $\,\xcM\,$ ({\it i.e.}, locally modelled on the same vector bundle over $\,|\xcM|\,$ in the sense of the Gaw\c{e}dzki--Batchelor Theorem of Refs.\,\cite{Gawedzki:1977pb,Batchelor:1979a}) but with the homological duals of the supersymmetric de Rham super-cocycles without supersymmetric primitives on $\,\xcM$.\ In particular, the new supertarget is anticipated to be compact in (some of) the Gra\ss mann-odd directions. When seen from this perspective, the geometrisation of the GS super-$p$-cocycles regains its purely topological nature. Independent motivation for the geometrisation was presented in \Rcite{Fiorenza:2013nha} where its formal aspects and its relation to the classification of consistent (supersymmetric) $p$-brane models were discussed at great length in the super-Minkowskian setting, in the context of the ubiquitous holographic principle (and so also in the context of the ${\rm AdS}$/CFT correspondence).

A constructive approach to the problem of geometrisation was initiated in Part I. The proposed scheme exploited the classical relation between the Cartan--Eilenberg cohomology $\,{\rm CaE}^\bullet(\txG)\equiv H^\bullet(\txG)^\txG\,$ of the (supersymmetry) Lie supergroup $\,\txG\,$ and the Chevalley--Eilenberg cohomology $\,{\rm CE}^\bullet(\ggt,\bR)\,$ of its Lie superalgebra $\,\ggt\,$ (with values in the trivial module $\,\bR$), in conjunction with the correspondence between the second cohomology group $\,{\rm CE}^2(\ggt,\bR)\,$ of the latter and equivalence classes of supercentral extensions of the Lie superalgebra $\,\ggt$.\ Its underlying idea was to use the extended supersymmetry groups obtained through exponentiation of the supercentral extensions determined by the GS super-$(p+2)$-cocycles (for $\,p\in\{0,1,2\}$) over the Lie \emph{supergroup} $\,\bR^{d,1\,\vert\,D_{d,1}}\equiv{\rm sMink}^{d,1\,\vert\,D_{d,1}}$,\ in a manner originally discussed by de Azc\'arraga {\it et al.} in Refs.\,\cite{Chryssomalakos:2000xd}, in an explicit construction of (or sometimes even directly as) the surjective submersions entering the definition of the supergeometric objects, dubbed (\textbf{Green--Schwarz}) \textbf{super-$p$-gerbes} over $\,{\rm sMink}^{d,1\,\vert\,D_{d,1}}\,$ in Part I. The construction proceeds in full structural analogy with the by now standard geometrisation scheme of cohomological descent for de Rham cocycles, due to Murray \cite{Murray:1994db}.

The super-$p$-gerbes were subsequently shown to possess the expected supersymmetry-($\Ad_\cdot$-)equi\-vari\-ant structure, in perfect analogy with their bosonic counterpart for $\,p=1$,\ {\it cp} Refs.\,\cite{Gawedzki:2010rn,Gawedzki:2012fu,Suszek:2011,Suszek:2012ddg,Suszek:2013}, whose appearance in this picture follows from the identification of the GS super-$3$-cocycle on $\,{\rm sMink}^{d,1\,\vert\,D_{d,1}}\,$ as a super-variant of the canonical Cartan 3-form on a Lie group, and that of the associated super-$\si$-model in the Polyakov formulation as the super-variant of the well-known Wess--Zumino--Witten $\si$-model of Refs.\,\cite{Witten:1983ar,Gawedzki:1990jc,Gawedzki:1999bq,Gawedzki:2001rm}. Finally, the geometrisation scheme was adapted to the setting of the equivalent Hughes--Polchinski formulation of \Rcite{Hughes:1986dn} of the same GS super-$\si$-model, whereby another supergeometric object, dubbed \textbf{the extended Green--Schwarz super-$p$-gerbe} in Part I, was associated with the super-$\si$-model. The object unifies the metric and topological (gerbe-theoretic) data of the previously considered Nambu--Goto formulation. The passage to the Hughes--Polchinski formulation opened the possibility for a straightforward geometrisation of the $\k$-symmetry of the GS super-$\si$-model, {\it i.e.}, the linearised gauge supersymmetry discovered in Refs.\,\cite{deAzcarraga:1982njd,Siegel:1983hh,Siegel:1983ke}, known to effectively implement suppersymmetric balance between the bosonic and fermionic degrees of freedom in the field theories under consideration. The geometrisation assumed the form of an incomplete $\k$-equivariant structure on the extended super-$p$-gerbe, defined in analogy with its (complete) bosonic counterpart of Ref.\,\cite{Gawedzki:2010rn,Gawedzki:2012fu,Suszek:2012ddg}, derived explicitly for $\,p\in\{0,1\}\,$ and termed \textbf{the weak $\k$-equivariant structure}.

The geometrisation scheme developed in Part I exploited largely the exceptional tractability of the super-Minkowskian superbackground, as well as its Cartan-geometric description as a homogeneous space
\qq\nn
{\rm sMink}^{d,1\,\vert\,D_{d,1}}\cong{\rm sISO}(d,1\,\vert\,D_{d,1})/{\rm SO}(d,1)
\qqq
of the super-Poincar\'e supergroup
\qq\nn
{\rm sISO}(d,1\,\vert\,D_{d,1})(d,1)\equiv\bR^{d,1\,\vert\,D_{d,1}}\rx{\rm SO}(d,1)\,,
\qqq
and -- indeed -- as a Lie supergroup of supertranslations,
\qq\nn
{\rm sMink}^{d,1\,\vert\,D_{d,1}}\cong\bR^{d,1\,\vert\,D_{d,1}}\,.
\qqq
The latter allowed to rephrase the differential calculus on the flat supertarget entirely in terms of the components of the supersymmetry-(left-)invariant Maurer--Cartan super-1-form and of the dual supersymmetry-(left-)invariant vector fields on\footnote{The Hughes--Polchinski formulation naturally calls for the full supersymmetry group $\,{\rm s}\xcP(d,1)$.} $\,\bR^{d,1\,\vert\,D_{d,1}}\,$ and work with the Chevalley--Eilenberg model of the Lie-superalgebra cohomology for the Lie superalgebra $\,\gt{smink}^{d,1\,\vert\,D_{d,1}}\,$ of the Lie supergroup $\,{\rm sMink}^{d,1\,\vert\,D_{d,1}}\,$  -- hence the algebraisation of the supersymmetric de Rham cohomology of $\,{\rm sMink}^{d,1\,\vert\,D_{d,1}}$,\ explaining the definition of the geometrisation scheme in terms of extensions of the supertarget Lie superalgebra.

\medskip

The choice of the supertarget made in Part I masks a variety of topological and algebraic problems that arise over a generic homogeneous space $\,\txG/\txH\,$ of a supersymmetry group $\,\txG\,$ (with a Lie subgroup $\,\txH\subset\txG$), such as, {\it e.g.}, a non-trivial topology and metric curvature of the body, absence of a Lie-supergroup structure on $\,\txG/\txH$,\ induction of a highly non-linear realisation of supersymmetry and inheritance of the associated supersymmetric differential calculus on $\,\txG/\txH\,$ from the Lie supergroup $\,\txG\,$ along a family of \emph{locally} smooth sections of the principal $\txH$-bundle $\,\txG\too\txG/\txH$,\ in the spirit of the theory of nonlinear realisations of (super)symmetries, developed in Refs.\,\cite{Schwinger:1967tc,Weinberg:1968de,Coleman:1969sm,Callan:1969sn,Salam:1969rq,Salam:1970qk,Isham:1971dv,Volkov:1972jx,Volkov:1973ix,Ivanov:1978mx,Lindstrom:1979kq,Uematsu:1981rj,Ivanov:1982bpa,Samuel:1982uh,Ferrara:1983fi,Bagger:1983mv} and recently revived in the string-theoretic context in Refs.\,\cite{McArthur:1999dy,West:2000hr,Gomis:2006xw,McArthur:2010zm}. In the present paper, we make the first step in this general direction by extending our geometrisation scheme to the supermanifold with the topologically non-trivial and metrically curved body
\qq\nn
{\rm AdS}_5\x\bS^5
\qqq
and with the structure of a homogeneous space
\qq\nn
\txG/\txH\equiv{\rm SU}(2,2\,\vert\,4)/\bigl({\rm SO}(4,1)\x{\rm SO}(5)\bigr)=:{\rm s}\bigl({\rm AdS}_5\x\bS^5\bigr)
\qqq
of the supersymmetry group $\,\txG\equiv{\rm SU}(2,2\,\vert\,4)\,$ with the body $\,{\rm SO}(5,1)\x{\rm SO}(6)$.\ The homogeneous space is devoid of any (obvious) Lie-supergroup structure and embedded in $\,{\rm SU}(2,2\,\vert\,4)\,$ patchwise smoothly by a collection of local sections of the principal ${\rm SO}(4,1)\x{\rm SO}(5)$-bundle
\qq\label{eq:AdSprinc}
{\rm SU}(2,2\,\vert\,4)\too{\rm SU}(2,2\,\vert\,4)/\bigl({\rm SO}(4,1)\x{\rm SO}(5)\bigr)\,.
\qqq
The supertarget is further endowed with a GS super-3-cocycle induced by the ${\rm SO}(4,1)\x{\rm SO}(5)$-basic \textbf{Metsaev--Tseytlin super-3-cocycle}
\qq\nn
\underset{\tx{\ciut{(3)}}}{\chi}^{\rm MT}=-\bigl(\widehat C\,\widehat\G_{\widehat a}\ox\si_3\bigr)_{\widehat\a\widehat\b}\,\theta_{\rm L}^{\widehat\a}\wedge\theta_{\rm L}^{\widehat a}\wedge\theta_{\rm L}^{\widehat\b}
\qqq
on $\,{\rm SU}(2,2\,\vert\,4)$,\ first introduced in \Rcite{Metsaev:1998it} and written in terms of components $\,\theta_{\rm L}^{\widehat\a}\,$ and $\,\theta_{\rm L}^{\widehat a}\,$ of the $\gt{su}(2,2\,\vert\,4)$-valued Maurer--Cartan super-1-form along the direct-sum complement $\,\tgt\,$ of the Lie subalgebra $\,\hgt\equiv\gt{so}(4,1)\oplus\gt{so}(5)\,$ of the isotropy Lie subgroup $\,\txH\equiv{\rm SO}(4,1)\x{\rm SO}(5)\subset{\rm SU}(2,2\,\vert\,4)\,$ defining a reductive decomposition
\qq\nn
\gt{su}(2,2\,\vert\,4)\equiv\ggt=\tgt\oplus\hgt\,,\qquad\qquad[\hgt,\tgt]\subset\tgt\,,
\qqq
and of certain $\txH$-invariant tensors $\,\bigl(\widehat C\,\widehat\G_{\widehat a}\ox\si_3\bigr)_{\widehat\a\widehat\b}\,$ defined in Sec.\,\ref{sec:ssextaAdSS}. Altogether, these form, according to the general rules discovered and elucidated in the original literature on the subject of nonlinear realisations of (super)symmetries, cited above, and reviewed in Section \ref{sec:Cartsgeom},  a super-3-form on $\,\txG\,$ that descends (through pullback) to the quotient supermanifold $\,{\rm s}\bigl({\rm AdS}_5\x\bS^5\bigr)\,$ along the aforementioned local sections of \eqref{eq:AdSprinc}, with local pullbacks glueing smoothly over intersections of their domains. This time, the choice of the supertarget is motivated by the critical relevance of the associated super-$\si$-model, postulated by Metsaev and Tseytlin in \Rcite{Metsaev:1998it}, to the formulation and study of the celebrated ${\rm AdS}$/CFT correspondence of Refs.\,\cite{Maldacena:1997re,Maldacena:1998im}, of much significance in the research on the quantum dynamics of strongly coupled systems with a non-abelian gauge symmetry, such as, {\it e.g.}, the quark-gluon plasma. The \textbf{superbackground}
\qq\nn
\bigl({\rm s}\bigl({\rm AdS}_5\x\bS^5\bigr),\underset{\tx{\ciut{(3)}}}{\chi}^{\rm MT}\bigr)
\qqq
is also directly related to the formerly scrutinised flat superbackground
\qq\nn
\bigl({\rm sMink}^{9,1\,\vert\,32},\underset{\tx{\ciut{(3)}}}{\chi}^{\rm GS}\bigr)
\qqq
through the flattening limit
\qq\label{eq:Rtoinf}
R\to\infty
\qqq
of the \emph{common} radius $\,R\,$ of the generating 1-cycle of $\,{\rm AdS}_5\cong\bS^1\x\bR^{\x 4}\,$ and that of the 5-sphere in the body of the supertarget, with a dual algebraic realisation in the form of an \.In\"on\"u--Wigner contraction
\qq\label{eq:IWcontrsAdS}
\gt{su}(2,2\,\vert\,4)\xrightarrow[{\rm rescaling}]{\ R-{\rm dependent}\ }\gt{su}(2,2\,\vert\,4)_R\xrightarrow{\ R\to\infty\ }\gt{smink}^{d,1\,\vert\,D_{d,1}}\,.
\qqq
It is worth emphasising that the asymptotic relation between the two superbackgrounds was one of the basic guiding principles on which Metsaev and Tseytlin founded their construction of the two-dimensional super-$\si$-model for $\,{\rm s}({\rm AdS}_5\x\bS^5)$,\ and so it is apposite to expect that it should lift to the sought-after geometrisation of the GS super-3-cocycle descended from the Metsaev--Tseytlin (MT) super-3-cocycle -- this is the premise upon which much of the work reported herein has been based.

The latter super-3-cocycle admits a manifestly supersymmetric primitive $\,\underset{\tx{\ciut{(2)}}}{\b}$,\ found by Roiban and Siegel in \Rcite{Roiban:2000yy}, that also descends to the supertarget $\,{\rm s}({\rm AdS}_5\x\bS^5)\,$ along the local sections of \eqref{eq:AdSprinc}. Thus, according to the geometrisation scheme formulated in Part I, the super-3-cocycle gives rise to a trivial GS super-1-gerbe over $\,{\rm s}({\rm AdS}_5\x\bS^5)$,\ which we construct explicitly in Section \ref{sec:trsAdSSext}. However, $\,\underset{\tx{\ciut{(2)}}}{\b}\,$ does \emph{not} reproduce the (non-supersymmetric) primitive of the GS super-3-cocycle $\,\underset{\tx{\ciut{(3)}}}{\chi}^{\rm GS}\,$ on $\,{\rm sMink}^{9,1\,\vert\,32}\,$ in the limit \eqref{eq:Rtoinf}. Consequently, the trivial super-1-gerbe does \emph{not} contract to its super-Minkowskian counterpart. In Sections \ref{sec:KamSakext} and \ref{sec:KostRabdef}, we systematically examine deformations of the supersymmetry algebra $\,\gt{su}(2,2\,\vert\,4)\,$ within the category of Lie superalgebras which could trivialise -- through a mechanism originally devised by de Azc\'arraga {\it et al.} in the super-Minkowskian setting in \Rcite{Chryssomalakos:2000xd} and applied successfully in the same setting in Part I -- the super-3-cocycle $\,\underset{\tx{\ciut{(3)}}}{\chi}^{\rm MT}\,$ in a manner compatible with the \.In\"on\"u--Wigner contraction \eqref{eq:IWcontrsAdS}. As the study of \emph{all} possible such deformations is well beyond the scope of the present work, we take guidance from an explicit asymptotic analysis of the charge deformation of the Poisson algebra of the Noether charges of supersymmetry in the MT super-$\si$-model, based -- in the sense made precise in Section \ref{sec:wrapanom} -- on the class of the supersymmetric primitive $\,\underset{\tx{\ciut{(2)}}}{\b}$.\  The analysis has been carried out in Section \ref{sec:ssextaAdSS} and prepared by an abstract discussion of the r\^ole of the topological WZ term
in the action functional of the (super-)$\si$-model in the said deformation, presented in Section \ref{sec:wrapanom}. Its results shed light on the significance of the so-called pseudo-invariance of the WZ term for the existence and structure of the deformation. When concretised in the setting of the GS super-$\si$-model for $\,{\rm sMink}^{d,1\,\vert\,D_{d,1}}\,$ in Sections \ref{subsect:spartextMink} and \ref{subsect:sstringextsMink}, they provide direct evidence of the central r\^ole of the Kosteleck\'y--Rabin charges in the geometrisation of the Cartan--Eilenberg cohomology on the supersymmetry group through Lie-superalgebra extensions determined by the GS super-$(p+2)$-cocycles. In particular, we reobtain the superstring extension of the super-Minkowski superalgebra $\,\gt{smink}^{d,1\,\vert\,D_{d,1}}\,$ of Part I as a deformation of the Poisson algebra of the Noether charges of supersymmetry of the two-dimensional GS super-$\si$-model for $\,{\rm sMink}^{d,1\,\vert\,D_{d,1}}\,$ by the Gra\ss mann-odd Kosteleck\'y--Rabin charge. This, in conjunction with the knowledge of the topology of the body of the supertarget of main interest, $\,{\rm s}({\rm AdS}_5\x\bS^5)$,\ allows us to organise our search for extensions of $\,\gt{su}(2,2\,\vert\,4)\,$ into two natural directions:
\ben
\item a Gra\ss mann-even central extension deforming (exclusively) the anticommutator of the supercharges in an arbitrary manner, contemplated with view to recovering the desired non-supersymmetric correction to the Roiban--Siegel primitive of the Metsaev--Tseytlin super-3-cocycle as the leading term in an asymptotic expansion of a supersymmetric super-2-form on the resultant extended supersymmetry group, and
\item a generic extension determined by a Gra\ss mann-odd deformation of the commutator $\,[Q_{\a\a'I},P_{\widehat a}]\,$ engineered so as to allow for a trivialisation, on the resultant extended supersymmetry group, of a super-2-cocycle asymptoting to the Kosteleck\'y--Rabin super-2-cocycle of Section \ref{subsect:sstringextsMink} in the limit of an infinite radius of $\,{\rm AdS}_5\x\bS^5$.
\een
The search returns negative results, and so -- at this stage -- the non-contractible trivial super-1-gerbe with curvature $\,\underset{\tx{\ciut{(3)}}}{\chi}^{\rm MT}\,$ associated with the supersymmetric primitive $\,\underset{\tx{\ciut{(2)}}}{\b}\,$ remains as the sole consistent geometrisation of the MT super-3-cocycle. The paper concludes with a discussion of an alternative approach to geometrisation founded on the assumption that it is \emph{not} the asymptotic relation between the fixed Green--Schwarz super-3-cocycles: $\,\underset{\tx{\ciut{(3)}}}{\chi}^{\rm MT}\,$ and $\,\underset{\tx{\ciut{(2)}}}{\chi}^{\rm GS}\,$ (the latter on $\,{\rm Mink}^{9,1}$) but rather the asymptotic relation between the supertarget geometries: $\,{\rm s}({\rm AdS}_5\x\bS^5)\,$ and $\,{\rm Mink}^{9,1}$,\ modelled by the contraction mechanism for the (extended) super-${\rm AdS}_5\x\bS^5\,$ Lie superalgebra, that ought to be regarded as fundamental. The approach takes as the point of departure a pair of extended Lie superalgebras: a Gra\ss mann-odd deformation of the super-${\rm AdS}_5\x\bS^5\,$ superalgebra and the superstring deformation of the super-Minkowskian superalgebra, assumed to be related by an \.In\"on\"u--Wigner contraction to begin with, and \emph{defines} the Green--Schwarz super-3-cocycle on $\,{\rm s}({\rm AdS}_5\x\bS^5)\,$ as the exterior derivative of the manifestly (extended-)supersymmetric super-2-form on the Lie supergroup $\,\widetilde{{\rm SU}(2,2\,\vert\,4)}\,$ (integrating the former extended Lie superalgebra) constructed as a direct structural counterpart of the known supersymmetric primitive of the GS super-3-cocycle on the extended super-Minkowski superspace used in Part I. This logical possibility, and its potential consequences for the very definition of the curved supertarget, are analysed in the toy model of the super-${\rm AdS}$ Lie superalgebra with a built in Gra\ss mann-odd deformation. The analysis indicates an interesting and promising direction of the unfinished quest for the mechanism of trivialisation of the physically relevant (class in the) Chevalley--Eilenberg cohomology of $\,\gt{su}(2,2\,\vert\,4)\,$ compatible with the \.In\"on\"u--Wigner contraction and the associated geometric transition
\qq\nn
{\rm s}\bigl({\rm AdS}_5\x\bS^5\bigr)\xrightarrow[{\rm rescaling}]{\ R-{\rm dependent}\ }{\rm s}\bigl({\rm AdS}_5(R)\x\bS^5(R)\bigr)\xrightarrow{\ R\to\infty\ }{\rm sMink}^{9,1\,\vert\,32}\,,
\qqq
to which we hope to return in a future work.

\medskip

The paper is organised as follows:
\bit
\item in Section \ref{sec:Cartsgeom}, we review systematically the logic of the construction of a lagrangean field theory on a homogeneous space $\,\txG/\txH\,$ of a (super)symmetry Lie (super)group $\,\txG\,$ (corresponding to a reductive decomposition $\,\ggt=\tgt\oplus\hgt\,$ of the supersymmetry algebra $\,\ggt$) using suitable elements of the Cartan differential calculus on $\,\txG$,\ and the ensuing induction of a non-linear realisation of supersymmetry;
\item in Section \ref{sec:wrapanom}, we discuss at length the canonical description of a field-theoretic realisation, in terms of the Poisson algebra of the relevant Noether charges, of supersymmetry in a super-$\si$-model with a pseudo-invariant Wess--Zumino term, whereby we discover the wrapping anomaly that quantifies the departure of that realisation from the original supersymmetry algebra $\,\ggt\,$ -- this we take as the germ of a physically motivated (normal) extension of $\,\ggt\,$ studied subsequently;
\item in Section \ref{sec:sMinkext}, we identify the field-theoretic source of the super-central extensions of the super-Minkowski Lie superalgebra encountered in Part I (and giving rise to the super-$p$-gerbes, for $\,p\in\{0,1\}$,\ associated with the Green--Schwarz super-$\si$-models with that supertarget) and provide concrete evidence of the r\^ole played in these deformations by the winding charges measuring the monodromy of the Gra\ss mann-odd coordinates along the non-contractible cycles in the Kosteleck\'y--Rabin quotient of the super-Minkowski superspace that topologises the non-trivial Cartan--Eilenberg cohomology of the underlying Lie supergroup;
\item in Section \ref{sec:ssextaAdSS}, we introduce the Metsaev--Tseytlin superbackground over the supertarget $\,{\rm s}({\rm AdS}_5\x\bS^5)\,$ and analyse in great detail the (super)geometric nature of the corresponding wrapping anomaly, including its asymptotics in the flat limit $\,R\to\infty\,$ -- this determines natural paths of deformation of the relevant supersymmetry algebra $\,\gt{su}(2,2\,\vert\,4)\,$ that we pursue in later sections;
\item in Section \ref{sec:trsAdSSext}, we describe the trivial Metsaev--Tseytlin super-1-gerbe associated with the manifestly supersymmetric Roiban--Siegel primitive of the Metsaev--Tseytlin super-3-cocycle, noting the incompatibility of its structure with the \.In\"on\"u--Wigner contraction;
\item in Section \ref{sec:KamSakext}, we study a class of super-central extensions of the supersymmetry algebra $\,\gt{su}(2,2\,\vert\,4)\,$ obtained through a Gra\ss mann-even deformation of the anticommutator of the supercharges, only to find out that the admissible ones do not allow for a supersymmetry-equivariant trivialisation of the Metsaev--Tseytlin super-3-cocycle compatible with the \.In\"on\"u--Wigner contraction;
\item in Section \ref{sec:KostRabdef}, we consider two classes of natural associative deformations of the supersymmetry algebra $\,\gt{su}(2,2\,\vert\,4)\,$ engendered by a Gra\ss mann-odd deformation, of the Kosteleck\'y--Rabin type, of the commutator of the supercharge and the momentum -- the deformations are proven algebraically inconsistent;
\item in Section \ref{sec:MTaway}, drawing inspiration from the previously encountered failures of geometrisation schemes compatible with the \.In\"on\"u--Wigner contraction, we conceive an alternative geometrisation scenario that takes the asymptotic relation between the two relevant supergeometries: $\,{\rm sMink}^{9,1\,\vert\,32}\,$ and $\,{\rm s}({\rm AdS}_5\x\bS^5)$,\ and between the respective supersymmetry algebras as the organising principle; the general idea is illustrated and tested on a toy example of a super-${\rm AdS}_d$ superspace and the associated supersymmetry algebra $\,\gt{sso}(d-1,2)$.
\item in Section \ref{sec:C&O}, we recapitulate the work reported in the present paper and indicate possible directions of future research that it motivates;
\item in the Appendices, we present the relevant conventions on and facts regarding the Clifford algebras employed in the paper, and gather various technical calculations, including proofs of the propositions and theorems stated in the main text.
\eit

\section{The Cartan geometry of homogeneous superspaces and super-$\si$-models thereon}\label{sec:Cartsgeom}

Let $\,\txG\,$ be a Lie supergroup, with the Lie superalgebra $\,\ggt\,$ as defined in Part I, to be referred to as {\bf the supersymmetry group} (resp.\ {\bf the supersymmetry algebra}) in what follows and let $\,\xcM\,$ be a supermanifold endowed with a transitive (left) action of $\,\txG$
\qq\nn
\la_\cdot\ :\ \txG\x\xcM\too\xcM\ :\ (g,m)\longmapsto g\lact m\equiv\la_g(m)\,,
\qqq
so that there exists a $\txG$-equivariant diffeomorphism
\qq\nn
\mu\ :\ \xcM\xrightarrow{\ \cong\ }\txG/\txH\,,
\qqq
where $\,\txH\equiv\txG_m\,$ is the isotropy group $\,\txG_m\,$ of an arbitrary point $\,m\in\xcM$.\ The manifold $\,\xcM\,$ shall be modelled on the homogeneous space $\,\txG/\txH\,$ henceforth, the latter being realised locally as a section of the principal bundle\footnote{The first arrow denotes the free and transitive action of the structure group on the fibre.}
\qq\nn
\alxydim{@C=1cm@R=1cm}{\txH \ar[r] & \txG \ar[d]^{\pi_{\txG/\txH}} \\ & \txG/\txH}
\qqq
with the structure group $\,\txH$,\ a closed Lie subgroup of $\,\txG$.\ Thus, we shall work with a family of submanifolds embedded in $\,\txG\,$ by the respective (local) sections\footnote{That is, equivalently, by a collection of local trivialisations.}
\qq\nn
\si_i\ :\ \cO_i\too\txG\ :\ g\txH\longmapsto g\cdot h_i(g)\,,\quad i\in I\,,
\qqq
of the submersive projection on the base $\,\pi_{\txG/\txH}$,\ associated with a trivialising cover $\,\cO=\{\cO_i\}_{i\in I}\,$ of the latter,
\qq\nn
\txG/\txH=\bigcup_{i\in I}\,\cO_i\,.
\qqq
The redundancy of such a realisation over any non-empty intersection, $\,\cO_{ij}\equiv\cO_i\cap\cO_j\neq\emptyset$,\ is accounted for by a collection of locally smooth (transition) maps
\qq\nn
h_{ij}\ :\ \cO_{ij}\too\txH\subset\txG
\qqq
fixed by the condition
\qq\label{eq:Htransmap}
\forall_{x\in\cO_{ij}}\ :\ \si_j(x)=\si_i(x)\cdot h_{ij}(x)\,.
\qqq
The original action $\,\la_\cdot$,\ with the simple model on $\,\txG/\txH$
\qq\label{eq:cosetlact}
[\la]_\cdot\ :\ \txG\x\txG/\txH\too\txG/\txH\ :\ (g',g\txH)\longmapsto(g'\cdot g)\txH\,,
\qqq
is transcribed, through the $\,\si_i^\txK$,\ into a geometric realisation of $\,\txG\,$ on the image of $\,\txG/\txK\,$ within $\,\txG$,\ with the same obvious redundancy. Indeed, consider a point $\,x\in\cO_i\,$ and an element $\,g\in\txG$.\ Upon choosing an \emph{arbitrary} index $\,j\in I\,$ with the property
\qq\label{eq:cosetactpt}
\widetilde x(x;g'):=\pi_{\txG/\txH}\bigl(g'\cdot\si_i(x)\bigr)\in\cO_j\,,
\qqq
we find a unique $\,\unl h_{ij}(x,g')\in\txH\,$ defined (on some open neighbourhood of $\,(x,g')$) by the condition
\qq\label{eq:cosetact}
g'\cdot\si_i(x)=\si_j\bigl(\widetilde x(x;g')\bigr)\cdot\unl h_{ij}(x;g')^{-1}\,.
\qqq
Note that for $\,\widetilde x(x;g')\in\cO_{jk}\,$ we have
\qq\nn
\unl h_{ik}(x;g')=\unl h_{ij}(x;g')\cdot h_{jk}\bigl(\widetilde x(x;g')\bigr)\,,
\qqq
so that the two realisations of the action are related by a compensating transformation from the structure group $\,\txH$.

The realisation of the homogeneous space $\,\txG/\txH\,$ within $\,\txG\,$ described above enables us to reconstruct the differential calculus on the former space (and so also on $\,\xcM$) from that on the Lie supergroup $\,\txG$.\ To this end, we decompose the (super)vector space $\,\ggt\,$ as
\qq\nn
\ggt=\tgt\oplus\hgt
\qqq
into the Lie algebra $\,\hgt\supset[\hgt,\hgt]\,$ of $\,\txH\,$ and its direct vector-space complement $\,\tgt$,\ assuming, furthermore, the decomposition to be \emph{reductive}, in the convention of Sec.\,I.3 (with $\,\rgt\equiv\hgt$), so that $\,\tgt\,$ acquires the status of a super-graded $\hgt$-module,
\qq\nn
[\hgt,\tgt]\subset\tgt\,.
\qqq
We make a choice of the basis of $\,\ggt\,$ compatible with the splitting, and -- accordingly -- mark the generators of $\,\tgt\,$ by an underline: $\,\{t_{\unl A}\}_{\unl A\in\ovl{1,\dim\,\tgt}}$,\ and denote those of $\,\hgt\,$ as $\,\{J_\k\}_{\k\in\ovl{1,\dim\,\hgt}}$.\ The generators are taken to be homogeneous with respect to the super-grading
\qq\nn
\ggt=\ggt^{(0)}\oplus\ggt^{(1)}
\qqq
in which $\,\ggt^{(0)}\,$ is the Gra\ss mann-even Lie subalgebra of $\,\ggt$,\ containing $\,\hgt$,\ and $\,\ggt^{(1)}\,$ is the Gra\ss mann-odd $\ad$-module thereof,
\qq\nn
[\ggt^{(0)},\ggt^{(1)}]\subset\ggt^{(1)}\,.
\qqq
The latter decomposition divides the set $\,\{t_A\}_{A\in\ovl{1,\dim\,\ggt}}\,$ of the homogeneous generators of $\,\ggt$,\ satisfying the defining supercommutation relations
\qq\nn
[t_A,t_B\}=f_{AB}^{\ \ C}\,t_C\,,
\qqq
into subsets with fixed Gra\ss mann parity: even ($\{B_a\}_{a\in\ovl{1,\dim\,\ggt^{(0)}}}$) and odd ($\{F_{\widehat\a}\}_{\widehat\a\in\ovl{1,\dim\,\ggt^{(1)}}}$), subject to relations
\qq\nn
[B_a,B_b]=f_{ab}^{\ \ c}\,B_c\,,\qquad\qquad\{F_{\widehat\a},F_{\widehat\b}\}=f_{\widehat\a\widehat\b}^{\ \ c}\,B_c\,,\qquad\qquad[B_a,F_{\widehat\a}]=f_{a\widehat\a}^{\ \ \widehat\b}\,F_{\widehat\b}\,.
\qqq
Accordingly, the generators of the module $\,\tgt\,$ further split into subsets: the Gra\ss mann-even ones $\,\{P_{\widehat{a}}\}_{\widehat{a}\in\ovl{1,\dim\,\tgt^{(0)}}}\,$ and the Gra\ss mann-odd ones $\,\{Q_{\widehat\a}\}_{\widehat\a\in\ovl{1,\dim\,\tgt^{(1)}}}$,\ where $\,\tgt^{(1)}\equiv\ggt^{(1)}\,$ in our considerations.

We may now span the tangent sheaf of $\,\txG/\txH\,$ over $\,\cO_i\ni x\,$ on (restrictions of) the fundamental vector fields of $\,[\la]_\cdot$.\ These correspond to certain point-dependent (over $\,\txG/\txH$) linear combinations $\,\cK_X\,$ of the right- and left-invariant vector fields on $\,\si_i(\cO_i)\subset\txG$,\ that is of
\qq\nn
\cR_X\bigl(\si_i(x)\bigr)=\tfrac{\sfd\ }{\sfd t}\rstr_{t=0}\,\bigl(\ee^{tX}\cdot\si_i(x)\bigr)\,,\quad X\in\ggt
\qqq
and
\qq\nn
\ceL_{Y_i(X;x)}\bigl(\si_i(x)\bigr)=\tfrac{\sfd\ }{\sfd t}\rstr_{t=0}\,\bigl(\si_i(x)\cdot\ee^{tY_i(X;x)}\bigr)\,,\quad Y_i(X;x)\in\hgt\,,
\qqq
respectively. The relevant combinations are readily read off from \Reqref{eq:cosetact}. Indeed, the left-regular translation of a point $\,\si_i(x),\ x\in\cO_i\,$ by $\,\ee^{tX},\ X\in\ggt\,$ fixes the right-invariant component of the (modelling) fundamental vector field in the form $\,\cR_X(\si_i(x))$,\ whereas the compensating r\^ole of the right-regular translation $\,\unl h_{ii}(x,\ee^{tX})\equiv\ee^{tY_i(X;x)}\,$ along $\,\txH\,$ identifies the corresponding left-invariant component, through imposition of the constraints
\qq\label{eq:Yid}
\ee^{tX}\cdot\si_i(x)\cdot\ee^{tY_i(X;x)}=\si_i\circ\cG_t^{\Xi_X}(x)\,,
\qqq
written for $\,t\in]-\vep,\vep[,\ \vep\gtrapprox 0\,$ and the fundamental vector field $\,\Xi_X\,$ associated with $\,X\in\ggt\,$ in the standard manner as
\qq\nn
\Xi_X(x)=\tfrac{\sfd\ }{\sfd t}\rstr_{t=0}\,[\la]_{\ee^{tX}}(x)\,,\qquad x\in\txG/\txH\,,
\qqq
with the (local) flow
\qq\nn
\cG_\cdot^{\Xi_X}\ :\ ]-\vep,\vep[\too\Diff_{\rm loc}(\txG/\txH)\,.
\qqq
We shall write out $\,\Xi_X\,$ in the local coordinate basis of the tangent bundle engendered by local coordinates $\,Z_i^{\unl A}\equiv(X_i^{\widehat a},\theta_i^{\widehat\a})\,$ corresponding to a charting of $\,\cO_i\,$ (and so also of $\,\si_i(\cO_i)\subset\txG$) by flows of left-invariant vector fields on $\,\txG\,$ along $\,\tgt$ (in the standard manner, familiar from constructive proofs of the Frobenius Theorem) as
\qq\nn
\Xi_X\rstr_{\cO_i}=\d_X Z_i^{\unl A}\,\tfrac{\p\ }{\p Z^{\unl A}_i}\,.
\qqq
Differentiating both sides of \Reqref{eq:Yid} at $\,t=0$,\ we then obtain the result
\qq\nn
\Xi_X\con E(Z_i)=\sfT_e\Ad_{\si_i(Z_i)^{-1}}(X)+Y_i(X;x)\,,
\qqq
expressed in terms of the pullbacks
\qq\label{eq:pullbackMC}\qquad\qquad\qquad
\si_i^*\theta_{\rm L}(Z_i)=\sfd Z_i^{\unl A}\,E_{\unl A}^{\ A}(Z_i)\ox t_A\equiv E^A(Z_i)\ox t_A\equiv\sfd Z_i^{\unl A}\,E_{\unl A}(Z_i)\equiv E(Z_i)\in\sfT^*_{Z_i}(\txG/\txH)\ox\ggt\,,
\qqq
of the left-invariant $\ggt$-valued Maurer--Cartan super-1-form
\qq\nn
\theta_{\rm L}=\theta_{\rm L}^A\ox t_A
\qqq
on $\,\txG$,\ the latter being fixed by the standard condition
\qq\nn
\ceL_X\con\theta_{\rm L}=X\,.
\qqq
It is these pullbacks of certain distinguished -- through the analysis that follows -- linear combinations of components of the Maurer--Cartan super-1-form that descend to the dual sheaf of (super)differential forms on $\,\txG/\txH$.

Our result translates into a pair of equations:
\qq\label{eq:leftvarfiel}
\d_X Z_i^{\unl A}\,E_{\unl A}^{\ \unl B}(Z_i)&=&\sfT_e\Ad_{\si_i(Z_i)^{-1}}(X)^{\unl B}\,,\\\cr
Y_i^\k(X;x)&=&\d_X Z_i^{\unl A}\,E_{\unl A}^{\ \k}(Z_i)-\sfT_e\Ad_{\si_i(Z_i)^{-1}}(X)^\k \nn
\qqq
of which the former determines (components of) the field $\,\Xi_{X;i}$,\ whereas the latter subsequently uses them to determine the compensating translation $\,Y_i(X;x)$.\ Note, in particular, that the dependence of the latter upon $\,X\,$ is linear, as expected, {\it i.e.},
\qq\label{eq:YlinX}
Y_i(X;x)=X^A\,y_{i\,A}(x)
\qqq
for some (smooth) maps
\qq\nn
y_{i\,A}\ :\ \cO_i\too\hgt\,,\qquad A\in\ovl{1,\dim\,\ggt}\,.
\qqq
The ensuing local vector fields
\qq\nn
\cK_{i\,X}\bigl(\si_i(x)\bigr)=\cR_X\bigl(\si_i(x)\bigr)+\ceL_{Y_i(X;x)}\bigl(\si_i(x)\bigr)
\qqq
are, by definition, the pushforwards, along $\,\sfT\si_i$,\ of the fundamental vector fields for the \emph{left} action $\,[\la]_\cdot\,$ of $\,\txG\,$ on $\,\txG/\txH\,$ restricted to $\,\cO_i$,
\qq\label{eq:KXaspfXiX}
\cK_{i\,X}\bigl(\si_i(x)\bigr)\equiv\sfT_x\si_i\bigl(\Xi_X(x)\bigr)\,,
\qqq
and so they satisfy the relations
\qq\label{eq:KKKh}
[\cK_{i\,X_1},\cK_{i\,X_2}]=-\cK_{i\,[X_1,X_2]}\,.
\qqq

Our hitherto considerations, in conjunction with the assumptions made, pave the way to the standard construction of supersymmetric ({\it i.e.}, globally $\txG$-invariant) lagrangean field theories with the (typical) fibre of the covariant configuration bundle given by (or, to put it differently, with fields in the lagrangean density taking values in) $\,\xcM\cong\txG/\txH$.\ Indeed, denote, with view to subsequent analyses  and for $\,i\,$ and $\,j\,$ as above (in a mild abuse of the notation),
\qq\label{eq:cosetactptcoord}
Z_j\circ\pi_{\txG/\txH}\circ\la_{g'}\bigl(\si_i(Z_i)\bigr)=:\widetilde Z_j(Z_i,g')
\qqq
and
\qq\label{eq:cosetactcoord}
\bigl(g'\cdot\si_i(Z_i)\bigr)^{-1}\cdot\si_j\bigl(\widetilde Z_j(Z_i,g')\bigr)=:\unl h_{ij}(Z_i,g')\,.
\qqq
We then compute
\qq
E^A\bigl(\widetilde Z_j(Z_i;g')\bigr)\ox t_A&\equiv&\si_j^*\theta_{\rm L}\bigl(\widetilde Z_j(Z_i,g')\bigr)\cr\cr
&=&E^{\unl B}(Z_i)\,\G_{\unl B}^{\ \unl A}\bigl(Z_i,g';j\bigr)\ox t_{\unl A}+\bigl(E^\la(Z_i)\,\G_\la^{\ \k}\bigl(Z_i,g';j\bigr)+\bigl[\unl h_{ij}^*\theta_{\rm L}\bigl(Z_i,g'\bigr)\bigr]^\k\bigr)\ox J_\k \label{eq:thetgtens}
\qqq
in terms of the matrices
\qq\nn
\G_A^{\ B}\bigl(Z_i,g';j\bigr)\,t_B:=\sfT_e\Ad_{\unl h_{ij}(Z_i,g')^{-1}}(t_A)
\qqq
and the $\hgt$-valued 1-forms
\qq\nn
\bigl[\unl h_{ij}^*\theta_{\rm L}\bigl(Z_i,g'\bigr)\bigr]^\k\ox J_\k\equiv\unl h_{ij}^*\theta_{\rm L}\bigl(Z_i,g'\bigr)\,.
\qqq
Thus, components of the Maurer--Cartan super-1-form along $\,\hgt\,$ transform under the induced action as a connection 1-form for $\,\txG\too\txG/\txH$,\ whereas those along $\,\tgt\,$ undergo tensorial transformations under the compensating right $\txH$-translations. Consequently, linear combinations
\qq\nn
\widehat T\equiv T_{\unl A_1\unl A_2\ldots\unl A_p}\,\theta_{\rm L}^{\unl A_1}\wedge\theta_{\rm L}^{\unl A_2}\wedge\cdots\wedge\theta_{\rm L}^{\unl A_p}
\qqq
of (wedge products of) the latter with $\txH$-invariant tensors $\,T_{\unl A_1\unl A_2\ldots\unl A_p}\equiv T_{[\unl A_1\unl A_2\ldots\unl A_p]}\in\bC\,$ as coefficients can be used as building blocks of the sought-after supersymmetric lagrangean densities. Indeed, they are not only (right-)$\txH$-invariant (by construction), but also $\txH$-horizontal as
\qq\nn
\forall_{\unl A\in\ovl{1,\dim\,\tgt}}\ \forall_{Y\in\hgt}\ :\ \ceL_Y\con\theta_{\rm L}^{\unl A}=0\,,
\qqq
whence -- altogether -- $\txH$-basic, and so they descend to the orbit supermanifold $\,\txG/\txH$.\ It ought to be emphasised that their pullbacks to $\,\txG/\txH\,$ do \emph{not} depend on the choice of the local section $\,\si_i\,$ over $\,\cO_i\,$ (independently of the choice of the local coordinates $\,Z_i$) and hence, most importantly, glue smoothly over non-empty intersections $\,\cO_{ij}\,$ to \emph{global} superdifferential forms on $\,\txG/\txH\,$ in consequence of the tensorial properties of the $\,\theta_{\rm L}^{\unl A}\,$ with respect to right translations by elements of $\,\txH$.\void{The inherent redundancy of the realisation of the homogeneous space in $\,\txG\,$ in terms of sections of the principal $\txH$-bundle $\,\txG\too\txG/\txH\,$ does not affect the ensuing invariant differential calculus on $\,\txG/\txH$.\ Nevertheless, we may remove it for a \emph{given} family of local sections $\,\xcR=\{\si_i\}_{i\in I}\,$ covering the entire base $\,\txG/\txH\,$ ({\it i.e.}, for a complete family of local trivialisations of the principal $\txH$-bundle) by fixing a map
\qq\label{eq:trivindexmap}
\jmath_\xcR\ :\ \txG/\txH\too I
\qqq
with the property
\qq\nn
x\in\cO_{\jmath_\xcR(x)}\,.
\qqq
We shall employ this map in our later considerations.}\ Furthermore, and importantly, they satisfy, for arbitrary $\,X\in\ggt$,\ the identities
\qq\nn
\cK_{i\,X}\con\widehat T\equiv\cR_X\con\widehat T\,.
\qqq

We conclude this part of our discussion by taking a closer look at the pullbacks \eqref{eq:pullbackMC} along the distinguished sections
\qq\label{eq:embsec}
\si_i(Z_i)=\unl g_i\cdot g_i(X_i)\cdot\ee^{\Theta_i(Z_i)}\,,\qquad\qquad\Theta_i(Z_i)=\Theta_i^{\widehat\a}(Z_i)\,Q_{\widehat\a}
\qqq
with
\qq\nn
g_i(X_i)=\ee^{X_i^{\widehat a}\,P_{\widehat a}}
\qqq
and
\qq\nn
\Theta_i^{\widehat\a}(Z_i)=\theta_i^{\widehat\b}\,f_{i\,\widehat\b}^{\ \ \widehat\a}(X_i)
\qqq
in general depending upon the Gra\ss mann-even coordinate (through some functions $\,f_{i\,\widehat\b}^{\ \ \widehat\a}$). Here, $\,\unl g_i\in\txG\,$ defines a reference point $\,\unl g_i\,\txH\,$ in the neighbourhood $\,\cO_i\,$ on which the local coordinate system $\,Z_i\,$ is centred. In this setting, to be encountered in the sequel, explicit functional formul\ae ~for the restricted Vielbeine $\,E_{\unl A}^{\ A}(Z)\,$ are derived with the help of the retraction
\qq\nn
\si_{i\,\cdot}\ :\ \txG/\txH\x[0,1]\too\txG\ :\ (Z_i,t)\longmapsto \unl g_i\cdot g_i(X_i)\cdot\ee^{t\,\Theta_i(Z_i)}\equiv\si_{i\,t}(Z_i)
\qqq
of $\,\si_i(\cO_i)\,$ to the image of its body within $\,\txG$.\ The result is stated in
\berop\label{prop:Vielder}
Components of the pullback of the $\ggt$-valued Maurer--Cartan super-1-form along the local section \eqref{eq:embsec}, defined in \Reqref{eq:pullbackMC}, take the form
\qq\nn
\left(\begin{smallmatrix} E^{\widehat\a} \\ E^{\widehat a} \end{smallmatrix}\right)(Z_i)=\vec\cE_0(Z_i)+\tfrac{2\,{\rm sh}^2\tfrac{M_i(Z_i)}{2}}{M_i(Z_i)^2}\,\left(\begin{smallmatrix} 0 \\ [\sfD\Theta_i(Z_i),\Theta_i(Z_i)]^{\widehat a} \end{smallmatrix}\right)+\tfrac{{\rm sh}M_i(Z_i)}{M_i(Z_i)}\,\left(\begin{smallmatrix} \sfD\Theta_i^{\widehat\a}(Z_i) \\ 0 \end{smallmatrix}\right)\,,
\qqq
where
\qq\nn
\vec\cE_0(Z_i):=\left(\begin{smallmatrix} 0 \\ e^{\widehat a}(X_i) \end{smallmatrix}\right)
\qqq
is the value of the super-1-form at $\,\theta=0$,
\qq\nn
\sfD\Theta_i(Z_i)\equiv\sfD\Theta_i^A(Z_i)\ox t_A:=\bigl(\sfd\Theta_i^A(Z_i)+f_{\widehat a\widehat\b}^{\ \ A}\,e^{\widehat a}(X_i)\,\Theta_i^{\widehat\b}(Z_i)\bigr)\ox t_A
\qqq
and
\qq\nn
M_i(Z_i):=\left(\begin{smallmatrix} 0 & -\Theta_i^{\widehat\g}(Z_i)\,f_{\widehat\g b}^{\ \ \widehat\a} \\ \Theta_i^{\widehat\g}(Z_i)\,f_{\widehat\g\widehat\b}^{\ \ a} & 0 \end{smallmatrix}\right)
\qqq

\eerop
\beroof
A proof is given in App.\,\ref{app:Vielder}
\eroof

By now, we have all the tools necessary for the definition and subsequent study of the two-dimensional supersymmetric lagrangean field theory of interest, that is the two-dimensional Green--Schwarz super-$\si$-model of smooth embeddings\footnote{Here, we regard the target supermanifold $\,\txG/\txH\equiv\xcM\,$ as (the total space of) a Grassmann bundle of a vector bundle over a given base (body) $\,|\txG/\txH|$,\ in the spirit of the fundamental Gaw\c{e}dzki--Batchelor Theorem of Refs.\,\cite{Gawedzki:1977pb,Batchelor:1979a}.}
\qq\nn
\xi\ :\ \Om_2\too\txG/\txH
\qqq
of the compact worldsheet $\,\Om_2\,$ in the homogeneous space $\,\txG/\txH\,$ of the Lie supergroup $\,\txG$.\ The model uses components of the Maurer--Cartan super-1-form along $\,\tgt\,$ (contracted with suitable $\txH$-invariant tensors) which we pull back to the embedded worldsheet $\,\xi(\Om_2)\,$ along the previously considered family of (restrictions of) local sections of the principal $\txH$-bundle $\,\txG\too\txG/\txH\,$ of the (local-)coordinate form
\qq\nn
\si_i\ :\ \cO_i\too\txG\ :\ Z_i\equiv(\theta_i,X_i)\longmapsto\unl g_i\cdot g_i(X_i)\cdot\ee^{\Theta_i(Z_i)}\,,\qquad i\in I\,,
\qqq
supported over elements of the trivialising open cover $\,\cO=\{\cO_i\}_{i\in I}\,$ of the base $\,\txG/\txH\,$ of the principal bundle $\,\txG\too\txG/\txH$.\ Here, the right-hand side is to be understood as the unital(-time) flow of the group element $\,\unl g_i\,$ first along the integral lines of the left-invariant vector field engendered by the Lie-superalgebra element $\,X_i^{\widehat a}\,P_{\widehat a}\,$ and subsequently along the one associated with $\,\Theta_i^{\widehat\a}(Z_i)\,Q_{\widehat\a}$.\ Next, we take an arbitrary tesselation $\,\triangle(\Om_2)\,$ of $\,\Om_2\,$ subordinate, for a given map $\,\xi$,\ to the open cover $\,\cO$,\ as reflected by the existence of a map $\,i_\cdot\ :\ \triangle(\Om_2)\too I\,$ with the property
\qq\nn
\forall_{\z\in\triangle(\Om_2)}\ :\ \xi(\z)\subset\cO_{i_\z}\,.
\qqq
Let $\,\Pgt\subset\triangle(\Om_2)\,$ be the set of plaquettes of the tesselation,
\qq\nn
\Om_2=\bigcup_{\t\in\Pgt}\,\t\,.
\qqq
Given such data, we may write, in the Polyakov formulation,
\qq\label{eq:supersimod}
&S_{{\rm GS}}[\xi]=\sum_{\t\in\Pgt}\,S^{(\t)}_{{\rm GS}}[\xi_\t]\,,\qquad\qquad\xi_\t:=\xi\rstr_\t&\\ \cr
&S^{(\t)}_{{\rm GS}}[\xi_\t]=-\tfrac{1}{2}\,\int_\t\,\Vol(\Om_2)\,\sqrt{\vert\det\,\txg\vert}\,\txg^{-1\,ij}\,\txg_{\widehat a\widehat b}\,\bigl(\p_i\con(\si_{i_\t}\circ\xi_\t)^*\theta_{\rm L}^{\widehat a}\bigr)\,\bigl(\p_j\con(\si_{i_\t}\circ\xi_\t)^*\theta_{\rm L}^{\widehat b}\bigr)+S^{(\t)}_{{\rm WZ},{\rm MT}}[\xi_\t]\,,&\nn
\qqq
in which the Wess--Zumino term
\qq\nn
S^{(\t)}_{{\rm WZ},{\rm MT}}[\xi_\t]=\int_\t\,\xi_\t^*\sfd^{-1}\bigl(\si_{i_\t}^*\underset{\tx{\ciut{(3)}}}{\chi}^{\rm MT}\bigr)
\qqq
has as its integrand a global primitive of the relevant Green--Schwarz super-3-cocycle on $\,\txG/\txH\,$ with local restrictions
\qq\nn
\underset{\tx{\ciut{(3)}}}{\txH}\rstr_{\cO_i}\equiv\si_i^*\underset{\tx{\ciut{(3)}}}{\chi}\,,
\qqq
where $\,\underset{\tx{\ciut{(3)}}}{\chi}\,$ is a super-3-cocycle on $\,\txG\,$ defined as a linear combination
\qq\nn
\underset{\tx{\ciut{(3)}}}{\chi}=\g_{\unl A\unl B\unl C}\,\theta_{\rm L}^{\unl A}\wedge\theta_{\rm L}^{\unl B}\wedge \theta_{\rm L}^{\unl C}
\qqq
of the distinguished components of the Maurer--Cartan super-1-form along $\,\tgt\,$ with coefficients given by certain $\txH$-invariant tensors $\,\g_{\unl A\unl B\unl C}$.\ The latter super-3-cocycle is assumed to be a de Rham coboundary on the $\,\si_i(\cO_i)\,$ in what follows\footnote{In this manner, we isolate the difficulty in the construction of the Green--Schwarz super-$\si$-model resulting from the assumption of (nonlinearly realised) supersymmetry from that associated with the (de Rham-)cohomological non-triviality of a generic super-3-cocycle.} and we further presuppose the corresponding primitives $\,\underset{\tx{\ciut{(2)}}}{\txB}{}_i:=\sfd^{-1}(\si_i^*\underset{\tx{\ciut{(3)}}}{\chi})\,$ to form under the induced action of the supersymmetry group a pseudo-invariant family in the sense of the relation
\qq\label{eq:pseudolocB}
[\la]_g^*\underset{\tx{\ciut{(2)}}}{\txB}{}_j(x)=\underset{\tx{\ciut{(2)}}}{\txB}{}_i(x)+\sfd\underset{\tx{\ciut{(1)}}}{\D}{}^g_{ij}(x)
\qqq
valid for all $\,(g,x)\in\txG\x\cO_i$,\ for $\,j\in I\,$ such that $\,[\la]_g(x)\in\cO_j$,\ and for some $\,\underset{\tx{\ciut{(1)}}}{\D}{}^g_{ij}\in\Om^1(\cO_i)$,\ such that the action functional on the \emph{closed} worldsheet is effectively invariant under that action.

\section{Super-$\si$-model extensions of supertarget algebras by wrapping charges}\label{sec:wrapanom}

In the setting of the preceding section, with $\,\xcM\cong\txG/\txH\,$ realised within $\,\txG\,$ and in the notation
\qq\nn
\si(\txG/\txH):=\bigsqcup_{i\in I}\,\si_i(\cO_i)\,,
\qqq
assume given a $\txH$-basic representative
\qq\nn
\underset{\tx{\ciut{(p+2)}}}{\chi}\rstr_{\si(\txG/\txH)}=\sfd\underset{\tx{\ciut{(p+1)}}}{\b}
\qqq
of a class $\,[\underset{\tx{\ciut{(p+2)}}}{\chi}]\in{\rm CaE}^{p+2}(\txG)\,$ of the Cartan--Eilenberg cohomology of the Lie supergroup $\,\txG$ restricted to the image of $\,\txG/\txH\,$ within $\,\txG$,\ and -- further -- that the $\txH$-horizontal de Rham primitive $\,\underset{\tx{\ciut{(p+1)}}}{\b}\,$ of $\,\underset{\tx{\ciut{(p+2)}}}{\chi}\rstr_{\si(\txG/\txH)}\,$ (whose very existence is part of our assumptions) is \emph{pseudo}-invariant with respect to the transformations \eqref{eq:cosetact} of $\,\txG\,$ in the sense rendered precise by the identity
\qq\nn
\pLie{\cK_X}\underset{\tx{\ciut{(p+1)}}}{\b}=\sfd\underset{\tx{\ciut{(p)}}}{\G_X}\,,
\qqq
to be satisfied for the vector field $\,\cK_X\,$ on $\,\si(\txG/\txH)\,$ with restrictions $\,\cK_X\rstr_{\si_i(\cO_i)}:=\cK_{i\,X}\,$ associated with an arbitrary element $\,X\in\ggt$,\ and for some $\,\underset{\tx{\ciut{(p)}}}{\G_X}\in\Om^p(\si(\txG/\txH))$.\ The specific representative of the whole class of super-$p$-forms defined by the latter condition that we use hereunder is a \emph{choice}. In order to be able to proceed with our analysis, we consider henceforth, for the sake of transparency, (super-)linear combinations $\,X\,$ of the generators of $\,\ggt\,$ in which the coefficient in front of a given generator has the same Gra\ss man parity as the generator itself. Clearly, this does not, in any manner, affect the validity of our conclusions drawn from the ensuing analysis as the coefficients may always be removed at any stage of the analysis, with the sole effect that commutators become supercommutators, {\it e.g.}, $\,[\vep_1^{\widehat\a}\,Q_{\widehat\a},\vep_2^{\widehat\b}\,Q_{\widehat\b}]=-\vep_1^{\widehat\a}\,\vep_2^{\widehat\b}\,\{Q_{\widehat\a},Q_{\widehat\b}\}$.\ In this notation, we find, as a consequence of the pseudo-invariance, that the super-$p$-forms
\qq\nn
\underset{\tx{\ciut{(p)}}}{\a_{X_1,X_2}}:=\pLie{\cK_{X_1}}\underset{\tx{\ciut{(p)}}}{\G_{X_2}}-\pLie{\cK_{X_2}}\underset{\tx{\ciut{(p)}}}{\G_{X_1}}+\underset{\tx{\ciut{(p)}}}{\G_{[X_1,X_2]}}\,,
\qqq
with the $\,X_1,X_2\,$ given by the super-linear combinations described above, are closed, and so whenever $\,H^p_{\rm dR}(\txG/\txH)=0$,\ {\it i.e.}\footnote{We are invoking the Kostant Theorem of \Rcite{Kostant:1975}.}, whenever $\,H^p_{\rm dR}(\vert\txG/\txH\vert)=0$,\ there exist smooth super-$(p-1)$-forms $\,\sfd^{-1}\underset{\tx{\ciut{(p)}}}{\a_{X_1,X_2}}\,$ on the image of $\,\txG/\txH\,$ within $\,\txG\,$ such that
\qq\label{eq:gammaXdel}
\underset{\tx{\ciut{(p)}}}{\a_{X_1,X_2}}=\sfd\bigl(\sfd^{-1}\underset{\tx{\ciut{(p)}}}{\a_{X_1,X_2}}\bigr)\,.
\qqq

Consider, next, the previously defined Green--Schwarz super-$\si$-model with $\,\txG/\txH\,$ as the supertarget and with $\,\underset{\tx{\ciut{(p+2)}}}{\chi}\,$ as the Green--Schwarz super-$(p+2)$-cocycle. The corresponding (pre)symplectic form, derived in the first-order formalism of Refs.\,\cite{Gawedzki:1972ms,Kijowski:1973gi,Kijowski:1974mp,Kijowski:1976ze,Szczyrba:1976,Kijowski:1979dj} and evaluated on Cauchy data $\,(\si_{i_\cdot}\circ\unl\xi_\cdot,\pi_\cdot)\,$ (a degenerate canonical pair, with $\,\pi_{\cdot\,\widehat\a}\equiv 0$) of a classical configuration along a Cauchy hypersurface $\,\xcC_p\subset\Om_{p+1},\ \p\xcC_p=\emptyset$, with restrictions $\,\si_{i_\t}\circ\unl\xi_\t\equiv\si_{i_\t}\circ\xi\rstr_{\xcC_p\cap\t}\,$ and $\,\pi_\t\equiv\pi\rstr_{\xcC_p\cap\t}$,\ reads
\qq\label{eq:Omsip}\qquad\qquad
\Om^{({\rm NG})}_{{\rm GS},p}[\unl\xi,\pi]=\sum_{\t\in\Pgt}\,\bigg[\int_{\xcC_p\cap\t}\,\Vol(\xcC_p)\,\d\bigl(\pi_{\t\,\unl A}(\cdot)\,\theta^{\unl A}_{\rm L}\bigl(\si_{i_\t}\circ\unl\xi_\t(\cdot)\bigr)\bigr)+\int_{\xcC_p\cap\t}\,\ev^*\underset{\tx{\ciut{(p+2)}}}{\chi}\bigl(\si_{i_\t}\circ\unl\xi_\t(\cdot)\bigr)\bigg]\,,
\qqq
and so we find the canonical lifts of the fundamental vector fields
\qq\nn
\widetilde\cK_X[\unl\xi,\pi]=\sum_{\t\in\Pgt}\,\int_{\xcC_p\cap\t}\,\Vol(\xcC_p)\,\bigl[\cK_X\bigl(\si_{i_\t}\circ\unl\xi_\t(\cdot)\bigr)+\D_{\t\,\unl A}\bigl(X;\si_{i_\t}\circ\unl\xi_\t(\cdot)\bigr)\,\tfrac{\d\ }{\d\pi_{\t\,\unl A}}(\cdot)\bigr]\,,
\qqq
with the correction $\,\D_{\t\,\unl A}\,$ determined by the condition
\qq\label{eq:canlifunK}
\pLie{\widetilde\cK_X}\vartheta\must 0\,,\qquad\qquad\vartheta[\unl\xi,\pi]:=\sum_{\t\in\Pgt}\,\int_{\xcC_p\cap\t}\,\Vol(\xcC_p)\,\pi_{\t\,\unl A}(\cdot)\,\theta^{\unl A}_{\rm L}\bigl(\si_{i_\t}\circ\unl\xi_\t(\cdot)\bigr)\,.
\qqq
We have
\berop\label{prop:canlifunK}
The canonical lift defined by \Reqref{eq:canlifunK} admits a solution
\qq\nn
\D_{\t\,\unl A}\bigl(X;\si_{i_\t}\circ\unl\xi_\t(\cdot)\bigr)=\pi_{\t\,\unl B}(\cdot)\,\bigl(\ad_{Y_{i_\t}(X;\unl\xi_\t(\cdot))}\bigr)^{\unl B}_{\ \unl A}\,.
\qqq
\eerop
\beroof
A proof is given in App.\,\ref{app:canlifunK}.
\eroof
The Noether hamiltonians (charges) corresponding to the above-defined lifts read
\qq
&&h_X[\unl\xi,\pi]\cr\cr
&=&\sum_{\t\in\Pgt}\,\bigg[\int_{\xcC_p\cap\t}\,\Vol(\xcC_p)\,\pi_{\t\,\unl A}(\cdot)\,\bigl(\cK_X\con\theta^{\unl A}_{\rm L}\bigr)\bigl(\si_{i_\t}\circ\unl\xi_\t(\cdot)\bigr)+\int_{\xcC_p\cap\t}\,\ev^*\bigl(\cK_X\con\underset{\tx{\ciut{(p+1)}}}{\b}-\underset{\tx{\ciut{(p)}}}{\G_X}\bigr)\bigl(\si_{i_\t}\circ\unl\xi_\t(\cdot)\bigr)\bigg]\cr\cr
&\equiv&\sum_{\t\in\Pgt}\,\bigg[\int_{\xcC_p\cap\t}\,\Vol(\xcC_p)\,\pi_{\t\,\unl A}(\cdot)\,\bigl(\cR_X\con\theta^{\unl A}_{\rm L}\bigr)\bigl(\si_{i_\t}\circ\unl\xi_\t(\cdot)\bigr)+\int_{\xcC_p\cap\t}\,\ev^*\bigl(\cR_X\con\underset{\tx{\ciut{(p+1)}}}{\b}-\underset{\tx{\ciut{(p)}}}{\G_X}\bigr)\bigl(\si_{i_\t}\circ\unl\xi_\t(\cdot)\bigr)\bigg]\,.\cr && \label{eq:Noesusy}
\qqq
As we want these hamiltonians to be well-defined, independently of the arbitrary choices of the representatives $\,\underset{\tx{\ciut{(p+1)}}}{\b}\,$ and $\,\underset{\tx{\ciut{(p)}}}{\G_X}\,$ of the respective classes of superdifferential forms, we should replace the latter with
\qq\nn
\underset{\tx{\ciut{(p)}}}{\G_X}\longmapsto\underset{\tx{\ciut{(p)}}}{\G_X}+\cK_X\con\underset{\tx{\ciut{(p+1)}}}{\D\b}
\qqq
whenever the former is replaced by
\qq\nn
\underset{\tx{\ciut{(p+1)}}}{\b}\longmapsto\underset{\tx{\ciut{(p+1)}}}{\b}+\underset{\tx{\ciut{(p+1)}}}{\D\b}\,,
\qqq
with, of necessity,
\qq\nn
\underset{\tx{\ciut{(p+1)}}}{\D\b}\in Z^{p+1}_{\rm dR}(\txG)\,.
\qqq
In general, the hamiltonians are \emph{not} in involution under the Poisson bracket induced by $\,\Om^{({\rm NG})}_{{\rm GS},p}$.\ Instead, we establish
\berop\label{prop:PoiNoe}
The Poisson bracket of the Noether charges \eqref{eq:Noesusy} associated with the (pre)symplectic form \eqref{eq:Omsip} is given by the formula
\qq\nn
&&\{h_{X_1},h_{X_2}\}_{\Om^{({\rm NG})}_{{\rm GS},p}}[\unl\xi,\pi]\cr\cr
&=&h_{-[X_1,X_2]}[\unl\xi,\pi]+\sum_{\t\in\Pgt}\,\bigg[\int_{\xcC_p\cap\t}\,\ev^*\sfd\bigl(\cK_{X_1}\con\underset{\tx{\ciut{(p)}}}{\G_{X_2}}-\cK_{X_2}\con\underset{\tx{\ciut{(p)}}}{\G_{X_1}}+\cK_{X_2}\con\cK_{X_1}\con\underset{\tx{\ciut{(p+1)}}}{\b}\bigr)\bigl(\si_{i_\t}\circ\unl\xi_\t(\cdot)\bigr)\cr\cr
&&-\int_{\xcC_p\cap\t}\,\ev^*\underset{\tx{\ciut{(p)}}}{\a_{X_1,X_2}}\bigl(\si_{i_\t}\circ\unl\xi_\t(\cdot)\bigr)\bigg]\,.
\qqq
\eerop
\beroof
A proof is given in App.\,\ref{app:PoiNoe}.
\eroof
From the above, we read off possible \emph{classical} field-theoretic corrections to the supertarget Lie superalgebra $\,\ggt$:\ First of all, the corrections vanish \emph{on the level of the Poisson algebra of Noether charges}\footnote{They survive on the level of the underlying Poisson algebra of Noether currents.} for all field configurations with a vanishing monodromy around the embedded Cauchy $p$-cycle $\,\bigcup_{\tau\in\Pgt}\,\si_{i_\t}\circ\unl\xi_\t(\t\cap\xcC_p)$.\ In particular, they are -- apparently -- absent whenever $\,H_p(\txG/\txH)=0$.\ Thus, the corrections are sourced by field configurations wrapping non-contractible $p$-cycles in $\,\txG/\txH\,$ -- the charges are none other than the wrapping ({\it e.g.}, winding) numbers of the classical configuration $\,\unl\xi$.\ Secondly, we may engineer such corrections without affecting the topology of the body $\,\vert\txG/\txH\vert\,$ of the supertarget $\,\txG/\txH\,$ by considering the Green--Schwarz super-$\si$-model -- in conformity with the interpretation of the Chevalley--Eilenberg cohomology proposed by Rabin and Crane in \Rcite{Rabin:1984rm} -- as a model of super-$p$-brane dynamics on the quotient of $\,\txG/\txH\,$ by the discrete Kosteleck\'y--Rabin supersymmetry group of \Rcite{Kostelecky:1983qu}, and -- consequently -- by taking into account field configurations with a non-vanishing integral monodromy in the Gra\ss mann-odd directions, {\it i.e.}, the twisted sector of the super-$\si$-model on the quotient. In what follows, this will serve to demonstrate, quite naturally, the wrapping nature of the charges that define the physically relevant deformations ({\it e.g.}, super-central extensions) of the supertarget algebras on $\,{\rm sMink}^{d,1\,\vert\,D_{d,1}}\,$ and $\,{\rm s}({\rm AdS}_5\x\bS^5)\,$ encoded by the Green--Schwarz super-$(p+2)$-cocycles, and through these -- also the associated Cartan--Eilenberg super-$p$-gerbes. To this end, we shall study at length the specific examples in which the said super-central extensions are known ({\it cp} Part I) to integrate to surjective submersions of the super-$p$-gerbes, {\it i.e.}, those with $\,p\in\{0,1\}\,$ on $\,{\rm sMink}^{d,1\,\vert\,D_{d,1}}$.\ We shall also consider, from this point of view, geometrisations of the Metsaev--Tseytlin super-3-cocycle defining the superstring model on $\,{\rm s}({\rm AdS}_5\x\bS^5)$.\ In so doing, we shall employ the computationally less cumbersome\footnote{Note that we merely have to compute the supersymmetry-variation super-$p$-forms $\,\G_X\,$ for $\,X\in[\ggt,\ggt\}$.} formula for the wrapping anomaly:
\qq\nn
\xcW_{X_1,X_2}[\unl\xi]&:=&\sum_{\t\in\Pgt}\,\int_{\xcC_p\cap\t}\,\ev^*\bigl(\sfd\bigl(\cK_{X_1}\con\underset{\tx{\ciut{(p)}}}{\G_{X_2}}-\cK_{X_2}\con\underset{\tx{\ciut{(p)}}}{\G_{X_1}}+\cK_{X_2}\con\cK_{X_1}\con\underset{\tx{\ciut{(p+1)}}}{\b}\bigr)-\underset{\tx{\ciut{(p)}}}{\a_{X_1,X_2}}\bigr)\bigl(\si_{i_\t}\circ\unl\xi_\t(\cdot)\bigr)\cr\cr
&=&\sum_{\t\in\Pgt}\,\int_{\xcC_p\cap\t}\,\ev^*\bigl(\bigl(\pLie{\cK_{X_1}}\underset{\tx{\ciut{(p)}}}{\G_{X_2}}-\pLie{\cK_{X_2}}\underset{\tx{\ciut{(p)}}}{\G_{X_1}}-\underset{\tx{\ciut{(p)}}}{\a_{X_1,X_2}}+\underset{\tx{\ciut{(p)}}}{\G_{[X_1,X_2]}}\bigr)\cr\cr
&&\hspace{-.5cm}-\cK_{X_1}\con\pLie{\cK_{X_2}}\underset{\tx{\ciut{(p+1)}}}{\b}+\cK_{X_2}\con\pLie{\cK_{X_1}}\underset{\tx{\ciut{(p+1)}}}{\b}+\sfd\bigl(\cK_{X_2}\con\cK_{X_1}\con\underset{\tx{\ciut{(p+1)}}}{\b}\bigr)-\underset{\tx{\ciut{(p)}}}{\G_{[X_1,X_2]}}\bigr)\bigl(\si_{i_\t}\circ\unl\xi_\t(\cdot)\bigr)\cr\cr
&=&\sum_{\t\in\Pgt}\,\int_{\xcC_p\cap\t}\,\ev^*\bigl(-\cK_{X_1}\con\pLie{\cK_{X_2}}\underset{\tx{\ciut{(p+1)}}}{\b}+\cK_{X_2}\con\pLie{\cK_{X_1}}\underset{\tx{\ciut{(p+1)}}}{\b}+\pLie{\cK_{X_2}}\bigl(\cK_{X_1}\con\underset{\tx{\ciut{(p+1)}}}{\b}\bigr)\cr\cr
&&-\cK_{X_2}\con\pLie{\cK_{X_1}}\underset{\tx{\ciut{(p+1)}}}{\b}+\cK_{X_2}\con\cK_{X_1}\con\underset{\tx{\ciut{(p+2)}}}{\chi}-\underset{\tx{\ciut{(p)}}}{\G_{[X_1,X_2]}}\bigr)\bigl(\si_{i_\t}\circ\unl\xi_\t(\cdot)\bigr)\cr\cr
&=&\sum_{\t\in\Pgt}\,\int_{\xcC_p\cap\t}\,\ev^*\bigl(\cK_{X_2}\con\cK_{X_1}\con\underset{\tx{\ciut{(p+2)}}}{\chi}+\cK_{[X_1,X_2]}\con\underset{\tx{\ciut{(p+1)}}}{\b}-\underset{\tx{\ciut{(p)}}}{\G_{[X_1,X_2]}}\bigr)\bigl(\si_{i_\t}\circ\unl\xi_\t(\cdot)\bigr)\,.
\qqq
Taking into account the $\txH$-horizontality of the forms involved, we may further reduce the above to
\qq\nn
\xcW_{X_1,X_2}[\unl\xi]=\sum_{\t\in\Pgt}\,\int_{\xcC_p\cap\t}\,\ev^*\bigl(\cR_{X_2}\con\cR_{X_1}\con\underset{\tx{\ciut{(p+2)}}}{\chi}+\cR_{[X_1,X_2]}\con\underset{\tx{\ciut{(p+1)}}}{\b}-\underset{\tx{\ciut{(p)}}}{\G_{[X_1,X_2]}}\bigr)(\si_{i_\t}\circ\unl\xi_\t(\cdot)\bigr)\,.
\qqq
It deserves to be emphasised that the anomaly thus defined does \emph{not} depend on the choice of the primitive $\,\underset{\tx{\ciut{(p+1)}}}{\b}\,$ of the Green--Schwarz super-$(p+2)$-cocycle $\,\underset{\tx{\ciut{(p+2)}}}{\chi}\,$ and of the corresponding variance super-$p$-form $\,\underset{\tx{\ciut{(p)}}}{\G_X}\,$ for precisely the same reason as that for the well-definedness of the hamiltonians.

Let us extract from the above formula the contribution to the wrapping anomaly of a \emph{supersymmetric} (component of a) primitive $\,\underset{\tx{\ciut{(p+1)}}}{\b}\,$ obtained through restriction from a super-$(p+1)$-form on $\,\txG$,\ which we write in the form
\qq\nn
\underset{\tx{\ciut{(p+1)}}}{\b}=\theta_{\rm L}^{A_1}\wedge\theta_{\rm L}^{A_2}\wedge\cdots\wedge\theta_{\rm L}^{A_{p+1}}\,b_{A_1 A_2\ldots A_{p+1}}\rstr_{\si(\txG/\txH)}\,,
\qqq
with functional coefficients $\,b_{A_1 A_2\ldots A_{p+1}}\,$ (of Gra\ss mann parity $\,\sum_{k=1}^{p+1}\,|A_k|$). As
\qq\nn
\forall_{X\in\ggt} \ :\ \pLie{\cK_X}\underset{\tx{\ciut{(p+1)}}}{\b}=0\,,
\qqq
we may take
\qq\nn
\underset{\tx{\ciut{(p)}}}{\G_X}=0=\underset{\tx{\ciut{(p)}}}{\a_{X_1,X_2}}\,.
\qqq
We then obtain
\qq\label{eq:wrappaninvB}
\xcW^{({\rm inv})}_{X_1,X_2}[\unl\xi]=\sum_{\t\in\Pgt}\,\int_{\xcC_p\cap\t}\,\ev^*\sfd\bigl(\cK_{X_2}\con\cK_{X_1}\con\underset{\tx{\ciut{(p+1)}}}{\b}\bigr)\bigl(\si_{i_\t}\circ\unl\xi_\t(\cdot)\bigr)
\qqq
Upon invoking \Reqref{eq:leftvarfiel}, we may render the last expression even more explicit, to wit,
\qq\nn
&&\xcW_{X_1,X_2}^{({\rm inv})}[\unl\xi]\cr\cr
&=&p(p+1)\,\sum_{\t\in\Pgt}\,\int_{\xcC_p\cap\t}\,\ev^*\sfd\bigl(\bigl(\cK_{X_1}\con\theta_{\rm L}^A\bigr)\,\bigl(\cK_{X_2}\con\theta_{\rm L}^B\bigr)\,\theta_{\rm L}^{A_1}\wedge\theta_{\rm L}^{A_2}\wedge\cdots\wedge\theta_{\rm L}^{A_{p-1}}\,b_{A B A_1 A_2\ldots A_{p-1}}\bigr)\bigl(\si_{i_\t}\circ\unl\xi_\t(\cdot)\bigr)\cr\cr
&=&p(p+1)\,\sum_{\t\in\Pgt}\,\int_{\xcC_p\cap\t}\,\ev^*\sfd\bigl(\bigl(\d_{X_1}Z^{\unl A}\,E_{\unl A}^A\bigr)\,\bigl(\d_{X_2}Z^{\unl B}\,E_{\unl B}^B\bigr)\,\theta_{\rm L}^{A_1}\wedge\theta_{\rm L}^{A_2}\wedge\cdots\wedge\theta_{\rm L}^{A_{p-1}}\,b_{A B A_1 A_2\ldots A_{p-1}}\bigr)\bigl(\si_{i_\t}\circ\unl\xi_\t(\cdot)\bigr)\,.
\qqq
Thus, whenever $\,\underset{\tx{\ciut{(p+1)}}}{\b}\,$ is $\txH$-horizontal,
\qq\label{eq:Binvhor}
\underset{\tx{\ciut{(p+1)}}}{\b}=\theta_{\rm L}^{\unl A_1}\wedge\theta_{\rm L}^{\unl A_2}\wedge\cdots\wedge\theta_{\rm L}^{\unl A_{p+1}}\,b_{\unl A_1\unl A_2\ldots\unl A_{p+1}}\rstr_{\si(\txG/\txH)}\,,
\qqq
with, this time, constant $\txH$-invariant tensors $\,b_{\unl A_1\unl A_2\ldots\unl A_{p+1}}\,$ as coefficients,
we have
\qq\nn
\xcW_{X_1,X_2}^{({\rm inv})}[\unl\xi]&=&p(p+1)\,\sum_{\t\in\Pgt}\,\int_{\xcC_p\cap\t}\,\ev^*\sfd\bigl(\sfT_e\Ad_{\si_{i_\t}\circ\unl\xi_\t(\cdot)^{-1}}(X_1)^{\unl A}\,\sfT_e\Ad_{\si_{i_\t}\circ\unl\xi_\t(\cdot)^{-1}}(X_2)^{\unl B}\cr\cr
&&\hspace{3.75cm}\theta_{\rm L}^{\unl A_1}\wedge\theta_{\rm L}^{\unl A_2}\wedge\cdots\wedge\theta_{\rm L}^{\unl A_{p-1}}\bigl(\si_{i_\t}\circ\unl\xi_\t(\cdot)\bigr)\,b_{\unl A\unl B\unl A_1\unl A_2\ldots\unl A_{p-1}}\bigr)\,.
\qqq
A word of comment is due at this point. Given that the $\,\cK_{i\,X_\a},\ \a\in\{1,2\}\,$ are the local pushforwards \eqref{eq:KXaspfXiX} of the corresponding fundamental vector fields $\,\Xi_{X_\a}\,$ on $\,\txG/\txH\,$ and that the primitive $\,\underset{\tx{\ciut{(p+1)}}}{\b}\,$ of \Reqref{eq:Binvhor} is the pullback, along $\,\pi_{\txG/\txH}$,\ of a globally smooth super-$(p+1)$-form $\,\underset{\tx{\ciut{(p+1)}}}{\txB}\,$ on $\,\txG/\txH$,\ we readily see that the local super-$(p-1)$-form 
\qq\nn
\si_i^*\bigl(\cK_{X_2}\con\cK_{X_1}\con\underset{\tx{\ciut{(p+1)}}}{\b}\bigr)
\qqq
is, in fact, a (locally continuous) restriction of the super-$(p-1)$-form 
\qq\nn
\txB_{X_1,X_2}\equiv\Xi_{X_2}\con\Xi_{X_1}\con\underset{\tx{\ciut{(p+1)}}}{\txB}
\qqq
to $\,\cO_i$.\ This observation justifies our former identification of the anomaly $\,\xcW_{X_1,X_2}^{({\rm inv})}\,$ as the wrapping charge trapped by a non-contractible Cauchy $p$-cycle $\,\xi(\xcC_p)\,$ to which the latter super-$(p-1)$-form couples, or -- more formally -- the monodromy of $\,\txB_{X_1,X_2}\,$ along $\,\xi(\xcC_p)$.

In particular for $\,p=1\,$ we readily establish that the presence of the wrapping anomaly indicates the physical relevance of a deformation of the original Lie superalgebra $\,\ggt\,$ of the general structure
\qq\label{eq:wrapcentrext}
[t_A,t_B\}^\sim=f_{AB}^{\ \ C}\,t_C+Z_{AB}\,,\qquad\qquad Z_{BA}=-(-1)^{|A|\cdot|B|}\,Z_{AB}\qquad A,B\in\ovl{1,\dim\,\ggt}\,,
\qqq
where the (wrapping) charges $\,Z_{AB}\,$ are induced by the monodromies
\qq\label{eq:monofAB}
\mu\bigl(f_{AB};\xi(\xcC_1)\bigr)
\qqq
around $\,\xi(\xcC_1),\ \xcC_1\cong\bS^1$,\ of functions $\,f_{AB}\,$ with local representations 
\qq\label{eq:fAB}
f_{AB}\rstr_{\cO_i}\equiv(-1)^{|\unl A|\cdot |B|+1}\,2\bigl(\sfT_e\Ad_{\si_i(\cdot)^{-1}}\bigr)^{\unl A}_{\ A}\,\bigl(\sfT_e\Ad_{\si_i(\cdot)^{-1}}\bigr)^{\unl B}_{\ B}\,b_{\unl A\unl B}\bigl(\si_i(\cdot)\bigr)\,.
\qqq
The deformation may now take on the form of a super-central or some more general \emph{associative}\footnote{We do not consider non-associative deformations in which the super-Jacobi identity might fail. The lack of associativity renders the Cartan calculus, central to our considerations, ill-defined. While the alternative of working with an algebra loop with inverses seems an interesting theoretical possibility, we do not consider it in what follows, and, instead, \emph{impose} the super-Jacobi identity in the deformed algebra with the given, physically motivated `germ'.} extension of $\,\ggt\,$ with the `germ' \eqref{eq:wrapcentrext}. The idea behind it, advocated in \Rcite{Suszek:2017xlw}, is a trivialisation, in the spirit of Prop.\,I.C.4, of the class $\,[\varpi_p]\in{\rm CE}^2(\txG)\,$ in the Chevalley--Eilenberg cohomology associated with it as per
\qq\nn
\varpi_p(\ceL_{X_1},\ceL_{X_2})=\xcW_{X_1,X_2}\,.
\qqq
This is attained on the Lie supergroup $\,\widetilde\txG\,$ that integrates the new Lie superalgebra with the vector-space structure
\qq\nn
\widetilde\ggt:=\ggt\oplus\corr{Z_{AB}\,\vert\,A,B\in\ovl{1,\dim\,\ggt}}_\bC\equiv\ggt\oplus\zgt
\qqq
and a Lie superbracket
\qq\nn
[\cdot,\cdot\}^\sim\ :\ \widetilde\ggt\x\widetilde\ggt\too\widetilde\ggt
\qqq
determined (whenever possible, and then typically non-uniquely) by the imposition of the super-Jacobi constraints on the superalgebra with the `germ' specified. Given the straightforward geometric and physical interpretation of the Lie subalgebra $\,\txH\,$ which we want to preserve under the extension, we are led to constrain the admissible deformations so that the commutation relations of its elements with the rest of $\,\ggt\subset\widetilde\ggt\,$ remain unchanged,
\qq\nn
\forall_{(\k,A)\in\ovl{1,\dim\,\hgt}\x\ovl{1,\dim\,\ggt}}\ :\ [J_\k,t_A]^\sim=f_{\k A}^{\ \ \ B}\,t_B\,.
\qqq
Moreover, we take the charges $\,Z_{AB}\,$ to transform linearly under the isotropy group $\,\txH$,\ as do the remaining generators. In particular, $\,\zgt\,$ is assumed to be an $\ad$-module of the Lie algebra $\,\hgt$,
\qq\nn
[\hgt,\zgt]^\sim\subset\zgt\,,
\qqq
so that the new decomposition
\qq\nn
\widetilde\ggt=\widetilde\tgt\oplus\hgt\,,\qquad\qquad\widetilde\tgt\equiv\tgt\oplus\zgt
\qqq
is reductive just as the original one. Accordingly, we may repeat the previous constructions over the new Lie supergroup which we, once again, regard as the total space of the principal $\txH$-bundle
\qq\nn
\alxydim{@C=1cm@R=1cm}{\txH \ar[r] & \widetilde\txG \ar[d]^{\pi_{\widetilde\txG/\txH}} \\ & \widetilde\txG/\txH}\,.
\qqq
Its base, given by the homogeneous space $\,\widetilde\txG/\txH$,\ is locally charted by flows of the left-invariant vector fields on $\,\widetilde\txG\,$ along $\,\widetilde\tgt$,\ whence the coordinates $\,(Z_i^{\unl A},\z_i^{AB})$,\ where the $\,Z_i^{\unl A}\,$ are as before and where the new ones, $\,\z_i^{AB}\equiv-(-1)^{|A|\cdot|B|}\,\z_i^{BA}$,\ of the same Gra\ss mann parity as the corresponding (independent) charges, $\,|\z_i^{AB}|\equiv |Z_{AB}|=|A|+|B|\,$ correspond to the left-invariant vector fields from $\,\zgt$.\ It is realised within $\,\widetilde\txG\,$ by means of a collection of the distinguished local sections (the last sum is over the independent charges)
\qq\nn
\widetilde\si_i\ :\ \widetilde\cO_i\too\widetilde\txG\ :\ \widetilde Z_i\equiv(Z_i,\z)\longmapsto\si_i(Z_i)\cdot \ee^{\z_i^{AB}\,Z_{AB}}\,.
\qqq
Here, by a mild abuse of the notation, we take the $\,\si_i\,$ to denote the same products of flows of right-invariant vector fields (along $\,\tgt^{(0)}\,$ and $\,\tgt^{(1)}$) as before, but with the understanding that they are now subject to the deformed structure equations ({\it i.e.}, $\,\tgt^{(0)},\tgt^{(1)}\subset\widetilde\ggt$) and initiate at some $\,\widetilde{\unl g}_i\in\widetilde\txG$.\ The choice of the embedding sections is the first step towards a reconstruction, in the same spirit as over $\,\txG/\txH$,\ of the differential calculus over the new homogeneous space $\,\widetilde\txG/\txH$.\ Inspection of the Schur--Poincar\'e formula for the fundamental Maurer--Cartan super-1-form,
\qq\nn
\widetilde\si_i^*\theta_{\rm L}(\widetilde Z_i)&=&\sum_{k,l=0}^\infty\,\tfrac{(-1)^{k+l}}{k!(l+1)!}\,\bigl(\id_{\sfT^*\cO_i}\ox\widetilde\ad^k_{\z_i^{AB}\,Z_{AB}}\bigr)\bigl(\sfd X_i^{\widehat a}\ox\sum_{m=0}^\infty\,\tfrac{(-1)^m}{m!}\,\widetilde\ad^m_{\Theta_i^{\widehat\b}(Z_i)\,Q_{\widehat\b}}\circ\widetilde\ad^l_{X_i^{\widehat b}\,P_{\widehat b}}(P_{\widehat a})\cr\cr
&&+\sfd\Theta_i^{\widehat\a}(Z_i)\ox\widetilde\ad^l_{\Theta_i^{\widehat\b}(Z_i)\,Q_{\widehat\b}}(Q_{\widehat\a})\bigr)+\sum_{m=0}^\infty\,\tfrac{(-1)^m}{(m+1)!}\,\sfd\z_i^{AB}\ox\widetilde\ad_{\z_i^{CD}\,Z_{CD}}^m(Z_{AB})\,,
\qqq
in conjunction with the Baker--Campbell--Dynkin--Hausdorff formula, helps to determine the admissible structure of the deformation $\,\widetilde\ggt$,\ with the `germ' constrained by the foregoing analysis in the form
\qq\nn
&\{Q_{\widehat\a},Q_{\widehat\b}\}^\sim=f_{\widehat\a\widehat\b}^{\ \ \ c}\,B_c+Z_{\widehat\a\widehat\b}\,,\qquad\qquad[Q_{\widehat\a},P_{\widehat a}]^\sim= f_{\widehat\a\widehat a}^{\ \ \ \widehat\b}\,Q_{\widehat\b}+Z_{\widehat\a\widehat a}\,,\qquad\qquad[P_{\widehat a},P_{\widehat b}]^\sim=f_{\widehat a\widehat b}^{\ \ \ c}\,B_c+Z_{\widehat a\widehat b}\,,&\cr\cr
&[J_\k,Z_{\widehat\a\widehat\b}]^\sim=\La_{\k\,\widehat\a\widehat\b}^{\ \ \ \ \widehat\g\widehat\d}\,Z_{\widehat\g\widehat\d}\,,\qquad\qquad[J_\k,Z_{\widehat\a\widehat a}]^\sim=\La_{\k\,\widehat\a\widehat a}^{\ \ \ \ \widehat\b\widehat b}\,Z_{\widehat\b\widehat b}&
\qqq
for some $\,\La_{\k\,\widehat\a\widehat\b}^{\ \ \ \widehat\g\widehat\d},\La_{\k\,\widehat\a\widehat a}^{\ \ \ \widehat\b\widehat b}\in\bC$,\ where $\,Z_{\widehat\a\widehat\b}\equiv Z_{\widehat\b\widehat\a},Z_{\widehat a\widehat b}=-Z_{\widehat b\widehat a}\in\zgt^{(0)}\,$ and $\,Z_{\widehat\a\widehat a}\in\zgt^{(1)}$,\ and where all other supercommutators are as in $\,\ggt$.\ Indeed, a deformation $\,\widetilde\ggt\,$ with $\,[\widetilde\ggt,\zgt\}^\sim\cap\ggt\neq 0\,$ leads to an alteration of the coordinate expressions for the components of the Maurer--Cartan super-1-form along $\,\tgt\,$ entering the definition of the original physical model, and also for the induced supersymmetry on $\,\txG/\txH$,\ which is physically untenable, given that it is the canonical analysis of the original model that sources the deformation in the first place. Therefore, we are led to assume that the deformation $\,\widetilde\ggt\,$ is, in fact, \textbf{normal} in the sense expressed by the identities
\qq\nn
[\widetilde\ggt,\zgt\}^\sim\subset\zgt\,.
\qqq
Let us denote the independent generators of the supervector space
\qq\nn
\zgt=\zgt^{(0)}\oplus\zgt^{(1)}
\qqq
as
\qq\nn
\zgt^{(0)}=\bigoplus_{\widetilde a=1}^{\dim\,\zgt^{(0)}}\,\corr{Z_{\widetilde a}}_\bC\,,\qquad\qquad\zgt^{(1)}=\bigoplus_{\widetilde\a=1}^{\dim\,\zgt^{(1)}}\,\corr{Z_{\widetilde\a}}_\bC\,.
\qqq
The Lie super-brackets of the normal extension $\,\widetilde\ggt\,$ of $\,\ggt\,$ may now be cast in the form
\qq\nn
&[P_{\widehat a},P_{\widehat b}]^\sim=f_{\widehat a\widehat b}^{\ \ \widehat c}\,P_{\widehat c}+f_{\widehat a\widehat b}^{\ \ \k}\,J_\k+Z_{\widehat a\widehat b}\,,\qquad\qquad[J_\k,J_\la]^\sim=f_{\k\la}^{\ \ \mu}\,J_\mu\,,&\cr\cr
&[J_\k,P_{\widehat a}]^\sim=f_{\k\widehat a}^{\ \ \widehat b}\,P_{\widehat b}\,,\qquad\qquad[J_\k,Q_{\widehat\a}]^\sim=f_{\k\widehat\a}^{\ \ \widehat\b}\,Q_{\widehat\b}\,,&\cr\cr
&\{Q_{\widehat\a},Q_{\widehat\b}\}^\sim=f_{\widehat\a\widehat\b}^{\ \ c}\,B_c+\D_{\widehat\a\widehat\b}^{\ \ \ \widetilde a}\,Z_{\widetilde a}\,,\qquad\qquad[Q_{\widehat\a},P_{\widehat a}]^\sim= f_{\widehat\a\widehat a}^{\ \ \widehat\b}\,Q_{\widehat\b}+\D_{\widehat\a\widehat a}^{\ \ \ \widetilde\b}\,Z_{\widetilde\b}\,,&\cr\cr
&[Q_{\widehat\a},Z_{\widetilde a}]^\sim=\D_{\widehat\a\widetilde a}^{\ \ \ \widetilde\b}\,Z_{\widetilde\b}\,,\qquad\qquad[P_{\widehat a},Z_{\widetilde b}]^\sim=\D_{\widehat a\widetilde b}^{\ \ \ \widetilde c}\,Z_{\widetilde c}\,,&\cr\cr
&\{Q_{\widehat\a},Z_{\widetilde\b}\}^\sim=\D_{\widehat\a\,\widetilde\b}^{\ \ \ \widetilde c}\,Z_{\widetilde c}\,,\qquad\qquad[P_{\widehat a},Z_{\widetilde\a}]^\sim=\D_{\widehat a\widetilde\a}^{\ \ \ \widetilde\b}\,Z_{\widetilde\b}\,,&\cr\cr
&[J_\k,Z_{\widetilde a}]^\sim=\D_{\k\widetilde a}^{\ \ \ \widetilde b}\,Z_{\widetilde b}\,,\qquad\qquad[J_\k,Z_{\widetilde\a}]^\sim=\D_{\k\widetilde\a}^{\ \ \ \widetilde\b}\,Z_{\widetilde\b}&
\qqq
for some $\,\D_{\widehat\a\widehat\b}^{\ \ \ \widetilde a},\D_{\widehat\a\widehat a}^{\ \ \ \widetilde\b},\D_{\widehat\a\widetilde a}^{\ \ \ \widetilde\b},\D_{\widehat a\widetilde b}^{\ \ \ \widetilde c},\D_{\widehat\a\widetilde\b}^{\ \ \ \widetilde c},\D_{\widehat a\widetilde\a}^{\ \ \ \widetilde\b},\D_{\k\widetilde a}^{\ \ \ \widetilde b},\D_{\k\widetilde\a}^{\ \ \ \widetilde\b}\in\bC$.\ Upon calculating the relevant commutators in the previous formula for the Maurer--Cartan super-1-form, we find the general structure
\qq
\widetilde\si_i^*\theta_{\rm L}(Z_i,\z)-\si_i^*\theta_{\rm L}^A(Z_i)\ox t_A&=&\bigl(e_{\widehat a}^{\ \widetilde a}(Z_i,\z_i)\,\sfd X_i^{\widehat a}+e_{\widehat\b}^{\ \widetilde a}(Z_i,\z_i)\,\sfd\theta_i^{\widehat\b}+e_{\widetilde b}^{\ \widetilde a}(Z_i,\z_i)\,\sfd\z_i^{\widetilde b}\bigr)\ox Z_{\widetilde a}\cr\cr
&&+\bigl(e_{\widehat a}^{\ \widetilde\a}(Z_i,\z_i)\,\sfd X_i^{\widehat a}+e_{\widehat\b}^{\ \widetilde\a}(Z_i,\z_i)\,\sfd\theta_i^{\widehat\b}+e_{\widetilde b}^{\ \widetilde\a}(Z_i,\z_i)\,\sfd\z_i^{\widetilde b}\bigr)\ox Z_{\widetilde\a}\,,\label{eq:MCsformext}
\qqq
where the second (subtracted) term on the left-hand side is identical (in the functional sense) with its counterpart on $\,\cO_i\,$ derived for $\,Z_{\widetilde a}=0=Z_{\widetilde\a}$,\ and where the nontrivial Vielbeine $\,e_x^{\widetilde a}\,$ and $\,e_x^{\widetilde\a},\ x\in\{\widehat a,\widehat\b,\widetilde b\}\,$ in the component of the super-1-form along $\,\zgt\,$ reflect the deformation.

\brem 
Our hitherto analysis gives us a particular choice of a surjective submersion over the original target superspace $\,\txG/\txH\,$ of the super-$\si$-model, to wit,
\qq
\pi_{\sfY(\txG/\txH)}\ &:&\ \sfY(\txG/\txH):=\bigsqcup_{i\in I}\,\widetilde\cO_i\too\txG/\txH\cr\cr
&:&\ \bigl(\pi_{\widetilde\txG/\txH}\circ\widetilde\si_j(\widetilde Z_j),j\bigr)\equiv(Z_j,\z_j,j)\longmapsto Z_j\equiv\pi_{\txG/\txH}\circ\si_j(Z_j)\,.\label{eq:surjsubmext}
\qqq
It is this surjective submersion that we might take as the point of departure of a construction of the geometric object (a super-gerbe) over $\,\txG/\txH\,$ presenting, in a manner that generalises the previously considered mechanism of geometrisation of Green--Schwarz super-$(p+2)$-cocycles on Lie supergroups to the setting of their homogeneous spaces, the supersymmetric Green--Schwarz super-$(p+2)$-cocycle $\,\underset{\tx{\ciut{(p+2)}}}{\txH}\,$ on $\,\txG/\txH\,$ pulling back to a given super-$(p+2)$-cocycle $\,\underset{\tx{\ciut{(p+2)}}}{\chi}\,$ on $\,\txG\,$ as per 
\qq\nn
\underset{\tx{\ciut{(p+2)}}}{\chi}=\pi_{\txG/\txH}^*\underset{\tx{\ciut{(p+2)}}}{\txH}\,.
\qqq
Instrumental in this construction is the non-linear realisation of the extended supersymmetry group $\,\widetilde\txG\,$ on the homogeneous space $\,\widetilde\txG/\txH\,$ with a definition, using the local sections $\,\widetilde\si_i$,\ fully analogous to that of the non-linear realisation of the original supersymmetry group $\,\txG\,$ on $\,\txG/\txH\cong\xcM$.\ In consequence of the assumptions made as to the nature of the extension $\,\widetilde\ggt\too\ggt$,\ integrating to a Lie-supergroup extension $\,\widetilde\pi\ :\ \widetilde\txG\too\txG$,\ the realisation employs the same locally smooth mappings $\,\unl h_{ij}\,$ as before, {\it cp} \Reqref{eq:cosetact}. With these tools in hand, we may subsequently examine invariance of the various components of the $\widetilde\ggt$-valued Maurer--Cartan super-1-forms on $\,\widetilde\txG\,$ restricted to the image of $\,\widetilde\txG/\txH\,$ within $\,\widetilde\txG\,$ under the family $\,\{\widetilde\si_i\}_{i\in I}\,$ of sections, whereby we conclude that it is the linear combinations of wedge products of components along $\,\widetilde\tgt\,$ with $\txH$-invariant tensors as coefficients that ought to be considered, upon pullback to $\,\sfY(\txG/\txH)\,$ (along the $\,\widetilde\si_i$), as admissible trivialisations of the pullback of the Green--Schwarz super-$(p+2)$-cocycles along $\,\pi_{\sfY(\txG/\txH)}$.\ The first step in the geometrisation would therefore consist in finding such a global primitive $\,\underset{\tx{\ciut{(p+1)}}}{\widetilde\b}\,$ of the pullback of $\,\underset{\tx{\ciut{(p+2)}}}{\chi}\,$ to the extended supersymmetry group $\,\widetilde\txG$,
\qq\nn
\widetilde\pi^*\underset{\tx{\ciut{(p+2)}}}{\chi}=\sfd\underset{\tx{\ciut{(p+1)}}}{\widetilde\b}\,,
\qqq
and checking that it be invariant under locally unique lifts to 
\qq\nn
\widetilde\si(\widetilde\txG/\txH)=\bigsqcup_{i\in I}\,\widetilde\si_i(\widetilde\cO_i)
\qqq
of the flows of the fundamental vector fields $\,\widetilde\Xi_{\widetilde X}\,$ for the transitive action of the extended supersymmetry group $\,\widetilde\txG\,$ on $\,\widetilde\txG/\txH$,\ or, equivalently, that the conditions
\qq\nn
\pLie{\widetilde\cK_{\widetilde X}}\underset{\tx{\ciut{(p+1)}}}{\widetilde\b}=0\,,\qquad\widetilde X\in\widetilde\ggt
\qqq
be satisfied for 
\qq\nn
\widetilde\cK_{\widetilde X}\rstr_{\widetilde\si_i(\widetilde\cO_i)}=\sfT\widetilde\si_i(\widetilde\Xi_{\widetilde X})\,.
\qqq 
A meaningful continuation of the thus initiated generalisation of the definition of a Cartan--Eilenberg super-1-gerbe over a Lie supergroup to the setting of homogeneous spaces (without any obvious Lie-supergroup structure) would require further insights into some working examples of supersymmetry-equivariant trivialisation of Green--Schwarz super-3-cocycles on such homogeneous spaces that we currently lack. Hence, we abandon our considerations at this stage and postpone them to a future investigation.
\erem

The above supersymmetric trivialisation mechanism will be implicit in the examples scrutinised in Sections \ref{sec:sMinkext} and the one suggested in Section \ref{sec:MTaway}.

\brem A particular setting in which an explicit contribution to the charge and a resultant deformation of (super)\-sym\-me\-try sourced by the pseudo-invariance of the global primitive $\,\underset{\tx{\ciut{(p+1)}}}{\b}\,$ of the Green--Schwarz super-$(p+2)$-cocycle $\,\underset{\tx{\ciut{(p+2)}}}{\chi}\rstr_{\si(\txG/\txH)}\,$ manifests itself is the realisation of supersymmetry on states from the Hilbert space $\,\ceH\,$ of the theory, induced, along the lines of Section I.2.2, in the standard procedure of geometric (pre)quantisation available in the field-theoretic setting in hand. In order to isolate and elucidate the phenomenon of interest, we make several simplifying assumptions. Thus, we consider a global primitive $\,\underset{\tx{\ciut{(p+1)}}}{\txB}\,$ of the supersymmetric super-$(p+2)$-cocycle $\,\underset{\tx{\ciut{(p+2)}}}{\txH}\,$ on $\,\txG/\txH$,\ the latter being descended from $\,\underset{\tx{\ciut{(p+2)}}}{\chi}\rstr_{\si(\txG/\txH)}$.\ We take its behaviour under the left action $\,[\la]_\cdot\,$ of \Reqref{eq:cosetlact} to be captured by the identities
\qq\nn
[\la]_g^*\underset{\tx{\ciut{(p+1)}}}{\txB}-\underset{\tx{\ciut{(p+1)}}}{\txB}=\sfd\underset{\tx{\ciut{(p)}}}{\jmath_g}\,,\qquad g\in\txG\,,
\qqq
with globally defined supersymmetry currents $\,\underset{\tx{\ciut{(p)}}}{\jmath_g}\in\Om^p(\txG/\txH)$.\ In the above, we recognise a simplified (global) variant of relations \eqref{eq:pseudolocB}. The said action $\,R\ :\ \txG\too{\rm U}(\ceH)$,\ is now readily derived, for any $\,g\in\txG$,\ in the form
\qq\nn
\bigl(R(g)\Psi\bigr)[\phi]:=c_g[\phi]\cdot\Psi\bigl[[\la]_{g^{-1}}\circ\phi\bigr]\,,\qquad\qquad c_g[\phi]:=\ee^{\sfi\,\int_{\xcC_{p,{\rm in}}}\,([\la]_{g^{-1}}\circ\phi)^*\underset{\tx{\ciut{(p)}}}{\jmath_g}}\,.
\qqq
Accordingly, we find, for arbitrary $\,g_1,g_2\in\txG$,\ the identity
\qq\nn
R(g_1)\circ R(g_2)=(\d_\txG c)_{g_1,g_2}[\cdot]\cdot R(g_1\cdot g_2)\,,
\qqq
with the homomorphicity 2-cocycle given by
\qq\nn
(\d_\txG c)_{g_1,g_2}[\phi]=\ee^{\sfi\,\int_{\xcC_{p,{\rm in}}}\,([\la]_{(g_1\cdot g_2)^{-1}}\circ\phi)^*(\d_\txG\underset{\tx{\ciut{(p)}}}{\jmath})_{g_1,g_2}}
\qqq
in terms of the current 2-cocycle
\qq\nn
(\d_\txG\underset{\tx{\ciut{(p)}}}{\jmath})_{g_1,g_2}:=[\la]_{g_2}^*\underset{\tx{\ciut{(p)}}}{\jmath_{g_1}}-\underset{\tx{\ciut{(p)}}}{\jmath_{g_1\cdot g_2}}+\underset{\tx{\ciut{(p)}}}{\jmath_{g_2}}\,.
\qqq
The latter defines a class in $\,H^p(\txG/\txH)\,$ as
\qq\nn
\sfd(\d_\txG\jmath)_{g_1,g_2}=[\la]_{g_2}^*\bigl([\la]_{g_1}^*\underset{\tx{\ciut{(p+1)}}}{\txB}-\underset{\tx{\ciut{(p+1)}}}{\txB}\bigr)-\bigl([\la]_{g_1\cdot g_2}^*\underset{\tx{\ciut{(p+1)}}}{\txB}-\underset{\tx{\ciut{(p+1)}}}{\txB}\bigr)+\bigl([\la]_{g_2}^*\underset{\tx{\ciut{(p+1)}}}{\txB}-\underset{\tx{\ciut{(p+1)}}}{\txB}\bigr)=0\,,
\qqq
and so the exponent of the homomorphicity 2-cocycle describes the standard pairing between the (de Rham) cohomology class $\,[([\la]_{(g_1\cdot g_2)^{-1}}\circ\phi)^*(\d_\txG\jmath)_{g_1,g_2}]\,$ of the pullback of that 2-cocycle and the homology class $\,[\xcC_{p,{\rm in}}]\,$ of the Cauchy $p$-cycle,
\qq\nn
(\d_\txG c)_{g_1,g_2}[\phi]\equiv\ee^{\sfi\,\corr{[([\la]_{(g_1\cdot g_2)^{-1}}\circ\phi)^*(\d_\txG\jmath)_{g_1,g_2}],[\xcC_{p,{\rm in}}]}}\,.
\qqq
This demonstrates that we are dealing with a wrapping-charge extension of $\,\txG$.\ The extension trivialises whenever the primitive $\,\underset{\tx{\ciut{(p+1)}}}{\b}\,$ is invariant on the nose as we may then choose $\,\jmath_g=0$,\ whereby we obtain $\,(\d_\txG c)_{g_1,g_2}[\phi]=1$.

Note that the extension is \emph{not} independent of the the choice of the representative of the class of primitives of the Green--Schwarz super-$(p+2)$-cocycle. It depends, though, on the primitive $\,\underset{\tx{\ciut{(p+1)}}}{\txB}\,$ solely up to a de Rham-exact correction. Indeed, upon replacement
\qq\nn
\underset{\tx{\ciut{(p+1)}}}{\txB}\longmapsto\underset{\tx{\ciut{(p+1)}}}{\txB}+\sfd\underset{\tx{\ciut{(p)}}}{\eta}\,,
\qqq
we may redefine the current as
\qq\nn
\underset{\tx{\ciut{(p)}}}{\jmath_g}\longmapsto\underset{\tx{\ciut{(p)}}}{\jmath_g}+\bigl(\d_\txG\underset{\tx{\ciut{(p)}}}{\eta}\bigr)_g\,,
\qqq
and so
\qq\nn
(\d_\txG c)_{g_1,g_2}\longmapsto(\d_\txG c)_{g_1,g_2}\,.
\qqq
In particular, and this is to be emphasised in the context of subsequent case studies, whenever there exists a $\txG$-invariant primitive, a correction by an exact super-$(p+1)$-form -- whether $\txG$-invariant or \emph{not} -- does not affect the homomorphicity 2-cocycle.

In summary, the dependence of the quantum realisation of the (super)symmetry group upon the choice of the primitive $\,\underset{\tx{\ciut{(p+1)}}}{\txB}\,$ is, as usual, stronger than the one present in the canonical setting of the previous section.
\erem

\section{The Kosteleck\'y--Rabin extensions of the super-Minkowskian algebra}\label{sec:sMinkext}

We begin our case-by-case analysis of the wrapping charges and the associated deformations of the supertarget Lie superalgebras by reconsidering the super-Minkowski space (essentially in the notation of Part I)
\qq\nn
{\rm sMink}^{d,1\,\vert\,D_{d,1}}\equiv{\rm sISO}(d,1\,\vert\,D_{d,1})/{\rm SO}(d,1)\,,
\qqq
a homogeneous space of the super-Poincar\'e supergroup
\qq\nn
{\rm sISO}(d,1\,\vert\,D_{d,1})\equiv\bR^{d,1\,\vert\,D_{d,1}}\rx{\rm SO}(d,1)
\qqq
with the Lie superalgebra
\qq\nn
\gt{siso}(d,1\,\vert\,D_{d,1})=\bigg(\bigoplus_{\widehat\a=1}^{D_{d,1}}\,\corr{Q_{\widehat\a}}_\bC\oplus\bigoplus_{\widehat a=0}^d\,\corr{P_{\widehat a}}_\bC\bigg)\oplus\bigoplus_{\widehat a,\widehat b=0}^d\,\corr{J_{\widehat a\widehat b}\equiv-J_{\widehat b\widehat a}}_\bC
\qqq
defined by the structure equations (here, $\,(\eta_{\widehat a\widehat b})\equiv\diag(-1,\underbrace{+1,+1,\ldots,+1}_{d\ {\rm times}})$)
\qq\nn
&[P_{\widehat a},P_{\widehat b}]=0\,,\qquad\qquad\{Q_{\widehat\a},Q_{\widehat\b}\}=\ovl\G^{\widehat a}_{\widehat\a\widehat\b}\,P_{\widehat a}\,,\qquad\qquad[P_{\widehat a},Q_{\widehat\a}]=0\,,&\cr\cr
&[J_{\widehat a\widehat b},J_{\widehat c\widehat d}]=\eta_{\widehat a\widehat d}\,J_{\widehat b\widehat c}-\eta_{\widehat a\widehat c}\,J_{\widehat b\widehat d}+\eta_{\widehat b\widehat c}\,J_{\widehat a\widehat d}-\eta_{\widehat b\widehat d}\,J_{\widehat a\widehat c}\,,&\cr\cr
&[J_{\widehat a\widehat b},P_{\widehat c}]=\eta_{\widehat b\widehat c}\,P_{\widehat a}-\eta_{\widehat a\widehat c}\,P_{\widehat b}\,,\qquad\qquad[J_{\widehat a\widehat b},Q_{\widehat\a}]=\tfrac{1}{2}\,\bigl(\G_{\widehat a\widehat b}\bigr)^{\widehat\b}_{\ \widehat\a}\,Q_{\widehat\b}\,.&
\qqq
The homogeneous space is embedded in the supersymmetry group $\,{\rm sISO}(d,1\,\vert\,D_{d,1})\,$ by a single section using the standard (global) coordinates $\,\{\theta^{\widehat\a},X^{\widehat a}\}^{\widehat\a\in\ovl{1,D_{d,1}},\ \widehat a\in\ovl{0,d}}\,$ on $\,{\rm sMink}^{d,1\,\vert\,D_{d,1}}$,
\qq\nn
\si\ :\ {\rm sMink}^{d,1\,\vert\,D_{d,1}}\too{\rm sISO}(d,1\,\vert\,D_{d,1})\ :\ (\theta,x)\longmapsto\ee^{\theta^{\widehat\a}\,Q_{\widehat\a}+X^{\widehat a}\,P_{\widehat a}}\,.
\qqq
It is a Lie supergroup itself, namely the supertranslation group $\,\bR^{d,1\,\vert\,D_{d,1}}$.\ We read off its binary operation from the above embedding using its Lie (sub-)superalgebra -- from the (Baker--Campbell--Dynkin--Hausdorff) identity
\qq\nn
\ee^{\theta_1^{\widehat\a}\,Q_{\widehat\a}+X_1^{\widehat a}\,P_{\widehat a}}\cdot\ee^{\theta_2^{\widehat\b}\,Q_{\widehat\b}+X_2^{\widehat b}\,P_{\widehat b}}&=&\ee^{(\theta_1^{\widehat\a}+\theta_2^{\widehat\a})\,Q_{\widehat\a}+(X_1^{\widehat a}+X_2^{\widehat a}+\frac{1}{2}\,[\theta_1^{\widehat\a}\,Q_{\widehat\a}+X_1^{\widehat a}\,P_{\widehat a},\theta_2^{\widehat\b}\,Q_{\widehat\b}+X_2^{\widehat b}\,P_{\widehat b}])^{\widehat c}\,P_{\widehat c}}\cr\cr
&=&\ee^{(\theta_1^{\widehat\a}+\theta_2^{\widehat\a})\,Q_{\widehat\a}+(X_1^{\widehat a}+X_2^{\widehat a}-\frac{1}{2}\,\ovl\theta_1\,\G^{\widehat a}\,\theta_2)\,P_{\widehat a}}
\qqq
we extract the explicit formula
\qq\nn
(\theta_1^{\widehat\a},X_1^{\widehat a})\cdot(\theta_2^{\widehat\b},X_2^{\widehat b})=\bigl(\theta_1^{\widehat\a}+\theta_2^{\widehat\a},X_1^{\widehat a}+X_2^{\widehat a}-\tfrac{1}{2}\,\ovl\theta_1\,\G^{\widehat a}\,\theta_2\bigr)\,.
\qqq
Given these, we take a closer look at the Green--Schwarz super-$(p+2)$-cocycles over the super-Minkowski space, for $\,p\in\{0,1\}$,\ (left-)invariant under the action of the super-Poincar\'e supergroup, alongside the corresponding non-supersymmetric primitives, found in Part I. For the latter supergroup, we compute the supersymmetry-variation super-$p$-forms $\,\underset{\tx{\ciut{(p)}}}{\G_X}$.\ In our computations, we use the coordinate form of the relevant right-invariant vector fields
\qq\nn
\cR_{(\vep,0)}(\theta,X)=\vep^{\widehat\a}\,\tfrac{\vec\p\ }{\p\theta^{\widehat\a}}-\tfrac{1}{2}\,\ovl\vep\,\G^{\widehat a}\,\theta\,\tfrac{\p\ }{\p X^{\widehat a}}\,,\qquad\qquad\cR_{(0,y)}(\theta,X)=y^{\widehat a}\,\tfrac{\p\ }{\p X^{\widehat a}}\,.
\qqq

\subsection{The superparticle extension of $\,{\rm sMink}^{9,1\,\vert\,D_{9,1}}$}\label{subsect:spartextMink}

The relevant ${\rm sISO}(9,1\,\vert\,D_{9,1})$-invariant Green--Schwarz super-2-cocycle reads
\qq\nn
\underset{\tx{\ciut{(2)}}}{\chi}(\theta,X)=\theta_{\rm L}^{\widehat\a}\wedge\ovl\G_{11\,\widehat\a\widehat\b}\,\theta^{\widehat\b}_{\rm L}(\theta,X)\equiv\ovl{\sfd\theta}\wedge\G_{11}\,\sfd\theta
\qqq
and admits a primitive
\qq\nn
\underset{\tx{\ciut{(1)}}}{\b}(\theta,X)=\ovl{\theta}\,\G_{11}\,\sfd\theta\,,
\qqq
for which we get
\qq\nn
\pLie{\cR_{(\vep,0)}}\underset{\tx{\ciut{(1)}}}{\b}(\theta,X)=\ovl\vep\,\G_{11}\,\sfd\theta\,,\qquad\qquad\pLie{\cR_{(0,y)}}\underset{\tx{\ciut{(1)}}}{\b}(\theta,X)=0\,,
\qqq
and so also
\qq\nn
\underset{\tx{\ciut{(0)}}}{\G_{(\vep,0)}}(\theta,X)=\ovl\vep\,\G_{11}\,\theta\,,\qquad\qquad\underset{\tx{\ciut{(0)}}}{\G_{(0,y)}}(\theta,X)=0\,,
\qqq
whence
\qq\nn
h_{(\vep,0)}(\theta,X,p)=-\ovl\vep\,(p_{\widehat a}\,\G^{\widehat a}+2\G_{11})\,\theta\,,\qquad\qquad h_{(0,y)}(\theta,X,p)=y^{\widehat a}\,p_{\widehat a}\,,
\qqq
and
\qq\nn
\xcW_{(\vep_1,0),(\vep_2,0)}[\theta,X]=2\ovl\vep_1\,\G_{11}\,\vep_2\,,\qquad\qquad\xcW_{(0,y_1),(0,y_2)}[\theta,X]=0=\xcW_{(\vep,0),(0,y)}[\theta,X]\,.
\qqq
The ensuing deformation of the super-Minkowski superalgebra (obtained, {\it e.g.}, through canonical quantisation of the super-centrally extended Poisson--Lie algebra of Noether charges, and the obvious sign flip) is the familiar Lie superalgebra
\qq\nn
\{Q_{\widehat\a},Q_{\widehat\b}\}^\sim=\ovl\G^{\widehat a}_{\widehat\a\widehat\b}\,P_{\widehat a}+2\,\ovl\G_{11\,\widehat\a\widehat\b}\,Z\,,\qquad\qquad[P_{\widehat a},P_{\widehat b}]^\sim=0=[Q_{\widehat\a},P_{\widehat a}]^\sim
\qqq
encountered in Part I, with an additional central generator $\,Z$.

The super-2-cocycle associated with the above anomaly,
\qq\nn
\varpi_0=\ovl{\si}\wedge\G_{11}\,\si\,,
\qqq
is precisely the GS super-2-cocycle whose trivialisation over the Lie supergroup that integrates the super-centrally extended $\,\bR^{9,1\,\vert\,D_{9,1}}\,$ determines the Green--Schwarz super-0-gerbe of Def.\,I.5.2.

\subsection{The superstring extension of $\,{\rm sMink}^{d,1\,\vert\,D_{d,1}}$}\label{subsect:sstringextsMink}

The  ${\rm sISO}(d,1\,\vert\,D_{d,1})$-invariant Green--Schwarz super-$3$-cocycle of interest can be written in the following form:
\qq\label{eq:sMinkGS3}
\underset{\tx{\ciut{(3)}}}{\chi}(\theta,X)=\theta_{\rm L}^{\widehat\a}\wedge\ovl\G_{{\widehat a}\,\widehat\a\widehat\b}\,\theta_{\rm L}^{\widehat\b}\wedge\theta_{\rm L}^{\widehat a}(\theta,X)\equiv\ovl{\sfd\theta}\wedge\G_{\widehat a}\,\sfd\theta\wedge\sfd X^{\widehat a}\,,
\qqq
which immediately gives a primitive
\qq\label{eq:sMincurv}
\underset{\tx{\ciut{(2)}}}{\b}(\theta,X)=\ovl{\theta}\,\G_{\widehat a}\,\sfd\theta\wedge\sfd X^{\widehat a}\,,
\qqq
satisfying
\qq\nn
\pLie{\cR_{(\vep,0)}}\underset{\tx{\ciut{(2)}}}{\b}(\theta,X)=\ovl\vep\,\G_{\widehat a}\,\sfd\theta\wedge\sfd X^{\widehat a}+\tfrac{1}{2}\,\ovl\vep\,\G_{\widehat a}\,\sfd\theta\wedge\ovl\theta\,\G^{\widehat a}\,\sfd\theta\,,\qquad\qquad\pLie{\cR_{(0,y)}}\underset{\tx{\ciut{(2)}}}{\b}(\theta,X)=0\,,
\qqq
so that we obtain
\qq\nn
\underset{\tx{\ciut{(1)}}}{\G_{(\vep,0)}}(\theta,X)=\ovl\vep\,\G_{\widehat a}\,\theta\,\bigl(\sfd X^{\widehat a}+\tfrac{1}{6}\,\ovl\theta\,\G^{\widehat a}\,\sfd\theta\bigr)\,,\qquad\qquad\underset{\tx{\ciut{(1)}}}{\G_{(0,y)}}(\theta,X)=0
\qqq
and the hamiltonians
\qq\nn
h_{(\vep,0)}[\theta,X,p]&=&-\int_0^{2\pi}\,\sfd\varphi\,(\ovl\vep\,\G_{\widehat a}\,\theta)\,\bigl(\eta^{\widehat a\widehat b}\,p_{\widehat b}+2\p_\varphi X^{\widehat a}-\tfrac{1}{3}\,\ovl\theta\,\G^{\widehat a}\,\p_\varphi\theta\bigr)(\varphi)\,,\cr\cr
h_{(0,y)}[\theta,X,p]&=&\int_0^{2\pi}\,\sfd\varphi\,y^{\widehat a}\,\bigl(p_{\widehat a}-\ovl\theta\,\G_{\widehat a}\,\p_\varphi\theta\bigr)(\varphi)\,.
\qqq
Accordingly, the wrapping anomaly reads
\qq\nn
&\xcW_{(\vep_1,0),(\vep_2,0)}[\theta,X]=\int_0^{2\pi}\,\sfd\varphi\,\p_\varphi\bigl(2\ovl\vep_1\,\G_{\widehat a}\,\vep_2\,X^{\widehat a}\bigr)(\varphi)=0\,,\qquad\qquad\xcW_{(0,y_1),(0,y_2)}[\theta,X]=0\,,&\cr\cr
&\xcW_{(\vep,0),(0,y)}[\theta,X]=\int_0^{2\pi}\,\sfd\varphi\,\p_\varphi\bigl(-2y^{\widehat a}\,\ovl\vep\,\G_{\widehat a}\,\theta\bigr)(\varphi)\,.&
\qqq
Owing to the topological triviality of the body $\,{\rm Mink}^{d,1}\equiv\vert{\rm sMink}^{d,1\,\vert\,D_{d,1}}\vert\,$ of the supertarget, the only non-vanishing contribution to the wrapping anomaly comes from the last term, and only if we take into account the Kosteleck\'y--Rabin states. Let $\,\nu\,$ be the (constant) Gra\ss mann-odd vector defining the Kosteleck\'y--Rabin lattice in $\,\bR^{0\,\vert\,D_{d,1}}$,\ so that we may consider winding states with
\qq\nn
\theta(2\pi)=\theta(0)+\nu\,.
\qqq
We may then rewrite the last component of the anomaly in the form
\qq\nn
\xcW_{(\vep,0),(0,y)}[\theta,X]=-2y^{\widehat a}\,\ovl\vep\,\G_{\widehat a}\,\nu\,.
\qqq
This produces ({\it e.g.}, through canonical quantisation, after a sign flip) -- as in \Reqref{eq:wrapcentrext} -- the super-centrally extended $\,{\rm sMink}^{d,1\,\vert\,D_{d,1}}\,$ algebra
\qq
&\{Q_{\widehat\a},Q_{\widehat\b}\}=\ovl\G^{\widehat a}_{\widehat\a\widehat\b}\,P_{\widehat a}\,,\qquad\qquad[P_{\widehat a},P_{\widehat b}]=0\,,\qquad\qquad[Q_{\widehat\a},P_{\widehat a}]=2\ovl\G_{{\widehat a}\,\widehat\a\widehat\b}\,Z^{\widehat\b}\,,&\nonumber \\ \label{eq:ssextsMink} \\
&[Z^{\widehat\a},\xi\}=0\,,\quad \xi\in\bR^{d,1\,\vert\,D_{d,1}}\,.&\nonumber
\qqq

The anomaly corresponds to the family of super-2-cocycles
\qq\nn
\varpi_{1\,\widehat\a}=2\si^{\widehat\b}\wedge\ovl\G_{\widehat a\,\widehat\b\widehat\a}\,e^{\widehat a}\,,
\qqq
whose trivialisation over the Lie supergroup that integrates the super-centrally extended $\,\bR^{d,1\,\vert\,D_{d,1}}\,$ determines (the surjective submersion of) the Green--Schwarz super-1-gerbe of Def.\,I.5.9.

\section{The superstring deformation of the super-${\rm AdS}_5\x\bS^5\,$ algebra}\label{sec:ssextaAdSS}

The point of departure of our subsequent considerations is the Metsaev--Tseytlin super-$\si$-model of superloop mechanics on the homogeneous space
\qq\nn
{\rm SU}(2,2\,\vert\,4)/({\rm SO}(4,1)\x{\rm SO}(5))\equiv{\rm s}\bigl({\rm AdS}_5\x\bS^5\bigr)
\qqq
of the Lie supergroup $\,{\rm SU}(2,2\,\vert\,4)$,\ with the body given by the homogeneous space
\qq\nn
{\rm SO}(4,2)/{\rm SO}(4,1)\x{\rm SO}(6)/{\rm SO}(5)\equiv{\rm AdS}_5\x\bS^5
\qqq
of the Lie group
\qq\nn
{\rm SU}(2,2)\x{\rm SU}(4)\cong{\rm SO}(4,2)\x{\rm SO}(6)\,.
\qqq
In the notation of the original paper \cite{Metsaev:1998it} and upon incorporation of the findings of Roiban and Siegel reported in \Rcite{Roiban:2000yy}, the model takes the form \eqref{eq:supersimod} with the supertarget (group) metric
\qq\nn
&(\txG_{\widehat a\widehat b})\equiv(\eta_{\widehat a\widehat b})=(\eta_{ab})\oplus(\d_{a'b'})\,,&\cr\cr &(\eta_{ab})\equiv\diag(-1,+1,+1,+1,+1)\,,\qquad\qquad(\d_{a'b'})\equiv\diag(+1,+1,+1,+1,+1)&
\qqq
and with the relevant Green--Schwarz super-3-cocycle
\qq\label{eq:MT3scocyc}
\underset{\tx{\ciut{(3)}}}{\chi}^{\rm MT}=-\Si_{\rm L}\wedge(\widehat C\ox\si_3)\,\cancel e\wedge\Si_{\rm L}\,,
\qqq
the latter being written, in the shorthand notation of Eqs.\,\eqref{eq:Gext1} and \eqref{eq:CCc}, in terms of the components
\qq\nn
\Si_{\rm L}^{\a\a' I}=\theta_{\rm L}^{\a\a' I}\,,\qquad\qquad\qquad\cancel e=(\widehat\G_{\widehat a}\ox\bd1)\,e^{\widehat a}\,,\quad e^{\widehat a}=\theta_{\rm L}^{\widehat a}
\qqq
of the left-invariant Maurer--Cartan super-1-form on $\,{\rm SU}(2,2\vert 4)\,$ along the vector-space complement of the isotropy Lie algebra $\,\gt{so}(4,1)\oplus\gt{so}(5)\,$ within $\,\gt{su}(2,2\,\vert\,4)$,\ contracted with ${\rm SO}(4,1)\x{\rm SO}(5)$-invariant tensors. Here, the superalgebra and the supergeometry of the model (and so also its field-theoretic content) have been cast in a form compatible with the decomposition of the body into its independent constituents: $\,{\rm AdS}_5\,$ and $\,\bS^5$,\ {\it i.e.}, we work with the Majorana--Weyl spinors of the product $\,{\rm Spin}\,$ group $\,{\rm Spin}(4,1)\x{\rm Spin}(5)\x{\rm Spin}(2,1)\,$ (the last factor accounts for the two species of chiral spinors entering the construction), whence the presence of the multi-indices $\,\widehat\a\equiv\a\a' I\,$ on them, with $\,\a,\a'\in\{1,2,3,4\}\,$ and $\,I\in\{1,2\}\,$ (and that of the tensor products of elements of the Clifford algebras of the quadratic spaces $\,\bR^{4,1},\bR^{5,0}\,$ and $\,\bR^{2,1}$), and with tensors of the product isotropy group $\,{\rm SO}(4,1)\x{\rm SO}(5)$,\ whence the two subsets of vector indices: $\,a\in\ovl{0,4}\,$ and $\,a'\in\ovl{5,9}$.\ Important properties of the distinguished representations of the said Clifford algebras entering the construction, including the fundamental Fierz identities satisfied by their generators, have been recapitulated in App.\,\ref{app:CliffAdSS}.

The Lie superalgebra $\,\mathfrak{su}(2,2\,\vert\,4)\,$ of the supersymmetry group has generators
\qq\nn
\gt{su}(2,2\,\vert\,4)&\equiv&\bigl(\tgt^{(0)}\oplus\tgt^{(1)}\bigr)\oplus\bigl(\gt{so}(4,1)\oplus\gt{so}(5)\bigr)\cr\cr
&=&\bigg(\bigl(\bigoplus_{a=0}^4\,\corr{P_a}_\bC\oplus\bigoplus_{a'=5}^9\,\corr{P_{a'}}_\bC\bigr)\oplus\bigoplus_{(\a,\a', I)\in\ovl{1,4}\x\ovl{1,4}\x\{1,2\}}\,\corr{Q_{\a\a' I}}_\bC\bigg)\cr\cr
&&\oplus\bigg(\bigoplus_{a,b=0}^4\,\corr{J_{ab}=-J_{ba}}_\bC\oplus\bigoplus_{a',b'=5}^9\,\corr{J_{a'b'}=-J_{b'a'}}_\bC\bigg)
\qqq
subject to the structure relations
\qq
\{Q_{\a\a' I},Q_{\b\b' J}\}=\sfi\,\bigl(-2(\widehat C\,\widehat\G^{\widehat a}\ox\bd1)_{\a\a'I\b\b'J}\,P_{\widehat a}+(\widehat C\,\widehat\G^{\widehat a\widehat b}\ox\si_2)_{\a\a'I\b\b'J}\,J_{\widehat a\widehat b}\bigr)\,,\cr\cr\cr
[P_{\widehat a},P_{\widehat b}]=\vep_{\widehat a\widehat b}\,J_{\widehat a\widehat b}\,,\qquad\vep_{\widehat a\widehat b}=\left\{ \barr{cl} +1 & \tx{if}\ \widehat a,\widehat b\in\ovl{0,4} \\
-1 & \tx{if}\ \widehat a,\widehat b\in\ovl{5,9} \\
0 & \tx{otherwise}\earr\right.\,,\cr\cr\cr
[J_{\widehat a\widehat b},J_{\widehat c\widehat d}]=\eta_{\widehat a\widehat d}\,J_{\widehat b\widehat c}-\eta_{\widehat a\widehat c}\,J_{\widehat b\widehat d}+\eta_{\widehat b\widehat c}\,J_{\widehat a\widehat d}-\eta_{\widehat b\widehat d}\,J_{\widehat a\widehat c}\,,\qquad\qquad[P_{\widehat a},J_{\widehat b\widehat c}]=\eta_{\widehat a\widehat b}\,P_{\widehat c}-\eta_{\widehat a\widehat c}\,P_{\widehat b}\,,\label{eq:AdSS}\\ \cr\cr
[Q_{\a\a' I},P_{\widehat a}]=-\tfrac{1}{2}\,(\widehat\G_{\widehat a}\ox\si_2)^{\b\b'J}_{\ \ \a\a'I}\,Q_{\b\b' J}\,,\qquad\qquad[Q_{\a\a' I},J_{\widehat a\widehat b}]=-\tfrac{1}{2}\,\vep_{\widehat a\widehat b}\,(\widehat\G_{\widehat a\widehat b}\ox\bd1)^{\b\b'J}_{\ \ \a\a'I}\,Q_{\b\b'J}\,.\nn
\qqq
Upon rescaling
\qq\label{eq:IWrescale}
\bigl(Q_{\a\a'I},P_{\widehat a},J_{\widehat a\widehat b}\bigr)\longmapsto\bigl(R^{\frac{1}{2}}\,Q_{\a\a'I},R\,P_{\widehat a},J_{\widehat a\widehat b}\bigr)\,,\quad R\in\bR
\qqq
the Lie superalgebra thus defined is readily seen to contract (in the sense of \.In\"on\"u and Wigner), in the limit $\,R\to\infty$,\ to the super-Minkowski algebra. The above structure relations are employed in the construction of the local sections that embed, in terms of local coordinates $\,\{\theta_i^{\a\a'I},X_i^{\widehat a}\}\,$ (around some $\,\unl g_i\,\txH$), elements $\,\cO_i\ni(\theta_i,X_i)\,$ of a trivialising cover $\,\{\cO_i\}_{i\in I}\,$ of (the base of) the principal ${\rm SO}(4,1)\x{\rm SO}(5)$-bundle $\,{\rm SU}(2,2\,\vert\,4)\too{\rm SU}(2,2\,\vert\,4)/({\rm SO}(4,1)\x{\rm SO}(5))\,$ in the total space of that bundle as per
\qq\qquad\qquad
\si_i\ :\ \cO_i\too{\rm SU}(2,2\,\vert\,4)\ :\ Z_i\equiv\bigl(\theta_i^{\a\a' I},x_i^a,x_i'{}^{a'}\bigr)\equiv\bigl(\theta_i^{\a\a'I},X_i^{\widehat a}\bigr)\longmapsto\unl g_i\cdot\ee^{X_i^{\widehat a}\,P_{\widehat a}}\cdot\ee^{\theta_i^{\a\a' I}\,Q_{\a\a' I}}\,. \label{eq:MTsect}
\qqq

The construction of the Wess--Zumino term of the Metsaev--Tseytlin super-$\si$-model calls for a global primitive of the Green--Schwarz super-3-cocycle \eqref{eq:MT3scocyc}. Among such primitives, we find the manifestly left-invariant one
\qq\label{eq:susypriMT}
\underset{\tx{\ciut{(2)}}}{\b}\equiv\sfd^{-1}\underset{\tx{\ciut{(3)}}}{\chi}^{\rm MT}=\sfi\,\ovl\Si_{\rm L}\wedge(\bd1\ox\si_1)\,\Si_{\rm L}\,.
\qqq
The latter may be corrected by the addition of arbitrary closed super-2-forms, in a manner motivated by various physical and geometric considerations. In the light of our previous findings, the presence of such de Rham cocycles does not affect the Poisson bracket of the Noether charges of the ensuing (super-centrally extended) field-theoretic realisation of the supersymmetry algebra. Consequently, we may launch a systematic study of the structure of field-theoretic deformations of the latter Lie superalgebra for a large class of super-2-form potentials using the particularly simple form \eqref{eq:susypriMT} of the primitive. We shall perform the study with view to identifying those deformations motivated by the simple (super-$\si$-model) mechanics of extended charged objects introduced in Sec.\,\ref{sec:Cartsgeom} whose associative completions (derived through imposition of the super-Jacobi identity) geometrise, upon integration to the deformed Lie supergroup, the Green--Schwarz super-3-cocycle in that they support supersymmetric super-2-forms which descend to the (corrected) primitives of $\,\underset{\tx{\ciut{(3)}}}{\chi}^{\rm MT}$.\ In order to better understand the topology quantified by these deformations through isolation of contributions from the non-trivial topology $\,\bR^4\x\bS^1\x\bS^5\,$ of the body of the super-target and those from the twisted sector of the Kosteleck\'y--Rabin quotient, the latter in direct relation to the previously identified deformations of the flat super-Minkowskian limit of the supergeometry under consideration, we shall rescale the (local) coordinates on the homogeneous space $\,{\rm SU}(2,2\,\vert\,4)/({\rm SO}(4,1)\x{\rm SO}(5))\,$ in a standard manner dual to \eqref{eq:IWrescale}, {\it i.e.}, uniformly as
\qq\label{eq:AdSSResc}
\bigl(\theta_i^{\a\a' I},X_i^{\widehat a}\bigr)\longmapsto\bigl(R^{-\frac{1}{2}}\,\theta_i^{\a\a' I},R^{-1}\,X_i^{\widehat a}\bigr)\,,\quad i\in I
\qqq
and write out the corresponding restrictions of the Maurer--Cartan super-1-form
\qq
\si_i^*\theta_{\rm L}(Z_i)&=&\sum_{k=0}^\infty\,\tfrac{(-1)^k}{(k+1)!\,R^{\frac{k+1}{2}}}\,\bigg(\tfrac{1}{R^{\frac{k+1}{2}}}\,\sfd X_i^{\widehat a}\ox\sfT_e\Ad_{\ee^{-R^{-\frac{1}{2}}\,\theta_i^{\a\a' I}\,Q_{\a\a' I}}}\circ{\rm ad}^k_{X_i^{\widehat b}\,P_{\widehat b}}(P_{\widehat a})\cr\cr
&&\hspace{2.5cm}+\sfd\theta_i^{\a\a' I}\ox{\rm ad}^k_{\theta_i^{\b\b' J}\,Q_{\b\b' J}}(Q_{\a\a' I})\bigg)\,. \label{eq:MCs1f}
\qqq
Upon taking into account the structure equations \eqref{eq:AdSS}, alongside the expansion
\qq\nn
\ee^{-R^{-\frac{1}{2}}\,\theta_i^{\a\a' I}\,Q_{\a\a' I}}&=&1-\frac{1}{R^{\frac{1}{2}}}\,\theta_i^{\a\a' I}\,Q_{\a\a' I}-\frac{1}{2R}\,\theta_i^{\a\a' I}\,\theta_i^{\b\b' J}\,Q_{\a\a' I}\,Q_{\b\b' J}+O\bigl(R^{-\frac{3}{2}}\bigr)\,,
\qqq
we may, next, study the asymptotics of the various components of the above pullback super-1-forms. We do that with view to understanding the asymptotic relation between the left-invariant primitive $\,\underset{\tx{\ciut{(2)}}}{\b}\,$ and the non-supersymmetric curving \eqref{eq:sMincurv} of the Green--Schwarz super-1-gerbe on ${\rm sMink}^{1,9\,\vert\,D_{1,9}}$,\ and -- in so doing -- to geometrising the Metsaev--Tseytlin super-3-cocycle in relation to the well-understood super-Minkowskian construction of Part I. Thus, we eventually arrive at the result
\qq\nn
\si_i^*\theta_{\rm L}(Z_i)=\tfrac{1}{R^{\frac{1}{2}}}\,\sfd\theta_i^{\a\a'I}\ox Q_{\a\a'I}+\tfrac{1}{R}\,\bigl[\bigl(\sfd X_i^{\widehat a}-\sfi\,\ovl\theta_i\,(\widehat\G^{\widehat a}\ox\bd1)\,\sfd\theta_i\bigr)\ox P_{\widehat a}+\tfrac{\sfi}{2}\,\ovl\theta_i\,(\widehat\G^{\widehat a\widehat b}\ox\si_2)\,\sfd\theta_i\ox J_{\widehat a\widehat b}\bigr]\cr\cr
+\tfrac{1}{2R^{\frac{3}{2}}}\,\bigl[\bigl(\sfd X_i^{\widehat a}-\tfrac{\sfi}{3}\,\ovl\theta_i\,(\widehat\G^a\ox\bd1)\,\sfd\theta_i\bigr)\,(\widehat\G_{\widehat a}\ox\si_2)^{\a\a'I}_{\ \b\b'J}+\tfrac{\sfi\,\vep_{\widehat a\widehat b}}{3!}\,\bigl(\ovl\theta_i\,(\widehat\G^{\widehat a\widehat b}\ox\si_2)\,\sfd\theta_i\,(\widehat\G_{\widehat a\widehat b}\ox\bd1)^{\a\a'I}_{\ \b\b'J}\bigr]\,\theta_i^{\b\b'J}\ox Q_{\a\a' I}\cr\cr
-\tfrac{1}{2R^2}\,\bigl[\bigl(\sfi\,\bigl(\sfd X_i^{\widehat b}-\tfrac{\sfi}{6}\,\ovl\theta_i\,\bigl(\widehat\G^{\widehat b}\ox\bd1\bigr)\,\sfd\theta_i\bigr)\,\ovl\theta_i\,\bigl(\{\widehat\G^{\widehat a},\widehat\G_{\widehat b}\}\ox\si_2\bigr)\,\theta_i-\tfrac{\vep_{\widehat b\widehat c}}{12}\,\ovl\theta_i\,\bigl(\widehat\G_{\widehat b\widehat c}\ox\si_2\bigr)\,\sfd\theta_i\cdot\ovl\theta_i\,\bigl(\widehat\G^{\widehat a}\,\widehat\G^{\widehat b\widehat c}\ox\bd1\bigr)\,\theta_i\bigr)\ox P_{\widehat a}\cr\cr
-\bigl(\tfrac{\sfi}{2}\,\bigl(\sfd X_i^{\widehat c}-\tfrac{\sfi}{3!}\,\ovl\theta_i\,\bigl(\widehat\G^{\widehat c}\ox\bd1\bigr)\,\sfd\theta_i\bigr)\,\ovl\theta_i\,\bigl(\widehat\G^{\widehat a\widehat b}\,\widehat\G_{\widehat c}\ox\bd1\bigr)\,\theta_i-\tfrac{\vep_{\widehat a\widehat b}}{4!}\,\ovl\theta_i\,\bigl(\widehat\G^{\widehat c\widehat d}\ox\si_2\bigr)\,\sfd\theta_i\cdot\ovl\theta_i\,\bigl(\widehat\G^{\widehat a\widehat b}\,\widehat\G_{\widehat c\widehat d}\ox\si_2\bigr)\,\theta_i\cr\cr
-\vep_{\widehat a\widehat b}\,X_i^{\widehat a}\,\sfd X_i^{\widehat b}\bigr)\ox J_{\widehat a\widehat b}\bigr]+O\bigl(R^{-\frac{5}{2}}\bigr)\,,
\qqq
written in the shorthand notation
\qq\nn
\widehat\G^{\widehat a}=\eta^{\widehat a\widehat b}\,\widehat\G_{\widehat b}\,.
\qqq
From the above, we read off, in particular, the $R^{-1}$-expansion of the spinorial component of the Maurer--Cartan super-1-form
\qq\nn
\si_i^*\Si^{\a\a'I}_{\rm L}(Z_i)&=&\tfrac{1}{R^{\frac{1}{2}}}\,\sfd\theta_i^{\a\a'I}+\tfrac{1}{2R^{\frac{3}{2}}}\,\bigl[\bigl(\sfd X_i^{\widehat a}-\tfrac{\sfi}{3}\,\ovl\theta_i\,(\widehat\G^a\ox\bd1)\,\sfd\theta_i\bigr)\,(\widehat\G_{\widehat a}\ox\si_2)^{\a\a'I}_{\ \b\b'J}\cr\cr
&&+\tfrac{\sfi\,\vep_{\widehat a\widehat b}}{3!}\,\bigl(\ovl\theta_i\,(\widehat\G^{\widehat a\widehat b}\ox\si_2)\,\sfd\theta_i\,(\widehat\G_{\widehat a\widehat b}\ox\bd1)^{\a\a'I}_{\ \b\b'J}\bigr]\,\theta_i^{\b\b'J}+O\bigl(R^{-\frac{5}{2}}\bigr)\,,
\qqq
and so also the asymptotics of the Green--Schwarz super-3-cocycle
\qq\nn
\si_i^*\underset{\tx{\ciut{(3)}}}{\chi}^{\rm MT}(Z_i)=\tfrac{1}{R^2}\,\bigl(\sfd X_i^{\widehat a}-\sfi\,\ovl\theta_i\,\bigl(\widehat\G^{\widehat a}\ox\bd1\bigr)\,\sfd\theta_i\bigr)\wedge\sfd\ovl\theta_i\wedge\bigl(\widehat\G_{\widehat a}\ox\si_3\bigr)\,\sfd\theta_i+O\bigl(R^{-3}\bigr)
\qqq
and of its left-invariant primitive
\qq\nn
\si_i^*\underset{\tx{\ciut{(2)}}}{\b}(Z_i)&=&\tfrac{\sfi}{R}\,\sfd\ovl\theta_i\wedge(\bd1\ox\si_1)\,\sfd\theta_i+\tfrac{1}{R^2}\,\bigl[\ovl\theta_i\,\bigl(\widehat\G_{\widehat a}\ox\si_3\bigr)\,\sfd\theta_i\wedge\bigl(\sfd X_i^{\widehat a}-\tfrac{\sfi}{3}\,\ovl\theta_i\,(\widehat\G^a\ox\bd1)\,\sfd\theta_i\bigr)\cr\cr
&&-\tfrac{\vep_{\widehat a\widehat b}}{3!}\,\ovl\theta_i\,\bigl(\widehat\G_{\widehat a\widehat b}\ox\si_1\bigr)\,\sfd\theta_i\wedge\ovl\theta_i\,\bigl(\widehat\G^{\widehat a\widehat b}\ox\si_2\bigr)\,\sfd\theta_i\bigr]+O\bigl(R^{-3}\bigr)\,.
\qqq
The latter two reveal the precise relation between the superdifferential forms on the curved (super)target and their flat-superspace counterparts when rewritten in the notation of App.\,\ref{app:CliffAdSS},
\qq\nn
\si_i^*\underset{\tx{\ciut{(3)}}}{\chi}^{\rm MT}(Z_i)&=&\tfrac{1}{R^2}\,\bigl(\sfd X_i^{\widehat a}-\sfi\,\theta_i^{\a\a'I}\,\bigl(\cC\,\unl\g^{\widehat a}\,\D^1_{\widehat a}\bigr)_{\a\a'I\b\b'J}\,\sfd\theta_i^{\b\b'J}\bigr)\wedge\sfd\theta_i^{\g\g'K}\wedge\bigl(\cC\,\unl\g_{\widehat a}\,\D^2_{\widehat a}\bigr)_{\g\g'K\d\d'L}\,\sfd\theta_i^{\d\d'L}\cr\cr
&&+O\bigl(R^{-3}\bigr)\,,\cr\cr\cr
\si_i^*\underset{\tx{\ciut{(2)}}}{\b}(Z_i)&=&\tfrac{\sfi}{R}\,\sfd\ovl\theta_i\wedge(\bd1\ox\si_1)\,\sfd\theta_i\cr\cr
&&+\tfrac{1}{R^2}\,\bigl[\theta_i^{\a\a'I}\,\bigl(\cC\,\unl\g_{\widehat a}\,\D^2_{\widehat a}\bigr)_{\a\a'I\b\b'J}\,\sfd\theta_i^{\b\b'J}\wedge\bigl(\sfd X_i^{\widehat a}-\tfrac{\sfi}{3}\,\theta_i^{\a\a'I}\,\bigl(\cC\,\unl\g^{\widehat a}\,\D^1_{\widehat a}\bigr)_{\a\a'I\b\b'J}\,\sfd\theta_i^{\b\b'J}\bigr)\cr\cr
&&-\tfrac{\vep_{\widehat a\widehat b}}{3!}\,\ovl\theta_i\,\bigl(\widehat\G_{\widehat a\widehat b}\ox\si_1\bigr)\,\sfd\theta_i\wedge\ovl\theta_i\,\bigl(\widehat\G^{\widehat a\widehat b}\ox\si_2\bigr)\,\sfd\theta_i\bigr]+O\bigl(R^{-3}\bigr)\,,
\qqq
with
\qq\nn
\D^1_{\widehat a}=\left\{ \barr{cl} \bd1\ox\bd1\ox\si_3 & \tx{if}\ \widehat a\in\ovl{0,4} \\
\bd1\ox\bd1\ox\bd1 & \tx{if}\ \widehat a\in\ovl{5,9}\earr\right.\,,\qquad\qquad\D^2_{\widehat a}=\left\{ \barr{cl} \bd1\ox\bd1\ox\bd1 & \tx{if}\ \widehat a\in\ovl{0,4} \\
\bd1\ox\bd1\ox\si_3 & \tx{if}\ \widehat a\in\ovl{5,9}\earr\right.
\qqq
Thus, while the leading asymptotics of the Green--Schwarz super-3-cocycle reproduces the super-Minkowskian object \eqref{eq:sMinkGS3} upon evaluation on an ${\rm SU}(2,2\,\vert\,4)$-spinor $\,\theta=\psi\ox\psi'\ox\binom{1}{0}\,$ of positive chirality, along the lines of \Rcite{Metsaev:1998it}, its primitive may do that only upon substraction of the leading term of its $R^{-1}$-expansion. This prompts us to consider, in what follows, the distinguished (local) primitives
\qq\label{eq:AdSprimcorr}\qquad
\underset{\tx{\ciut{(2)}}}{\txB_i}=\si_i^*\underset{\tx{\ciut{(2)}}}{\b}-\underset{\tx{\ciut{(2)}}}{\txD_i}\,,\qquad\qquad\underset{\tx{\ciut{(2)}}}{\txD_i}(Z_i)=\sfi\,\sfd\ovl\theta_i\wedge(\bd1\ox\si_1)\,\sfd\theta_i
\qqq
of the $\,\si_i^*\underset{\tx{\ciut{(3)}}}{\chi}^{\rm MT}\,$ (written out above in their coordinate form prior to the rescaling). These are manifestly non-invariant, and so it seems apposite to look for an extension of the original supersymmetry algebra whose integration would allow for a global supersymmetric trivialisation of the Metsaev--Tseytlin super-3-cocycle. Our quest starts with an analysis of the natural field-theoretic sources of a sought-after deformation of $\,\gt{su}(2,2\,\vert\,4)$. 

We begin our computation of the wrapping anomaly by noting that the ${\rm SU}(2,2\,\vert\,4)$-invariant primitive $\,\underset{\tx{\ciut{(2)}}}{\b}\,$ is ${\rm SO}(4,1)\x{\rm SO}(5)$-horizontal in the sense of \Reqref{eq:Binvhor}, and so
\qq\nn
\si_{i_\t}^*\bigl(\cK_{X_2}\con\cK_{X_1}\con\underset{\tx{\ciut{(2)}}}{\b}\bigr)=2\sfi\,\Ad_{\si_{i_\t}(\cdot)^{-1}}(X_1)^{\a\a' I}\,\Ad_{\si_{i_\t}(\cdot)^{-1}}(X_2)^{\b\b' J}\,(\si_1)_{IJ}\,C_{\a\b}\,C_{\a'\b'}'\,.
\qqq
Above, and in what follows, we are using the notation for the standard Pauli matrices
\qq\nn
(\si_A)_{IJ}\equiv\d_{IK}\,(\si_A)^K_{\ J}\equiv(\si_A)_I^{\ K}\,\d_{KJ}\,,\quad A\in\{1,2,3\}\,.
\qqq
In keeping with Eqs.\,\eqref{eq:wrappaninvB}, \eqref{eq:monofAB} and \eqref{eq:fAB}, we obtain the wrapping anomaly in the form
\qq\nn
&\xcW^{({\rm inv})}_{X_1,X_2}[\unl\xi]=\mu\bigl(f_{X_1,X_2};\xi(\xcC_1)\bigr)\,,&\cr\cr
&f_{X_1,X_2}\rstr_{\cO_i}\equiv 2\sfi\,\sfT_e\Ad_{\si_i(\cdot)^{-1}}(X_1)^{\a\a' I}\,\bigl(\widehat C\ox\si_1\bigr)_{\a\a'I\ \b\b'J}\,\sfT_e\Ad_{\si_i(\cdot)^{-1}}(X_2)^{\b\b' J}\,.&
\qqq
Explicit expressions for the $\,\Ad_{\si_{i_\t}(\cdot)^{-1}}(X)^{\a\a' I}\,$ corresponding to the various vectors
\qq\nn
\D\in\{\vep\equiv\vep^{\a\a' I}\,Q_{\a\a' I},Y\equiv Y^{\widehat a}\,P_{\widehat a}\equiv\bigl(y^a\,P_a,y'{}_{a'}\,P_{a'}\bigr),\La\equiv\La^{\widehat a\widehat b}\,J_{\widehat a\widehat b}\equiv\bigl(\la^{ab}\,J_{ab},\la'{}^{a'b'}\,J_{a'b'}\bigr)\}
\qqq
in the Lie superalgebra $\,\mathfrak{su}(2,2\,\vert\,4)\,$ may be readily obtained from those for $\,\unl g_i=e\,$ ({\it i.e.}, computed in the neighbourhood of the supergroup unit) gathered in \Rcite{Hatsuda:2002hz}. These split as
\qq\nn
\sfT_e\Ad_{\si_i(Z_i)^{-1}}(\D)^{\a\a' I}\equiv\sfT_e\Ad_{\si_i(Z_i)^{-1}}(\D)_Q^{\a\a' I}+\sfT_e\Ad_{\si_i(Z_i)^{-1}}(\D)_P^{\a\a' I}+\sfT_e\Ad_{\si_i(Z_i)^{-1}}(\D)_J^{\a\a' I}\,,
\qqq
with components that read
\qq\nn
\sfT_e\Ad_{\si_i(Z_i)^{-1}}(\D)_Q^{\a\a' I}&=&(\cos\Psi)^{\a\a' I}_{\ \b\b' J}(\theta_i)\,\exp\bigl(-\tfrac{1}{2}\,X_i^{\widehat a}\,\widehat\G_{\widehat a}\ox\si_2\bigr)^{\b\b' J}_{\ \g\g' K}\,\unl\G_{i\ A}^{\g\g' K}\,\D^A\,,\cr\cr
\sfT_e\Ad_{\si_i(Z_i)^{-1}}(\D)_P^{\a\a' I}&=&\tfrac{1}{4}\,\bigl(\tfrac{\sin\Psi}{\Psi}\bigr)^{\a\a' I}_{\ \b\b' J}(\theta_i)\,\bigl(2\widetilde Y_{i\ A}^{\widehat a}(X_i)\,\widehat\G_{\widehat a}\ox\si_2+\vep_{\widehat a\widehat b}\,\widetilde{\widetilde Y}_{i\ A}^{\widehat a\widehat b}(X_i)\,\widehat\G_{\widehat a\widehat b}\ox\bd1\bigr)^{\b\b'J}_{\ \g\g'K}\,\theta_i^{\g\g' K}\,\D^A\,,\cr\cr
\sfT_e\Ad_{\si_i(Z_i)^{-1}}(\D)_J^{\a\a' I}&=&\tfrac{1}{4}\,\bigl(\tfrac{\sin\Psi}{\Psi}\bigr)^{\a\a' I}_{\ \b\b' J}(\theta_i)\,\bigl(2\widetilde\La_{i\ A}^{\widehat a}(X_i)\,\widehat\G_{\widehat a}\ox\si_2+\vep_{\widehat a\widehat b}\,\widetilde{\widetilde\La}_{i\ A}^{\widehat a\widehat b}(X_i)\,\widehat\G_{\widehat a\widehat b}\ox\bd1\bigr)^{\b\b'J}_{\ \g\g'K}\,\theta_i^{\g\g' K}\,\D^A\,,
\qqq
where\footnote{It is to be noted that the functions with the matrix argument $\,\Psi\,$ appearing in the expressions listed can all be expressed as power series in $\,\Psi^2$.\ (In fact, these series are finite owing to the anticommutativity of the Gra\ss mann-odd coordinates.)}
\qq\nn
\bigl(\Psi^2\bigr)^{\a\a' I}_{\ \b\b' J}(\theta_i)&=&\sfi\,\bigl(\bigl(\widehat\G^{\widehat a}\ox\si_2\bigr)^{\a\a' I}_{\ \g\g' K}\,(\widehat C\,\widehat\G_{\widehat a}\ox\bd1)_{\d\d' L\ \b\b' J}\cr\cr
&&-\tfrac{1}{2}\,\vep_{\widehat a\widehat b}\,\bigl(\widehat\G^{\widehat a\widehat b}\ox\bd1\bigr)^{\a\a' I}_{\ \g\g' K}\,\bigl(\widehat C\,\widehat\G_{\widehat a\widehat b}\ox\si_2\bigr)_{\d\d' L\ \b\b' J}\bigr)\,\theta_i^{\g\g' K}\,\theta_i^{\d\d' L}
\qqq
and
\qq\nn
&\widetilde Y_i^{\widehat a}(X_i)\equiv\bigl(\widetilde y_i^a(x_i^d),\widetilde y_i'{}^{a'}(x_i'{}^{d'})\bigr)\,,\qquad\qquad\widetilde{\widetilde Y}_i^{\widehat a\widehat b}(X_i)\equiv\bigl(\widetilde{\widetilde y}_i^{ab}(x_i^c),\widetilde{\widetilde y}_i'{}^{a'b'}(x_i'{}^{c'})\bigr)\,,&\cr\cr
&\widetilde\La_i^{\widehat a}(X_i)\equiv\bigl(\widetilde\la_i^a(x_i^d),\widetilde\la_i'{}^{a'}(x_i'{}^{d'})\bigr)\,,\qquad\qquad\widetilde{\widetilde\La}_i^{\widehat a\widehat b}(X_i)\equiv\bigl(\widetilde{\widetilde\la}_i^{ab}(x_i^c),\widetilde{\widetilde\la}_i'{}^{a'b'}(x_i'{}^{c'})\bigr)\,,&
\qqq
with
\qq\nn
\widetilde y_i^a(x_i^d)&=&\unl\G^b_{i\ A}\bigl(\d^a_{\ b}+({\rm ch}\,x_i-1)\,\bigl(\d_{\ b}^a-\tfrac{\eta_{bc}\,x_i^c\,x_i^a}{x_i^2}\bigr)\bigr)\,,\qquad\qquad\widetilde{\widetilde y}^{ab}(x_i^c)=-x_i^{[a}\,\unl\G_{i\ A}^{b]}\,\tfrac{{\rm sh}\,x_i}{x_i}\,,\cr\cr\cr
\widetilde y_i'{}^{a'}(x_i'{}^{d'})&=&\unl\G^{b'}_{i\ A}\bigl(\d_{\ b'}^{a'}+(\cos x_i'-1)\,\bigl(\d_{\ b'}^{a'}-\tfrac{\d_{b'c'}\,x_i'{}^{c'}\,x_i'{}^{a'}}{x_i'{}^2}\bigr)\bigr)\,,\qquad\qquad\widetilde{\widetilde y}_i'{}^{a'b'}(x_i'{}^{c'})=x_i'{}^{[a'}\,\unl\G_{i\ A}^{b']}\,\tfrac{\sin x_i'}{x_i'}\cr\cr\cr
\widetilde\la_i^a(x_i^e)&=&\unl\G_{i\ A}^{ab}\,\eta_{bc}\,x_i^c\,\tfrac{{\rm sh}\,x_i}{x_i}\,,\qquad\qquad\widetilde{\widetilde\la}_i^{ab}(x_i^c)=\unl\G_{i\ A}^{ab}+\tfrac{{\rm ch}\,x_i-1}{x_i^2}\,x_i^{[a}\,\unl\G_{i\ A}^{b]c}\,\eta_{cd}\,x_i^d\,,\cr\cr\cr
\widetilde\la_i'{}^{a'}(x_i'{}^{d'})&=&\unl\G_{i\ A}^{a'b'}\,\d_{b'c'}\,x_i'{}^{c'}\,\tfrac{\sin x_i'}{x_i'}\,,\qquad\qquad\widetilde{\widetilde\la}_i'{}^{a'b'}(x_i'{}^{e'})=\unl\G_{i\ A}^{a'b'}+\tfrac{\cos x_i'-1}{x_i'{}^2}\,x_i'{}^{[a'}\,\unl\G_{i\ A}^{b']c'}\,\d_{c'd'}\,x_i'{}^{d'}\,,\cr\cr\cr
x_i&=&\sqrt{\eta_{ab}\,x_i^a\,x_i^b}\equiv\sqrt{x_i^2}\,,\qquad\qquad x_i'=\sqrt{\d_{a'b'}\,x_i'{}^{a'}\,x_i'{}^{b'}}\equiv\sqrt{x_i'{}^2}\,,
\qqq
and where the matrix elements
\qq\nn
\unl\G^A_{i\,\ B}\equiv\bigl(\sfT_e\Ad_{\unl g_i^{-1}}\bigr)^A_{\ B}
\qqq
capture conjugation of the variation vector by the reference supergroup element $\,\unl g_i$.\ The ensuing Poisson algebra of the Noether charges,
\qq\nn
\{h_{\D_1},h_{\D_2}\}_{\Om^{\rm (NG)}_{{\rm GS},1}}[X,\theta,\pi]=-h_{[\D_1,\D_2]}[X,\theta,\pi]+\mu\bigl(f_{\D_1,\D_2};(X,\theta)(\xcC_1)\bigr)\,,
\qqq
has a fairly complicated local coordinate (or, equivalently, current) presentation. In order to draw concrete qualitative conclusions as to the (super)geometric nature of the wrapping-charge deformation present in it and, in so doing, establish a structural relation with the formerly discussed superstring deformation of the super-Minkowski superalgebra, understood as the flat-superspace limit of the superalgebra under consideration, we shall rescale the coordinates as previously ({\it cp} \Reqref{eq:AdSSResc}) and study the asymptotics of the above expressions in the r\'egime of large $\,R$,\ which corresponds to a correlated flattening of both: the 1-cycle in $\,{\rm AdS}_5\,$ and the 5-sphere $\,\bS^5$,\ and an attendant uniform rescaling of the fibre of the Gra\ss mann bundle over them. We further zoom in on a neighbourhood of the group unit, that is we examine the local presentation of the functions $\,f_{\D_1,\D_2}\,$ over the (blown-up) coordinate patch $\,\cO_{i*}\,$ centred on $\,\unl g_{i*}\equiv e$.\ We thus obtain
\qq\nn
\Ad_{\si_{i*}(Z_{i*})^{-1}}(\vep)^{\a\a' I}&=&\vep^{\a\a' I}-\tfrac{1}{2R}\,\bigl(\bigl(\Psi^2\bigr)^{\a\a' I}_{\ \b\b' J}(\theta_{i*})+X_{i*}^{\widehat a}\,\bigl(\widehat\G_{\widehat a}\ox\si_2\bigr)^{\a\a'I}_{\ \b\b'J}\bigr)\,\vep^{\b\b'J}+O\bigl(R^{-2}\bigr)\,,\cr\cr\cr
\Ad_{\si_{i*}(Z_{i*})^{-1}}(Y)^{\a\a' I}&=&\tfrac{1}{2R^{\frac{1}{2}}}\,Y^{\widehat a}\,\bigl(\widehat\G_{\widehat a}\ox\si_2\bigr)^{\a\a'I}_{\ \b\b'J}\,\theta_{i*}^{\b\b' J}+O\bigl(R^{-\frac{3}{2}}\bigr)\,,\cr\cr\cr
\Ad_{\si_{i*}(Z_{i*})^{-1}}(\La)^{\a\a' I}&=&\tfrac{1}{4R^{\frac{1}{2}}}\,\vep_{\widehat a\widehat b}\,\La^{\widehat a\widehat b}\,\bigl(\widehat\G_{\widehat a\widehat b}\ox\bd1)^{\a\a'I}_{\ \b\b'J}\,\theta_{i*}^{\b\b' J}+O\bigl(R^{-\frac{3}{2}}\bigr)\,,
\qqq
whence also the leading local asymptotics\footnote{In its derivation, we used the symmetry properties: $\,C^{\rm T}=-C,\ C'{}^{\rm T}=-C'\,$ and $\,(C\,\G_a)^{\rm T}=-C\,\G_a,\ (C'\,\G_{a'})^{\rm T}=-C'\,\G_{a'}$,\ as well as the absence of winding states on $\,\bS^5$.} 
\qq\nn
f_{\vep_1,\vep_2}\rstr_{\cO_{i*}}-2\sfi\,\ovl\vep_1\,(\bd1\ox\si_1)\,\vep_2\cr\cr
=-\tfrac{2\sfi}{R}\,\vep_1^{\a\a' I}\,\vep_2^{\b\b' J}\,\bigl[\bigl(\widehat C\ox\si_1\bigr)_{\g\g'K\d\d'L}\,\d^{\g\g'K}_{\ (\a\a'I}\,\d^{\ep\ep'L}_{\ \b\b'J)}\,\bigl(\Psi^2\bigr)^{\d\d'L}_{\ \ep\ep'M}(\theta_{i*})+\sfi\,\bigl(\widehat C\,\widehat\G_{\widehat a}\ox\si_3\bigr)_{\a\a'I\b\b'J}\,X_{i*}^{\widehat a}\bigr]+O\bigl(R^{-2}\bigr)\cr\cr\cr
f_{Y_1,Y_2}\rstr_{\cO_{i*}}=\tfrac{\sfi}{2R}\,Y_1^{\widehat a}\,Y_2^{\widehat b}\,\ovl\theta_{i*}\,\bigl(\widehat\G_{\widehat a\widehat b}\ox\si_1\bigr)\,\theta_{i*}+O\bigl(R^{-2}\bigr)\,,\cr\cr\cr
f_{\La_1,\La_2}\rstr_{\cO_{i*}}=-\tfrac{\sfi}{8R}\,\vep_{\widehat a\widehat b}\,\vep_{\widehat c\widehat d}\,\La_1^{\widehat a\widehat b}\,\La_2^{\widehat c\widehat d}\,\ovl\theta_{i*}\,\bigl(\widehat\G_{\widehat a\widehat b}\,\widehat\G_{\widehat c\widehat d}\ox\si_1\bigr)\,\theta_{i*}+O\bigl(R^{-2}\bigr)\cr\cr
=-\tfrac{\sfi}{8R}\,\bigl(\la_1^{ab}\,\la_2^{cd}\,\ovl\theta_{i*}\,\bigl(\widehat\G_{ab}\,\widehat\G_{cd}\ox\si_1\bigr)\,\theta_{i*}+\la_1'{}^{a'b'}\,\la_2'{}^{c'd'}\,\ovl\theta_{i*}\,\bigl(\widehat\G_{a'b'}\,\widehat\G_{c'd'}\ox\si_1\bigr)\,\theta_{i*}\bigr)+O\bigl(R^{-2}\bigr)\,,\cr\cr\cr
f_{\vep,Y}\rstr_{\cO_{i*}}=-\tfrac{1}{R^{\frac{1}{2}}}\,Y^{\widehat a}\,\ovl\vep\,\bigl(\widehat\G_{\widehat a}\ox\si_3\bigr)\,\theta_{i*}+O\bigl(R^{-\frac{3}{2}}\bigr)\,,\cr\cr\cr
f_{\vep,\La}\rstr_{\cO_{i*}}=\tfrac{\sfi\,\vep_{\widehat a\widehat b}}{2R^{\frac{1}{2}}}\,\La^{\widehat a\widehat b}\,\ovl\vep\,\bigl(\widehat\G_{\widehat a\widehat b}\ox\si_1\bigr)\,\theta_{i*}+O\bigl(R^{-\frac{3}{2}}\bigr)\,,\cr\cr\cr
f_{Y,\La}\rstr_{\cO_{i*}}=-\tfrac{\vep_{\widehat b\widehat c}}{4R}\,Y^{\widehat a}\,\La^{\widehat b\widehat c}\,\ovl\theta_{i*}\,\bigl(\widehat\G_{\widehat a}\,\widehat\G_{\widehat b\widehat c}\ox\si_3\bigr)\,\theta_{i*}+O\bigl(R^{-2}\bigr)\,.
\qqq
Our asymptotic analysis identifies, on the qualitative level adopted, two independent sources of the anomaly: the Gra\ss mann-even winding modes that wrap (any representative of) the generator of $\,H_1({\rm AdS}_5\x\bS^5)\,$ that sits in $\,{\rm AdS}_5\cong\bR^{\x 4}\x\bS^1\,$ (represented by the local coordinate functions $\,X^a\,$ with nontrivial monodromy), and the Gra\ss man-odd Kosteleck\'y--Rabin states (represented by the coordinate functions $\,\theta^{\a\a'I}$). Clearly, the wrapping anomaly is sourced by the latter in the leading order (in $\,R$). On the other hand, the subleading contribution from the standard winding charges from the $\,{\rm AdS}_5\cong\bR^{\x 4}\x\bS^1\,$ sector represents an irremovable geometric effect, and so it cannot be dropped.

At this stage, there are at least two inequivalent paths that can be taken to arrive at a geometrisation of the Metsaev--Tseytlin super-3-cocycle conformable with the supersymmetry present and leading through a `superstringy' deformation of the original Lie superalgebra $\,\gt{su}(2,2\,\vert\,4)$,\ to wit, we may
\bit
\item[(i)] take the left-invariant primitive $\,\underset{\tx{\ciut{(2)}}}{\b}\,$ of the Metsaev--Tseytlin super-3-cocycle as the point of departure of the geometrisation and work with the undeformed Lie superalgebra $\,\gt{su}(2,2\,\vert\,4)\,$ upon dropping the wrapping charges altogether, or
\item[(ii)] replace $\,\underset{\tx{\ciut{(2)}}}{\b}\,$ with the non-invariantly corrected primitive \eqref{eq:AdSprimcorr} and subsequently look for deformations of $\,\gt{su}(2,2\,\vert\,4)\,$ corresponding, along the lines of Part I, to extended superspacetimes surjectively submersed over $\,{\rm s}({\rm AdS}_5\x\bS^5)\,$ on which left-invariant super-2-forms exist with the desired asymptotics of the corrections $\,\underset{\tx{\ciut{(2)}}}{\txD_i}$.
\eit
The first path is accessible owing to the invariant nature of the primitive $\,\underset{\tx{\ciut{(2)}}}{\b}\,$ but, as noted earlier, the resultant super-1-gerbe does not asymptote to the one over $\,{\rm sMink}^{1,9\,\vert\,D_{1,9}}\,$ constructed in Part I. The second one calls for a systematic exploration of deformations (or, indeed, extensions) admissible in the highly constrained category of Lie superalgebras in which we can draw hints from the previous studies of the mother supersymmetry (super)algebra $\,\gt{osp}(1\,\vert\,32)\,$ of the superstring and M-theory, {\it cp} \Rcite{Bergshoeff:2000qu}. We shall examine the former path in full detail, with our analysis culminating in a full-fledged definition of the relevant super-1-gerbe with curving $\,\underset{\tx{\ciut{(2)}}}{\b}$.\ As for the latter one, we shall explore a large class of physically motivated deformations, only to conclude that there are no deformations with the desired properties (within the class considered). These results lead us to contemplate the third path along which we
\bit
\item[(iii)] seek an alternative ($\neq\underset{\tx{\ciut{(2)}}}{\b}$) left-invariant trivialisation of $\,\underset{\tx{\ciut{(3)}}}{\chi}\,$ only on the total space of a surjective submersion over $\,{\rm s}({\rm AdS}_5\x\bS^5)\,$ that integrates a (physically motivated) associative deformation of $\,\gt{su}(2,2\,\vert\,4)$.
\eit
The last path is -- by far -- the wildest, with no more than the flat-superspace intuition as guidance at best. Partial negative results in this direction will be obtained below in the course of our study of consistent `superstring' deformations of the basis Lie superalgebra. These are, ultimately, conducive to a rethinking of the very definition\footnote{One might, {\it e.g.}, replace the Metsaev--Tseytlin super-3-cocycle with another one, having the same asymptotic behaviour in the flat limit $\,R\to\infty$.} of the Green--Schwarz super-3-cocycle of the super-$\si$-model of interest along the most general guidelines of the original paper \cite{Metsaev:1998it}, an idea that we pursue in Section \ref{sec:MTaway}. The investigation thus initiated will be taken up in an upcoming work.

\section{The trivial Green--Schwarz super-1-gerbe over the super-${\rm AdS}_5\x\bS^5\,$ space}\label{sec:trsAdSSext}

The first of the three paths outlined in the previous section begins with the earlier observation that the Metsaev--Tseytlin super-3-cocycle admits a global manifestly (left-)${\rm SU}(2,2\,\vert\,4)$-invariant primitive $\,\underset{\tx{\ciut{(2)}}}{\b}\in\Om^2({\rm SU}(2,2\,\vert\,4))\,$ pulling back along the distinguished local sections $\,\si_i\,$ of \Reqref{eq:MTsect} to elements of the trivialising cover $\,\{\cO_i\}_{i\in I}\subset\xcT({\rm s}({\rm AdS}_5\x\bS^5))\,$ of the base of the principal ${\rm SO}(4,1)\x{\rm SO}(5)$-bundle $\,{\rm SU}(2,2\,\vert\,4)\too{\rm SU}(2,2\,\vert\,4)/({\rm SO}(4,1)\x{\rm SO}(5))\equiv{\rm s}({\rm AdS}_5\x\bS^5)\,$ and thus defining restrictions
\qq\nn
\underset{\tx{\ciut{(2)}}}{\txB}\rstr_{\cO_i}:=\sfi\,\si_i^*\bigl(\ovl\Si_{\rm L}\wedge(\bd1\ox\si_1)\,\Si_{\rm L}\bigr)\,,\qquad i\in I
\qqq
of a globally smooth super-2-form $\,\underset{\tx{\ciut{(2)}}}{\txB}\,$ on $\,{\rm s}({\rm AdS}_5\x\bS^5)$.\ Consequently, we may take as the (trivial) surjective submsersion of the super-1-gerbe under reconstruction the supertarget itself,
\qq\nn
\pi_{\sfY{\rm s}({\rm AdS}_5\x\bS^5)}\equiv\id_{{\rm s}({\rm AdS}_5\x\bS^5)}\ :\ \sfY{\rm s}({\rm AdS}_5\x\bS^5):={\rm s}({\rm AdS}_5\x\bS^5)\too{\rm s}({\rm AdS}_5\x\bS^5)\,,
\qqq
with $\,\underset{\tx{\ciut{(2)}}}{\txB}\,$ as the corresponding curving. Over its fibred square
\qq\nn
\sfY^{[2]}{\rm s}({\rm AdS}_5\x\bS^5)\equiv\sfY{\rm s}({\rm AdS}_5\x\bS^5)\x_{{\rm s}({\rm AdS}_5\x\bS^5)}\sfY{\rm s}({\rm AdS}_5\x\bS^5\cong{\rm s}({\rm AdS}_5\x\bS^5)\,,
\qqq
equipped with the canonical projections $\,\pr_\a,\ \a\in\{1,2\}$,\ identifiable with $\,\id_{{\rm s}({\rm AdS}_5\x\bS^5)}$,\ we obtain the trivial identity
\qq\nn
(\pr_2^*-\pr_1^*)\underset{\tx{\ciut{(2)}}}{\txB}\equiv0
\qqq
which enables us to take the trivial principal $\bC^\x$-bundle
\qq\nn
\pi_\xcL\equiv\pr_1\ :\ \xcL:=\sfY^{[2]}{\rm s}({\rm AdS}_5\x\bS^5)\x\bC^\x\too{\rm s}({\rm AdS}_5\x\bS^5)\ :\ (Z_i,Z_i,z)\longmapsto Z_i
\qqq
with the trivial principal connection
\qq\nn
\nabla_\xcL:=\sfd\,,
\qqq
or, equivalently, with a principal connection super-1-form
\qq\nn
\cA\bigl(Z_i,Z_i,z\bigr):=\sfi\,\tfrac{\sfd z}{z}
\qqq
as the data of the super-1-gerbe. The latter is manifestly invariant under the component-wise (non-linear) action of the \emph{product} Lie supergroup
\qq\nn
\widetilde{{\rm SU}(2,2\,\vert\,4)}:={\rm SU}(2,2\,\vert\,4)\x\bC^\x
\qqq
which, in the notation of \Reqref{eq:cosetactptcoord}, takes the form
\qq\nn
\widetilde{[\la]}_\cdot\ :\ \widetilde{{\rm SU}(2,2\,\vert\,4)}\x\xcL\too\xcL\ :\ \bigl((g,\z),(Z_i,Z_i,z)\bigr)\longmapsto\bigl(\widetilde Z_j(Z_i,g),\widetilde Z_j(Z_i,g),\z\cdot z\bigr)\,.
\qqq
Hence, the triple $\,(\xcL,\pi_\xcL,\cA)\,$ could be called a (trivial) super-0-gerbe in what seems a natural generalisation of Def.\,I.5.4. to the setting of a homogeneous space of a Lie supergroup.

In the last step, we consider the cartesian cube of the above trivial surjective submersion fibred over $\,({\rm s}({\rm AdS}_5\x\bS^5)$,
\qq\nn
\sfY^{[3]}{\rm s}({\rm AdS}_5\x\bS^5)\equiv\sfY{\rm s}({\rm AdS}_5\x\bS^5)\x_{{\rm s}({\rm AdS}_5\x\bS^5)}\sfY{\rm s}({\rm AdS}_5\x\bS^5)\x_{{\rm s}({\rm AdS}_5\x\bS^5)}\sfY{\rm s}({\rm AdS}_5\x\bS^5)\cong{\rm s}({\rm AdS}_5\x\bS^5)\,,
\qqq
(with its canonical projections $\,\pr_{\a,\b}\equiv(\pr_\a,\pr_\b),\ (\a,\b)\in\{(1,2),(2,3),(1,3)\}\,$ to the previously considered fibred square), and, over it, take the connection-preserving isomorphism
\qq\nn
\mu_\xcL\ :\ \pr_{1,2}^*\xcL\ox\pr_{2,3}^*\xcL\xrightarrow{\ \cong\ }\pr_{1,3}^*\xcL\ :\ \left[\bigl(Z_i,Z_i,z_{1,2}\bigr),\bigl(Z_i,Z_i,z_{2,3}\bigr)\right]\longmapsto\bigl(Z_i,Z_i,z_{1,2}\cdot z_{2,3}\bigr)\,,
\qqq
where we have identified
\qq\nn
\bigl(Z_i,Z_i,z_{\a,\b}\bigr)\equiv\bigl(Z_i,(Z_i,Z_i,z_{\a,\b})\bigr)\in\pr_{\a,\b}^*\xcL\,.
\qqq
A fibre-bundle map thus defined trivially satisfies the standard groupoid identity over $\,\sfY^{[4]}{\rm s}({\rm AdS}_5\x\bS^5)\cong{\rm s}({\rm AdS}_5\x\bS^5)\,$ and conforms structurally with the description of a super-0-gerbe isomorphism given in Def.\,I.5.4.

Our analysis results in
\bedef\label{def:s1gerbeAdS}
The \textbf{trivial Metsaev--Tseytlin super-1-gerbe} over $\,{\rm s}({\rm AdS}_5\x\bS^5)\,$ of curvature $\,\underset{\tx{\ciut{(3)}}}{\chi}^{\rm MT}\,$ is the sextuple 
\qq\nn
\cI^{\rm MT}_{\underset{\tx{\ciut{(2)}}}{\txB}}:=\bigl(\sfY{\rm s}({\rm AdS}_5\x\bS^5),\pi_{\sfY{\rm s}({\rm AdS}_5\x\bS^5)},\underset{\tx{\ciut{(2)}}}{\txB},\xcL,\nabla_\xcL,\mu_\xcL\bigr)
\qqq
constructed in the preceding paragraphs.
\exdef

Clearly, the latter does \emph{not} reproduce the non-supersymmetric curving \eqref{eq:sMincurv} of the Green--Schwarz super-1-gerbe on ${\rm sMink}^{1,9\,\vert\,D_{9,1}}\,$ (not even in restriction to positive-chirality spinors).

\section{The Kamimura--Sakaguchi supercentral charge extensions of $\,\gt{su}(2,2\,\vert\,4)$}\label{sec:KamSakext}

The most natural class of deformations encountered along the second path indicated at the end of Sec.\,\ref{sec:ssextaAdSS}, and well known independently from a variety of field-theoretic contexts, consists of super-central extensions. In order to identify, in this setting, the potential source of a (supersymmetric) correction to the previously considered primitive $\,\underset{\tx{\ciut{(2)}}}{\b}$,\ we shall consider the most general such extension
\qq\nn
\widetilde{\gt{su}(2,2\,\vert\,4)}=\gt{su}(2,2\,\vert\,4)\oplus\corr{Z_{\a\a'I\,\b\b'J},Z_{\widehat a\,\widehat b},Z_{\widehat a\widehat b\,\widehat c\widehat d},Z_{\widehat a\,\widehat b\widehat c},Z_{\a\a'I\,\widehat a},Z_{\a\a'I\,\widehat a\widehat b}}_\bC
\qqq
of $\,\gt{su}(2,2\,\vert\,4)$,\ as in \Reqref{eq:wrapcentrext}, and study the asymptotics of the pullback of the Maurer--Cartan super-1-form from the Lie supergroup $\,\widetilde{{\rm SU}(2,2\,\vert\,4)}\,$ integrating the extension,
\qq\nn
\bd1\too\exp\bigl(\corr{Z_{\a\a'I\,\b\b'J},Z_{\widehat a\,\widehat b},Z_{\widehat a\widehat b\,\widehat c\widehat d},Z_{\widehat a\,\widehat b\widehat c},Z_{\a\a'I\,\widehat a},Z_{\a\a'I\,\widehat a\widehat b}}_\bC\bigr)\too\widetilde{{\rm SU}(2,2\,\vert\,4)}\xrightarrow{\ \widetilde\pi\ }{\rm SU}(2,2\,\vert\,4)\too\bd1\,,
\qqq
along a distinguished (local) section
\qq
\widetilde\si_i\ &:&\ \widetilde\cO_i\too\widetilde{{\rm SU}(2,2\,\vert\,4)}\cr\cr
&:&\ (Z_i,\z_i)\longmapsto\ee^{X_i^{\widehat a}\,P_{\widehat a}}\cdot\ee^{\theta_i^{\a\a' I}\,Q_{\a\a' I}}\cdot\ee^{\z_i^{\a\a'I\,\b\b'J}\,Z_{\a\a'I\,\b\b'J}+\z_i^{\widehat a\,\widehat b}\,Z_{\widehat a\,\widehat b}+\z_i^{\widehat a\widehat b\,\widehat c\widehat d}\,Z_{\widehat a\widehat b\,\widehat c\widehat d}+\z_i^{\widehat a\,\widehat b\widehat c}\,Z_{\widehat a\,\widehat b\widehat c}+\z_i^{\a\a'I\,\widehat a}\,Z_{\a\a'I\,\widehat a}+\z_i^{\a\a'I\,\widehat a\widehat b}\,Z_{\a\a'I\,\widehat a\widehat b}}\,.\cr \label{eq:distlocsecext}
\qqq
In so doing, we shall not be concerned with the potential linear dependence of the various super-central charges defining the extension, a feature of no bearing on our conclusions at our level of generality.

Calculating the leading asymptotics of expression \eqref{eq:MCs1f} once again, but this time for the super-centrally extended Lie superalgebra, we eventually arrive at
\qq\nn
\widetilde\si_i^*\theta_{\rm L}(Z_i,\z_i)=\sfd\z_i^{\a\a'I\,\b\b'J}\ox Z_{\a\a'I\,\b\b'J}+\sfd\z_i^{\widehat a\,\widehat b}\ox Z_{\widehat a\,\widehat b}+\sfd\z_i^{\widehat a\widehat b\,\widehat c\widehat d}\ox Z_{\widehat a\widehat b\,\widehat c\widehat d}+\sfd\z_i^{\widehat a\,\widehat b\widehat c}\ox Z_{\widehat a\,\widehat b\widehat c}+\sfd\z_i^{\a\a'I\,\widehat a}\ox Z_{\a\a'I\,\widehat a}\cr\cr
+\sfd\z_i^{\a\a'I\,\widehat a\widehat b}\ox Z_{\a\a'I\,\widehat a\widehat b}+\tfrac{1}{R^{\frac{1}{2}}}\,\sfd\theta_i^{\a\a'I}\ox Q_{\a\a'I}\cr\cr
+\tfrac{1}{R}\,\bigl[\bigl(\sfd X_i^{\widehat a}-\sfi\,\ovl\theta_i\,(\widehat\G^{\widehat a}\ox\bd1)\,\sfd\theta_i\bigr)\ox P_{\widehat a}+\tfrac{\sfi}{2}\,\ovl\theta_i\,(\widehat\G^{\widehat a\widehat b}\ox\si_2)\,\sfd\theta_i\ox J_{\widehat a\widehat b}+\tfrac{1}{2}\,\theta_i^{\a\a'I}\,\sfd\theta_i^{\b\b'J}\ox Z_{\a\a'I\,\b\b'J}\bigr]\cr\cr
+\tfrac{1}{R^{\frac{3}{2}}}\,\bigl\{\bigl[\tfrac{1}{2}\,\bigl(\sfd X_i^{\widehat a}-\tfrac{\sfi}{3}\,\ovl\theta_i\,(\widehat\G^a\ox\bd1)\,\sfd\theta_i\bigr)\,(\widehat\G_{\widehat a}\ox\si_2)^{\a\a'I}_{\ \b\b'J}+\tfrac{\sfi\,\vep_{\widehat a\widehat b}}{12}\,\bigl(\ovl\theta_i\,(\widehat\G^{\widehat a\widehat b}\ox\si_2)\,\sfd\theta_i\,(\widehat\G_{\widehat a\widehat b}\ox\bd1)^{\a\a'I}_{\ \b\b'J}\bigr]\,\theta_i^{\b\b'J}\ox Q_{\a\a' I}\cr\cr
-\theta_i^{\a\a'I}\,\bigl[\bigl(\sfd X_i^{\widehat a}-\tfrac{\sfi}{3}\,\ovl\theta_i\,\bigl(\widehat\G^{\widehat a}\ox\bd1\bigr)\,\sfd\theta_i\bigr)\ox Z_{\a\a'I\,\widehat a}+\tfrac{\sfi}{3!}\,\ovl\theta_i\,\bigl(\widehat\G^{\widehat a\widehat b}\ox\si_2\bigr)\,\sfd\theta_i\ox Z_{\a\a'I\,\widehat a\widehat b}\bigr]\bigr\}\cr\cr
-\tfrac{1}{2R^2}\,\bigl\{\bigl(\sfi\,\bigl(\sfd X_i^{\widehat b}-\tfrac{\sfi}{6}\,\ovl\theta_i\,\bigl(\widehat\G^{\widehat b}\ox\bd1\bigr)\,\sfd\theta_i\bigr)\,\ovl\theta_i\,\bigl(\{\widehat\G^{\widehat a},\widehat\G_{\widehat b}\}\ox\si_2\bigr)\,\theta_i-\tfrac{\vep_{\widehat b\widehat c}}{12}\,\ovl\theta_i\,\bigl(\widehat\G_{\widehat b\widehat c}\ox\si_2\bigr)\,\sfd\theta_i\cdot\ovl\theta_i\,\bigl(\widehat\G^{\widehat a}\,\widehat\G^{\widehat b\widehat c}\ox\bd1\bigr)\,\theta_i\bigr)\ox P_{\widehat a}\cr\cr
-\bigl(\tfrac{\sfi}{2}\,\bigl(\sfd X_i^{\widehat c}-\tfrac{\sfi}{3!}\,\ovl\theta_i\,\bigl(\widehat\G^{\widehat c}\ox\bd1\bigr)\,\sfd\theta_i\bigr)\,\ovl\theta_i\,\bigl(\widehat\G^{\widehat a\widehat b}\,\widehat\G_{\widehat c}\ox\bd1\bigr)\,\theta_i-\tfrac{\vep_{\widehat a\widehat b}}{4!}\,\ovl\theta_i\,\bigl(\widehat\G^{\widehat c\widehat d}\ox\si_2\bigr)\,\sfd\theta_i\cdot\ovl\theta_i\,\bigl(\widehat\G^{\widehat a\widehat b}\,\widehat\G_{\widehat c\widehat d}\ox\si_2\bigr)\,\theta_i-\vep_{\widehat a\widehat b}\,X_i^{\widehat a}\,\sfd X_i^{\widehat b}\bigr)\ox J_{\widehat a\widehat b}\cr\cr
-\tfrac{1}{2}\,\bigl[\bigl(\sfd X_i^{\widehat a}-\tfrac{\sfi}{6}\,\ovl\theta_i\,\bigl(\widehat\G^{\widehat a}\ox\bd1\bigr)\,\sfd\theta_i\bigr)\,\bigl(\widehat\G_{\widehat a}\ox\si_2\bigr)^{\b\b'J}_{\ \g\g'K}+\tfrac{\sfi\vep_{\widehat a\widehat b}}{12}\,\ovl\theta_i\,\bigl(\widehat\G^{\widehat a\widehat b}\ox\si_2\bigr)\,\sfd\theta_i\,\bigl(\widehat\G_{\widehat a\widehat b}\ox\bd1\bigr)^{\b\b'J}_{\ \g\g'K}\bigr]\,\theta_i^{\a\a'I}\,\theta_i^{\g\g'K}\ox Z_{\a\a'I\,\b\b'J}\cr\cr
+X^{\widehat a}_i\,\sfd X^{\widehat b}_i\ox Z_{\widehat a\widehat b}\bigr\}+O\bigl(R^{-\frac{5}{2}}\bigr)\,.
\qqq
Inspection of the above formula reveals that the unique left-invariant super-1-form whose exterior derivative possesses the desired asymptotics and could, consequently, be considered as the candidate for a supersymmertic extension of the corrections $\,\underset{\tx{\ciut{(2)}}}{\txD_i}\,$ on $\,\widetilde{{\rm SU}(2,2\,\vert\,4)}\,$ (and so also on $\,\widetilde{{\rm SU}(2,2\,\vert\,4)}/({\rm SO}(4,1)\x{\rm SO}(5))$) is the linear combination of the components of the Maurer--Cartan super-1-form associated with the wrapping charges $\,Z_{\a\a'I\ \b\b'J}$,
\qq\nn
\widetilde\si_i^*\theta_{\rm L}^{\a\a'I\ \b\b'J}(Z_i,\z_i)&=&\sfd\z_i^{\a\a'I\ \b\b'J}+\tfrac{1}{4R}\,\bigl(\theta_i^{\a\a' I}\,\sfd\theta_i^{\b\b' J}+\theta_i^{\b\b' J}\,\sfd\theta_i^{\a\a' I}\bigr)\cr\cr
&&+\tfrac{1}{4R^2}\,\bigl[\bigl(\sfd X_i^{\widehat a}-\tfrac{\sfi}{6}\,\ovl\theta_i\,\bigl(\widehat\G^{\widehat a}\ox\bd1\bigr)\,\sfd\theta_i\bigr)\,\bigl(\widehat\G_{\widehat a}\ox\si_2\bigr)^{(\b\b'J}_{\ \g\g'K}\,\theta_i^{\a\a'I)}\cr\cr
&&+\tfrac{\sfi\vep_{\widehat a\widehat b}}{12}\,\ovl\theta_i\,\bigl(\widehat\G^{\widehat a\widehat b}\ox\si_2\bigr)\,\sfd\theta_i\,\bigl(\widehat\G_{\widehat a\widehat b}\ox\bd1\bigr)^{(\b\b'J}_{\ \g\g'K}\,\theta_i^{\a\a'I)}\bigr]\,\theta_i^{\g\g'K}+O\bigl(R^{-3}\bigr)\,,
\qqq
with coefficients $\,2\sfi\,(\widehat C\ox\si_1)_{\a\a'I\b\b'J}$.\ Indeed, upon taking into account the Fierz identity that ensures the vanishing of the super-Jacobiator
\qq\label{eq:sJacQQQ}
{\rm sJac}(Q_{\a\a'I},Q_{\b\b'J},Q_{\g\g'K})=0
\qqq
of $\,\gt{su}(2,2\,\vert\,4)$,\ or -- explicitly --
\qq\nn
\vep_{\widehat a\widehat b}\,\bigl(\widehat C\,\widehat\G^{\widehat a\widehat b}\ox\si_2\bigr)_{(\a\a'I\b\b'J}\,\bigl(\widehat\G_{\widehat a\widehat b}\ox\bd1\bigr)^{\d\d'L}_{\ \g\g'K)}=2\bigl(\widehat C\,\widehat\G^{\widehat a}\ox\bd1\bigr)_{(\a\a'I\b\b'J}\,\bigl(\widehat\G_{\widehat a}\ox\si_2\bigr)^{\d\d'L}_{\ \g\g'K)}\,,
\qqq
and the resultant identity (obtained through contraction with $\,(\widehat C\ox\si_1)_{\ep\ep'M\d\d'L}\,\sfd\theta_i^{\a\a'I}\wedge\sfd\theta_i^{\g\g'K}\,\theta_i^{\b\b'J}\,\theta_i^{\ep\ep'M}$)
\qq\nn
&&\vep_{\widehat a\widehat b}\,\sfd\ovl\theta_i\wedge\bigl(\widehat\G^{\widehat a\widehat b}\ox\si_2\bigr)\,\sfd\theta_i\cdot\ovl\theta_i\,\bigl(\widehat\G_{\widehat a\widehat b}\ox\si_1\bigr)\,\theta_i\cr\cr
&=&-2\vep_{\widehat a\widehat b}\,\ovl\theta_i\bigl(\widehat\G^{\widehat a\widehat b}\ox\si_2\bigr)\,\sfd\theta_i\wedge\,\ovl\theta_i\,\bigl(\widehat\G_{\widehat a\widehat b}\ox\si_1\bigr)\,\sfd\theta_i+4i\ovl\theta_i\bigl(\widehat\G^{\widehat a}\ox\bd1\bigr)\,\sfd\theta_i\wedge\,\ovl\theta_i\,\bigl(\widehat\G_{\widehat a}\ox\si_3\bigr)\,\sfd\theta_i\cr\cr
&&+2i\sfd\ovl\theta_i\wedge\bigl(\widehat\G^{\widehat a}\ox\bd1\bigr)\,\sfd\theta_i\cdot\ovl\theta_i\,\bigl(\widehat\G_{\widehat a}\ox\si_3\bigr)\,\theta_i\cr\cr
&=&-2\vep_{\widehat a\widehat b}\,\ovl\theta_i\bigl(\widehat\G^{\widehat a\widehat b}\ox\si_2\bigr)\,\sfd\theta_i\wedge\,\ovl\theta_i\,\bigl(\widehat\G_{\widehat a\widehat b}\ox\si_1\bigr)\,\sfd\theta_i+4i\ovl\theta_i\bigl(\widehat\G^{\widehat a}\ox\bd1\bigr)\,\sfd\theta_i\wedge\,\ovl\theta_i\,\bigl(\widehat\G_{\widehat a}\ox\si_3\bigr)\,\sfd\theta_i\,,
\qqq
we readily establish the equality
\qq\nn
\widetilde\si_i^*\cE(Z_i,\z_i)&:=&2\sfi\,\bigl(\widehat C\ox\si_1)_{\a\a'I\b\b'J}\,\widetilde\si_i^*\theta_{\rm L}^{\a\a'I\ \b\b'J}(Z_i,\z_i)\cr\cr
&=&2\sfi\,\bigl(\widehat C\ox\si_1)_{\a\a'I\b\b'J}\,\sfd\z_i^{\a\a'I\b\b'J}+\tfrac{\sfi}{R}\,\ovl\theta_i\,(\bd1\ox\si_1)\,\sfd\theta_i\cr\cr
&&-\tfrac{1}{2R^2}\,\bigl[\bigl(\sfd X_i^{\widehat a}-\tfrac{\sfi}{6}\,\ovl\theta_i\,\bigl(\widehat\G^{\widehat a}\ox\bd1\bigr)\,\sfd\theta_i\bigr)\,\ovl\theta_i\,\bigl(\widehat\G_{\widehat a}\ox\si_3\bigr)\,\theta+\tfrac{\vep_{\widehat a\widehat b}}{12}\,\ovl\theta_i\,\bigl(\widehat\G^{\widehat a\widehat b}\ox\si_2\bigr)\,\sfd\theta_i\cdot\ovl\theta_i\,\bigl(\widehat\G_{\widehat a\widehat b}\ox\si_1\bigr)\,\theta_i\bigr]\cr\cr
&&+O\bigl(R^{-3}\bigr)\cr\cr
&=&2\sfi\,\bigl(\widehat C\ox\si_1)_{\a\a'I\b\b'J}\,\sfd\z_i^{\a\a'I\b\b'J}+\tfrac{\sfi}{R}\,\ovl\theta_i\,(\bd1\ox\si_1)\,\sfd\theta_i-\tfrac{\vep_{\widehat a\widehat b}}{24R^2}\,\ovl\theta_i\,\bigl(\widehat\G^{\widehat a\widehat b}\ox\si_2\bigr)\,\sfd\theta_i\cdot\ovl\theta_i\,\bigl(\widehat\G_{\widehat a\widehat b}\ox\si_1\bigr)\,\theta_i\cr\cr
&&+O\bigl(R^{-3}\bigr)
\qqq
whence also
\qq\nn
\underset{\tx{\ciut{(2)}}}{\widetilde\sfD_i}(Z_i,\z_i)&:=&\sfd\widetilde\si_i^*\cE(Z_i,\z_i)=\tfrac{\sfi}{R}\,\sfd\ovl\theta_i\wedge(\bd1\ox\si_1)\,\sfd\theta_i-\tfrac{\vep_{\widehat a\widehat b}}{24R^2}\,\sfd\ovl\theta_i\wedge\bigl(\widehat\G^{\widehat a\widehat b}\ox\si_2\bigr)\,\sfd\theta_i\cdot\ovl\theta_i\,\bigl(\widehat\G_{\widehat a\widehat b}\ox\si_1\bigr)\,\theta_i\cr\cr
&&+\tfrac{\vep_{\widehat a\widehat b}}{12R^2}\,\ovl\theta_i\,\bigl(\widehat\G^{\widehat a\widehat b}\ox\si_2\bigr)\,\sfd\theta_i\wedge\ovl\theta_i\,\bigl(\widehat\G_{\widehat a\widehat b}\ox\si_1\bigr)\,\sfd\theta_i+O\bigl(R^{-3}\bigr)\cr\cr
&=&\underset{\tx{\ciut{(2)}}}{\sfD}(Z_i)+\tfrac{\sfi}{6R^2}\,\ovl\theta_i\,\bigl(\widehat\G^{\widehat a}\ox\si_3\bigr)\,\sfd\theta_i\wedge\ovl\theta_i\,\bigl(\widehat\G_{\widehat a}\ox\bd1\bigr)\,\sfd\theta_i\cr\cr
&&-\tfrac{\vep_{\widehat a\widehat b}}{6R^2}\,\ovl\theta_i\,\bigl(\widehat\G^{\widehat a\widehat b}\ox\si_1\bigr)\,\sfd\theta_i\wedge\ovl\theta_i\,\bigl(\widehat\G_{\widehat a\widehat b}\ox\si_2\bigr)\,\sfd\theta_i+O\bigl(R^{-3}\bigr)\,.
\qqq
This yields the desired asymptotics
\qq\nn
\widetilde\si_i^*\widetilde\pi^*\underset{\tx{\ciut{(2)}}}{\b}(Z_i,\z_i)-\underset{\tx{\ciut{(2)}}}{\widetilde\sfD_i}(Z_i,\z_i)&=&\tfrac{1}{R^2}\,\ovl\theta_i\,\bigl(\widehat\G_{\widehat a}\ox\si_3\bigr)\,\sfd\theta_i\wedge\bigl(\sfd X_i^{\widehat a}-\tfrac{\sfi}{2}\,\ovl\theta_i\,(\widehat\G^a\ox\bd1)\,\sfd\theta_i\bigr)+O\bigl(R^{-3}\bigr)\,.
\qqq
The terms of order $\,R^{-2}\,$ independent of $\,X^{\widehat a}\,$ sum up to zero for the asymptotic ten-dimensional spinors of positive chirality -- this is none other than the super-Minkowskian Fierz identity ensuring the vanishing of the super-Jacobiator \eqref{eq:sJacQQQ} for the super-Poincar\'e algebra. Thus, we have reproduced the (manifestly non-supersymmetric) primitive of the super-Minkowskian Green--Schwarz super-3-cocycle (and, in so doing, the result of \Rcite{Hatsuda:2002hz}) through our $\widehat{{\rm SU}(2,2\,\vert\,4)}$-invariant analysis. It remains to be checked whether such a correction can be obtained from a super-central extension of the original Lie superalgebra $\,\gt{su}(2,2\,\vert\,4)$.\ One may also rephrase this question of internal consistency so as to readily answer it in the negative for a large class of extensions. We begin by noting that in the above reasoning, we assumed that \emph{all} the $\,Z_{\a\a'I\,\b\b'J}\,$ are allowed as Gra\ss mann-even charges extending the original Lie superalgebra $\,\gt{su}(2,2\,\vert\,4)\,$ in a consistent manner -- this is the assumption behind the left-invariance of the component super-1-forms which allows us to take the desired linear combination of the super-1-forms (with coefficients given by ${\rm SO}(4,1)\x{\rm SO}(5)$-invariant tensors) and obtain -- upon differentiation -- the sought-after left-invariant super-2-form. In fact, we need to assume less, to wit, that the specific linear combination $\,2\sfi\,(\widehat C\ox\si_1)_{\a\a'I\b\b'J}\,\theta_{\rm L}^{\a\a'I\b\b'J}\,$ be left-invariant, or -- equivalently -- that a consistent Lie-superalgebraic deformation of the form
\qq\nn
\{Q_{\a\a' I},Q_{\b\b' J}\}^\sim=\sfi\,\bigl(-2(\widehat C\,\widehat\G^{\widehat a}\ox\bd1)_{\a\a'I\b\b'J}\,P_{\widehat a}+(\widehat C\,\widehat\G^{\widehat a\widehat b}\ox\si_2)_{\a\a'I\b\b'J}\,J_{\widehat a\widehat b}\bigr)+2\sfi\,\bigl(\widehat C\ox\si_1\bigr)_{\a\a'I\,\b\b'J}\,Z
\qqq
be allowed as the left-invariant super-1-form associated with charge $\,Z\,$ is $\,\cE$.\ It is easy to see that such a supposition is untenable as it leads to a contradiction, at least as long as the anticommutator of the supercharges is the only place where $\,Z\,$ appears. Indeed, whenever this is the case, we find, in virtue of the Maurer--Cartan equation,
\qq\nn
\sfd\cE=\tfrac{1}{2}\,2\sfi\,\bigl(\widehat C\ox\si_1\bigr)_{\a\a'I\,\b\b'J}\,\theta^{\a\a'I}_{\rm L}\wedge\theta^{\b\b'J}_{\rm L}\equiv\widetilde\pi^*\underset{\tx{\ciut{(2)}}}{\b}\,,
\qqq
which is manifestly at variance with the result derived previously (and ruins our plan of correcting the asymptotics of $\,\underset{\tx{\ciut{(2)}}}{\b}\,$ through the substraction of $\,\sfd\cE$). Below, we methodically check that (and see why) the desired deformation of $\,\gt{su}(2,2\,\vert\,4)\,$ is inconsistent with the assumption of associativity in a large class of geometrically motivated central-extensions.

A full-blown cartography of the \emph{entire} space of associative deformations of $\,\gt{su}(2,2\,\vert\,4)$,\ and even of the subspace of \emph{all} super-central extensions (without any further constraints) goes beyond the scope of the current report. We shall, instead, explore various corners of those spaces, guided by the physical intuition and algebraic hints from the analysis of the asymptotics of the Metsaev--Tseytlin super-3-cocycle and previous studies of the super-Minkowskian setting. We begin by considering an \emph{arbitrary} deformation of the anti-commutator of super-charges, keeping in mind its irremovable geometric germ expressible as the monodromy of the coordinate function $\,X^a\,$ in the vicinity of the unital coset $\,{\rm SO}(4,1)\x{\rm SO}(5)$.\ Thus, we write
\qq\label{eq:KSextend}\qquad\qquad
\{Q_{\a\a' I},Q_{\b\b' J}\}^\sim=\sfi\,\bigl(-2(\widehat C\,\widehat\G^{\widehat a}\ox\bd1)_{\a\a'I\b\b'J}\,P_{\widehat a}+(\widehat C\,\widehat\G^{\widehat a\widehat b}\ox\si_2)_{\a\a'I\b\b'J}\,J_{\widehat a\widehat b}\bigr)+Z_{\a\a'I\,\b\b'J}\,,
\qqq
with the $\,Z_{\a\a'I\,\b\b'J}=Z_{(\a\a'I\,\b\b'J)}\,$ central, and keep all other super-commutators unchanged. A moment's thought convinces us that the only super-Jacobi identities to be imposed are the following ones:
\qq\nn
{\rm sJac}(Q_{\a\a'I},Q_{\b\b'J},P_{\widehat a})=0={\rm sJac}(Q_{\a\a'I},Q_{\b\b'J},J_{\widehat a\widehat b})\,.
\qqq
Upon invoking their undeformed counterparts, the former ones give us the following constraints:
\qq\nn
Z_{\a\a'I\,\g\g'K}\,\bigl(\widehat\G_{\widehat a}\ox\si_2\bigr)^{\g\g'K}_{\ \b\b'J}+Z_{\b\b'J\,\g\g'K}\,\bigl(\widehat\G_{\widehat a}\ox\si_2\bigr)^{\g\g'K}_{\ \a\a'I}=0\,,\quad\widehat a\in\ovl{0,9}\,,
\qqq
or, equivalently,
\qq\nn
0&=&Z_{\a\a'I\,\b\b'J}\,\bigl(\d^{\a\a'I}_{\ \g\g'K}\,\d^{\b\b'J}_{\d\d'L}+\bigl(-\widehat\G_{\widehat a}\ox\sfi\,\si_2\bigr)^{\a\a'I}_{\ \g\g'K}\,\bigl(\vep_{\widehat a}\,\widehat\G^{\widehat a}\ox\sfi\,\si_2\bigr)^{\b\b'J}_{\ \d\d'L}\bigr)\cr\cr
&\equiv&Z_{\a\a'I\,\b\b'J}\,\bigl(\bd1\ox\bd1+\bigl(\vep_{\widehat a}\,\widehat\G^{\widehat a}\ox\sfi\,\si_2\bigr)^{-1}\ox\bigl(\vep_{\widehat a}\,\widehat\G^{\widehat a}\ox\sfi\,\si_2\bigr)\bigr)^{\a\a'I\b\b'J}_{\ \g\g'K\d\d'L}\,,
\qqq
written for
\qq\nn
\vep_{\widehat a}=\left\{ \barr{cl} +1 & \tx{if}\ \widehat a\in\ovl{0,4} \\
-1 & \tx{if}\ \widehat a\in\ovl{5,9}\earr\right.
\qqq
(\emph{no} summation over the range of the spacetime index $\,\widehat a$!). The linear operators annihilating the central charges,
\qq\nn
\sfP_{\widehat a}:=\tfrac{1}{2}\,\bigl(\bd1\ox\bd1+\pi_{\widehat a}^{-1}\ox\pi_{\widehat a}\bigr)\,,\qquad\pi_{\widehat a}=\vep_{\widehat a}\,\widehat\G^{\widehat a}\ox\sfi\,\si_2\,,\qquad\qquad\widehat a\in\ovl{0,9}\,,
\qqq
are in fact projectors,
\qq\nn
\sfP_{\widehat a}\cdot\sfP_{\widehat a}=\sfP_{\widehat a}\,,
\qqq
and we may summarise our first result in the concise form
\qq\label{eq:ZinKerPs}
(Z_{\a\a'I\,\b\b'J})\in\bigcap_{\widehat a=0}^9\,\ker\,\sfP_{\widehat a}\,.
\qqq
Here, the kernels are to be understood as subspaces within the 528-dimensional space $\,\bC(32)^{\rm sym}\,$ of (complex) symmetric matrices of size 32, a subspace in the 1024-dimensional $\bC$-linear space
\qq\nn
\bC(32)\cong{\rm Cliff}\bigl(\bR^{9,1}\bigr)^\bC\,.
\qqq
The last isomorphism is not invoked accidetally -- indeed, it can actually be employed in a systematic anlysis of the problem in hand. To this end, we decompose the central charge $\,Z_{\a\a'I\,\b\b'J}\,$ in the Clifford basis \eqref{eq:C32symbas} of $\,\bC(32)^{\rm sym}\,$ as
\qq\nn
Z_{\a\a'I\,\b\b'J}=\cZ_\la\,(\cC\,\unl\G^\la)_{\a\a'I\,\b\b'J}\,,
\qqq
and subsequently use the various properties of the Clifford algebras involved, {\it cp} App.\,\ref{app:CliffAdSS}, to identify $\,\bigcap_{\widehat a=0}^9\,\ker\,\sfP_{\widehat a}$.\ Prior to that, however, we take a closer look at the remaining set of constraints, following from the nullification of the other set of super-Jacobiators. We obtain
\qq\nn
Z_{\a\a'I\,\g\g'K}\,\bigl(\widehat\G_{\widehat a\widehat b}\ox\bd1\bigr)^{\g\g'K}_{\ \b\b'J}+Z_{\b\b'J\,\g\g'K}\,\bigl(\widehat\G_{\widehat a\widehat b}\ox\bd1\bigr)^{\g\g'K}_{\ \a\a'I}=0\,,\qquad\widehat a\in\ovl{0,9}\,,
\qqq
or, equivalently,
\qq\nn
(Z_{\a\a'I\,\b\b'J})\in\bigcap_{\widehat a,\widehat b=0}^9\,\ker\,\sfP_{\widehat a\widehat b}\,,
\qqq
where
\qq\nn
\sfP_{\widehat a\widehat b}:=\tfrac{1}{2}\,\bigl(\bd1\ox\bd1+\bigl(\vep_{\widehat a\widehat b}\,\widehat\G^{\widehat a\widehat b}\ox\bd1\bigr)^{-1}\ox\bigl(\vep_{\widehat a\widehat b}\,\widehat\G^{\widehat a\widehat b}\ox\bd1\bigr)\bigr)\,,\qquad(\widehat a,\widehat b)\in\ovl{0,4}^{\x 2}\cup\ovl{5,9}^{\x 2}\,,\ \widehat a\neq\widehat b
\qqq
form another set of projectors, written in terms of the inverses (as previously, no summation over repeated indices)
\qq\nn
\bigl(\vep_{\widehat a\widehat b}\,\widehat\G^{\widehat a\widehat b}\ox\bd1\bigr)^{-1}=-\eta_{\widehat a\widehat a}\,\eta_{\widehat b\widehat b}\,\vep_{\widehat a\widehat b}\,\widehat\G^{\widehat a\widehat b}\ox\bd1\equiv-\vep_{\widehat a\widehat b}\,\widehat\G_{\widehat a\widehat b}\ox\bd1\,.
\qqq
We have, for $\,(\widehat a,\widehat b)\in\ovl{0,4}^{\x 2}\cup\ovl{5,9}^{\x 2}\,$ and $\,\widehat a\neq\widehat b$,\ the identities
\qq\nn
\sfP_{\widehat a}\cdot\sfP_{\widehat b}&=&\tfrac{1}{4}\,\bigl(\bd1\ox\bd1-\bigl(\widehat\G_{\widehat a}\ox\sfi\,\si_2\bigr)\ox\bigl(\vep_{\widehat a}\,\widehat\G^{\widehat a}\ox\sfi\,\si_2\bigr)-\bigl(\widehat\G_{\widehat b}\ox\sfi\,\si_2\bigr)\ox\bigl(\vep_{\widehat b}\,\widehat\G^{\widehat b}\ox\sfi\,\si_2\bigr)+\bigl(\widehat\G_{\widehat a}\,\widehat\G_{\widehat b}\ox\bd1\bigr)\ox\bigl(\vep_{\widehat a}\,\vep_{\widehat b}\,\widehat\G^{\widehat a}\,\widehat\G^{\widehat b}\ox\bd1\bigr)\bigr)\cr\cr
&=&\tfrac{1}{4}\,\bigl(\bd1\ox\bd1-\bigl(\widehat\G_{\widehat a}\ox\sfi\,\si_2\bigr)\ox\bigl(\vep_{\widehat a}\,\widehat\G^{\widehat a}\ox\sfi\,\si_2\bigr)-\bigl(\widehat\G_{\widehat b}\ox\sfi\,\si_2\bigr)\ox\bigl(\vep_{\widehat b}\,\widehat\G^{\widehat b}\ox\sfi\,\si_2\bigr)+\bigl(\widehat\G_{\widehat a}\,\widehat\G_{\widehat b}\ox\bd1\bigr)\ox\bigl(\widehat\G^{\widehat a}\,\widehat\G^{\widehat b}\ox\bd1\bigr)\bigr)\cr\cr
&=&\tfrac{1}{4}\,\bigl(\bd1\ox\bd1-\bigl(\widehat\G_{\widehat a}\ox\sfi\,\si_2\bigr)\ox\bigl(\vep_{\widehat a}\,\widehat\G^{\widehat a}\ox\sfi\,\si_2\bigr)-\bigl(\widehat\G_{\widehat b}\ox\sfi\,\si_2\bigr)\ox\bigl(\vep_{\widehat b}\,\widehat\G^{\widehat b}\ox\sfi\,\si_2\bigr)+\bigl(\vep_{\widehat a\widehat b}\,\widehat\G_{\widehat a\widehat b}\ox\bd1\bigr)\ox\bigl(\vep_{\widehat a\widehat b}\,\widehat\G^{\widehat a\widehat b}\ox\bd1\bigr)\bigr)\cr\cr
&=&\tfrac{1}{4}\,\bigl(\bd1\ox\bd1-\bigl(\widehat\G_{\widehat a}\ox\sfi\,\si_2\bigr)\ox\bigl(\vep_{\widehat a}\,\widehat\G^{\widehat a}\ox\sfi\,\si_2\bigr)\bigr)+\tfrac{1}{4}\,\bigl(\bd1\ox\bd1-\bigl(\widehat\G_{\widehat b}\ox\sfi\,\si_2\bigr)\ox\bigl(\vep_{\widehat b}\,\widehat\G^{\widehat b}\ox\sfi\,\si_2\bigr)\bigr)\cr\cr
&&-\tfrac{1}{4}\,\bigl(\bd1\ox\bd1-\bigl(\vep_{\widehat a\widehat b}\,\widehat\G_{\widehat a\widehat b}\ox\bd1\bigr)\ox\bigl(\vep_{\widehat a\widehat b}\,\widehat\G^{\widehat a\widehat b}\ox\bd1\bigr)\bigr)\equiv\tfrac{1}{2}\,\bigl(\sfP_{\widehat a}+\sfP_{\widehat b}-\sfP_{\widehat a\widehat b}\bigr)\,,
\qqq
and so
\qq\nn
2\bigl(\sfP_{\widehat a}+\sfP_{\widehat b}\bigr)-\bigl(\sfP_{\widehat a}+\sfP_{\widehat b}\bigr)^2=\sfP_{\widehat a}+\sfP_{\widehat b}-\sfP_{\widehat a}\cdot\sfP_{\widehat b}-\sfP_{\widehat b}\cdot\sfP_{\widehat a}=\sfP_{\widehat a\widehat b}\,.
\qqq
From the above, we infer
\qq\nn
\bigcap_{\widehat a=0}^9\,\ker\,\sfP_{\widehat a}\subset\bigcap_{\widehat a,\widehat b=0}^9\,\ker\,\sfP_{\widehat a\widehat b}\,,
\qqq
which leaves us with the original constraint \eqref{eq:ZinKerPs} as the only one to be imposed. This we do in the aforementioned Clifford basis.

We begin by noting the symmetricity of
\qq\nn
\cC\,\pi_{\widehat a}\equiv-\sfi\,\widetilde\G^{0\widehat a}\,,
\qqq
and from that we derive the identity
\qq\nn
\bigl(\cC\,\unl\G^\la\bigr)_{\a\a'I\b\b'J}\,\bigl(\pi_{\widehat a}^{-1}\bigr)^{\a\a'I}_{\ \g\g'K}\,\bigl(\pi_{\widehat a}\bigr)^{\b\b'J}_{\ \d\d'L}=-\bigl(\cC\,\pi_{\widehat a}\bigr)_{\ep\ep'M\d\d'L}\,\bigl(\unl\G^\la\,\pi_{\widehat a}^{-1}\bigr)^{\ep\ep'M}_{\ \g\g'K}=-\bigl(\cC\,\pi_{\widehat a}\,\unl\G^\la\,\pi_{\widehat a}^{-1}\bigr)_{\d\d'L\g\g'K}\,,
\qqq
which further yields
\qq\nn
2\bigl(\cC\,\unl\G^\la\bigr)_{\a\a'I\b\b'J}\,\bigl(\sfP_{\widehat a}\bigr)^{\a\a'I\b\b'J}_{\ \g\g'K\d\d'L}=\cZ_\la\,\bigl(\cC\,[\unl\G^\la,\pi_{\widehat a}]\,\pi_{\widehat a}^{-1}\bigr)_{\d\d'L\g\g'K}\,.
\qqq
Thus we may rephrase the original condition \eqref{eq:ZinKerPs} in the easily tractable form
\qq\nn
\cZ_\la\,[\unl\G^\la,\pi_{\widehat a}]=0\,,\qquad\widehat a\in\ovl{0,9}\,.
\qqq
As the $\,\pi_{\widehat a}\,$ are (up to a trivial rescaling) among the $\,\unl\G^\mu$,\ to wit,
\qq\nn
\sfi\,\pi_{\widehat a}\equiv\unl\G^{0\widehat a}\,,
\qqq
and the $\,\unl\G^\la\,$ span, as linearly independent generators, a Lie algebra over $\,\bC\,$ with certain structure constants $\,c^{\mu\nu}_{\ \ \k}\in\bC$,\ {\it i.e.},
\qq\nn
[\unl\G^\mu,\unl\G^\nu]=c^{\mu\nu}_{\ \ \k}\,\unl\G^\k\,,
\qqq
we may further rewrite the above condition as
\qq\nn
\forall_{(\widehat a,\mu)\in\ovl{0,9}\x\ovl{0,527}}\ :\ \cZ_\la\,c^{\la\,0\widehat a}_{\ \ \ \ \mu}=0\,.
\qqq
It now suffices to calculate the relevant structure constants to prove the fundamental
\berop\label{prop:KS}
In the notation introduced above,
\qq\nn
\bigcap_{\widehat a=0}^9\,\ker\,\sfP_{\widehat a}=\corr{\widetilde\G^0}_\bC\,,
\qqq
and so the only admissible central extension of the type discussed takes the form
\qq\nn
\{Q_{\a\a' I},Q_{\b\b' J}\}^\sim=\sfi\,\bigl(-2(\widehat C\,\widehat\G^{\widehat a}\ox\bd1)_{\a\a'I\b\b'J}\,P_{\widehat a}+(\widehat C\,\widehat\G^{\widehat a\widehat b}\ox\si_2)_{\a\a'I\b\b'J}\,J_{\widehat a\widehat b}\bigr)+\bigl(\widehat C\ox\bd1\bigr)_{\a\a'I\,\b\b'J}\,\cZ_0\,.
\qqq
\eerop
\beroof
A proof, based on an explicit computation of all the structure constants $\,c^{\la\,0\widehat a}_{\ \ \ \ \mu}$,\ is given in App.\,\ref{app:KSproof}.
\eroof

\brem
The central extension derived above is one of the two Gra\ss mann-even deformations of the Lie superalgebra $\,\gt{su}(2,2\,\vert\,4)\,$ containing an undeformed Lie algebra $\,\gt{so}(4,2)\oplus\gt{so}(6)\,$ of isometries of the body $\,{\rm AdS}_5\x\bS^5\,$ of the supertarget of interest as a subalgebra of its even subalgebra considered by Kamimura and Sakaguchi in \Rcite{Kamimura:2003rx}. Consequently, we propose to and do call it \textbf{the Kamimura--Sakaguchi central extension of} $\,\gt{su}(2,2\,\vert\,4)\,$ in what follows.
\erem

We have
\bethe
The Lie supergroup $\,\widetilde{{\rm SU}(2,2\,\vert\,4)}\,$ integrating the Kamimura--Sakaguchi central extension defined above does not support over the local sections $\,\widetilde\si_i\,$ of \Reqref{eq:distlocsecext} a left-invariant super-1-form whose exterior derivative would exhibit the asymptotics of $\,\underset{\tx{\ciut{(2)}}}{\sfD_i}$.\ Accordingly, there exists no central extension of $\,\gt{su}(2,2\,\vert\,4)\,$ with only the anticommutator of the super-charges deformed as in \Reqref{eq:KSextend} whose integration to a Lie supergroup would trivialise the Metsaev--Tseytlin super-3-cocycle in a manner compatible with both: supersymmetry and the super-Minkowskian asymptotics of Sec.\,\ref{subsect:sstringextsMink}.
\ethe
\beroof
The question of the existence of the relevant super-1-form may be rephrased as the question of the presence, among the central terms in the extension, now cast in the form
\qq\nn
\z_i^{\a\a'I\b\b'J}\,\cZ_{\a\a'I\,\b\b'J}\equiv\z_i^{\a\a'I\b\b'J}\,\bigl(\cC\,\unl\G^\la\bigr)_{\a\a'I\b\b'J}\,\cZ_\la=:\z_i^\la\,\cZ_\la
\qqq
compatible with the choice of the basis in $\,\bC(32)^{\rm sym}\,$ made in the proof of Prop.\,\ref{prop:KS}, of a non-zero term with index $\,\la=2\,$ corresponding to the generator $\,\unl\G^2\equiv\bd1\ox\g^2$.\ But the only non-zero term carries index $\,\la=0\,$ corresponding to the generator $\,\unl\G^0\equiv\bd1\ox\g^0$,\ whence the answer in the negative, and the thesis of the Theorem.
\eroof

\section{A no-go for a class of Kosteleck\'y--Rabin deformations}\label{sec:KostRabdef}

In the next step, we pass to consider a class of Gra\ss mann-odd deformations of the Lie superalgebra $\,\gt{su}(2,2\,\vert\,4)\,$ motivated by our former experience with the super-Minkowskian (and super-Poincar\'e) algebra, which seems particularly well-founded and natural in the present context of the asymptotic analysis of associative deformations of the curved-superspacetime (super)algebra. Such considerations land us unavoidably on the third path indicated at the end of Sec.\,\ref{sec:ssextaAdSS}.

Taking into account the structure of the superstring extension of  $\,{\rm sMink}^{d,1\,\vert\,D_{d,1}}\,$ recalled in Sec.\,\ref{subsect:sstringextsMink}, we postulate the structure relation
\qq\nn
[Q_{\a\a' I},P_{\widehat a}]^\sim=-\tfrac{1}{2}\,(\widehat\G_{\widehat a}\ox\si_2)^{\b\b'J}_{\ \ \a\a'I}\,Q_{\b\b' J}-\tfrac{\sfi\,\a}{2}\,\bigl(\widehat C\,\widehat\G_{\widehat a}\ox\si_3\bigr)_{\a\a'I\b\b'J}\,\cZ^{\b\b'J}\,,\qquad\a\in\bC^\x
\qqq
as the basis of the deformation. Given our previous identification of the source of the Gra\ss mann-odd charges (in the flat limit, and in the more general situation), the latter may rightly be dubbed a \textbf{Kosteleck\'y--Rabin deformation of} $\,\gt{su}(2,2\,\vert\,4)$.\ Indeed, in the 10-dimensional notation (recalled in App.\,\ref{app:CliffAdSS}), we have
\qq\nn
[Q_{\a\a' I},P_{\widehat a}]^\sim=-\tfrac{\sfi}{2}\,(\unl\g_{\widehat a}\,\D^1_{\widehat a})^{\b\b'J}_{\ \ \a\a'I}\,Q_{\b\b' J}-\tfrac{\sfi\,\a}{2}\,\bigl(\cC\,\unl\g_{\widehat a}\,\D^2_{\widehat a}\bigr)_{\a\a'I\b\b'J}\,\cZ^{\b\b'J}\,,
\qqq
so that if we rescale the positive-chirality spinors $\,\cZ^{\b\b'J}\,$ as
\qq\nn
\cZ^{\a\a'I}\longmapsto R^{\frac{3}{2}}\,\cZ^{\a\a'I}
\qqq
when performing the \.In\"on\"u--Wigner contraction \eqref{eq:IWrescale} the deformed relation asymptotes to the superstring deformation of the super-Minkowski superalgebra
\qq\nn
[Q_{\a\a' I},P_{\widehat a}]^\sim=-\tfrac{\sfi\,\a}{2}\,\bigl(\cC\,\unl\g_{\widehat a}\bigr)_{\a\a'I\b\b'J}\,\cZ^{\b\b'J}
\qqq
of \Reqref{eq:ssextsMink}. Incidentally, the above argument justifies our parametrisation of the deformation (by $\,\a$) \emph{independent} of the sector ($\widehat a\in\ovl{0,4}\,$ {\it vs} $\,\widehat a\in\ovl{5,9}$) in the decomposition of the body of the supertarget.

We still may, and -- speaking with hindsight -- actually need to accomodate the Gra\ss mann-even charge of the usual topological origin. As the latter corresponds to the presence of a single generator of $\,H_1({\rm AdS}_5\x\bS^5)$,\ we are confronted with a choice: either we treat the charge as a ${\rm SO}(4,1)\x{\rm SO}(5)$-scalar, or as a ${\rm SO}(4,1)\x{\rm SO}(5)$-vector. We shall analyse both possibilities, calling the former, defined by the additional structure relation
\qq\nn
\{Q_{\a\a' I},Q_{\b\b' J}\}^\sim=\sfi\,\bigl(-2(\widehat C\,\widehat\G^{\widehat a}\ox\bd1)_{\a\a'I\b\b'J}\,P_{\widehat a}+(\widehat C\,\widehat\G^{\widehat a\widehat b}\ox\si_2)_{\a\a'I\b\b'J}\,J_{\widehat a\widehat b}\bigr)+\b^{\unl\mu}_{\rm s}\,\bigl(\widehat C\ox\si_{\unl\mu}\bigr)_{\a\a'I\,\b\b'J}\,\cZ
\qqq
with $\,\b^{\unl\mu}_{\rm s}\in\bC\,$ for some \emph{fixed} $\,\unl\mu\in\{0,1,3\}$,\ \textbf{the directional spinor-scalar Kosteleck\'y--Rabin deformation of} $\,\gt{su}(2,2\,\vert\,4)$,\ and the latter, with the additional structure relation
\qq\nn
\{Q_{\a\a' I},Q_{\b\b' J}\}^\sim=\sfi\,\bigl(-2(\widehat C\,\widehat\G^{\widehat a}\ox\bd1)_{\a\a'I\b\b'J}\,P_{\widehat a}+(\widehat C\,\widehat\G^{\widehat a\widehat b}\ox\si_2)_{\a\a'I\b\b'J}\,J_{\widehat a\widehat b}\bigr)+\b^{\unl\mu}_{{\rm v}\,\widehat a}\,\bigl(\widehat C\,\widehat\G_{\widehat a}\ox\si_{\unl\mu}\bigr)_{\a\a'I\,\b\b'J}\,\cZ^{\widehat a}
\qqq
with $\,\b^{\unl\mu}_{{\rm v}\,\widehat a}\in\bC\,$ for some \emph{fixed} $\,\unl\mu\in\{0,1,3\}$,\ \textbf{the directional spinor-vector Kosteleck\'y--Rabin deformation of} $\,\gt{su}(2,2\,\vert\,4)$.\ Here, it is presupposed that the formerly discussed rescaling of the original generators of the supersymmetry algebra and of the Gra\ss mann-odd charges is accompanied by
\qq\nn
\cZ^{(\widehat a)}\longmapsto\cZ^{(\widehat a)}\,,
\qqq
which ensures that the anticommutator of the supercharges asymptotes to the super-Minkowskian one, sourced by the surviving
\qq\nn
-2\sfi\,\bigl(\widehat C\,\widehat\G^{\widehat a}\ox\bd1\bigr)\equiv-2\sfi\,\bigl(\cC\,\unl\g^{\widehat a}\,\D^1_{\widehat a}\bigr)\,,
\qqq
and we allow for two independent parameters $\,\b^{\unl\mu}_{{\rm v}\,\widehat a}\,$ for $\,\widehat a\in\ovl{0,4}\,$ ($\b^{\unl\mu}_{{\rm v}\,0}=\b^{\unl\mu}_{{\rm v}\,1}=\b^{\unl\mu}_{{\rm v}\,2}=\b^{\unl\mu}_{{\rm v}\,3}=\b^{\unl\mu}_{{\rm v}\,4}$) and $\,\widehat a\in\ovl{5,9}\,$ ($\b^{\unl\mu}_{{\rm v}\,5}=\b^{\unl\mu}_{{\rm v}\,6}=\b^{\unl\mu}_{{\rm v}\,7}=\b^{\unl\mu}_{{\rm v}\,8}=\b^{\unl\mu}_{{\rm v}\,9}$), bearing in mind that the symmetry group has the product structure $\,{\rm SO}(4,1)\x{\rm SO}(5)\,$ (it is \emph{not} $\,{\rm SO}(9,1)$).

We shall deal with the spinor-scalar deformation first. Thus, assuming $\,\{Q_{\a\a'I},P_{\widehat a},J_{\widehat a\widehat b},\cZ^{\a\a'I},\cZ\}\,$ to be the generating set of the deformation $\,\widetilde{\gt{su}(2,2\,\vert\,4)}$,\ and -- as advocated earlier -- the body subalgebra $\,\gt{so}(4,2)\oplus\gt{so}(6)\,$ to be undeformed, we may write out the remaining \emph{deformed}\footnote{The undeformed structure relations have been left out.} structure relations:
\qq\nn
&[\cZ^{\a\a'I},J_{\widehat a\widehat b}]^\sim=\tfrac{\vep_{\widehat a\widehat b}}{2}\,\bigl(\widehat\G_{\widehat a\widehat b}\ox\bd1\bigr)^{\a\a'I}_{\ \b\b'J}\,\cZ^{\b\b'J}\,,\qquad\qquad[\cZ,J_{\widehat a\widehat b}]^\sim=0\,,&\cr\cr\cr
&[\cZ^{\a\a'I},P_{\widehat a}]^\sim=\g^\mu_{\widehat a}\,\bigl(\widehat\G_{\widehat a}\,\widehat C^{-1}\ox\si_\mu\bigr)^{\a\a'I\b\b'J}\,Q_{\b\b'J}+\d^\mu_{\widehat a}\,\bigl(\widehat\G_{\widehat a}\ox\si_\mu\bigr)\,^{\a\a'I}_{\ \b\b'J}\,\cZ^{\b\b'J}\,,&\cr\cr
&\{\cZ^{\a\a'I},Q_{\b\b'J}\}^\sim=\vep^{\widehat a,\mu}\,\bigl(\widehat\G^{\widehat a}\ox\si_\mu\bigr)_{\ \b\b'J}^{\a\a'I}\,P_{\widehat a}+\z^{\widehat a\widehat b,\mu}\,\bigl(\widehat\G^{\widehat a\widehat b}\ox\si_\mu\bigr)_{\ \b\b'J}^{\a\a'I}\,J_{\widehat a\widehat b}+\eta^\mu\,\bigl(\bd1\ox\si_\mu\bigr)_{\ \b\b'J}^{\a\a'I}\,\cZ\,,&\cr\cr
&\{\cZ^{\a\a'I},\cZ^{\b\b'J}\}^\sim=\theta^{\widehat a,K}\,\bigl(\widehat\G^{\widehat a}\,\widehat C^{-1}\ox\si_K\bigr)^{\a\a'I\b\b'J}\,P_{\widehat a}+\iota^{\widehat a\widehat b}\,\bigl(\widehat\G^{\widehat a\widehat b}\,\widehat C^{-1}\ox\si_2\bigr)^{\a\a'I\b\b'J}\,J_{\widehat a\widehat b}&\cr\cr
&\hspace{-2.25cm}+\k^K\,\bigl(\widehat C^{-1}\ox\si_K\bigr)^{\a\a'I\b\b'J}\,\cZ\,,&\cr\cr\cr
&[\cZ,P_{\widehat a}]^\sim=\la_{\widehat a}\,P_{\widehat a}\,,\qquad\qquad[\cZ,Q_{\a\a'I}]^\sim=\mu^\nu\,\bigl(\bd1\ox\si_\nu\bigr)^{\b\b'J}_{\ \a\a'I}\,Q_{\b\b'J}+\nu^\mu\,\bigl(\widehat C\ox\si_\mu\bigr)_{\a\a'I\b\b'J}\,\cZ^{\b\b'J}\,,&\cr\cr
&[\cZ,\cZ]^\sim=\rho\,\cZ\,,&\cr\cr\cr
&[\cZ^{\a\a'I},\cZ]^\sim=\varsigma^\mu\,\bigl(\widehat C^{-1}\ox\si_\mu\bigr)^{\a\a'I\b\b'J}\,Q_{\b\b'J}+\tau^\mu\,\bigl(\bd1\ox\si_\mu\bigr)^{\a\a'I}_{\ \b\b'J}\,\cZ^{\b\b'J}\,,&
\qqq
in which $\,\g^\mu_{\widehat a},\d^\mu_{\widehat a},\vep^{\widehat a,\mu},\z^{\widehat a\widehat b,\mu},\eta^\mu,\theta^{\widehat a,K},\iota^{\widehat a\widehat b},\k^K,\la_{\widehat a},\mu^\nu,\nu^\mu,\rho,\varsigma^\mu,\tau^\mu\in\bC\,$ are parameters, with indices $\,\mu,\nu\in\{0,1,2,3\},K\in\{0,1,3\}\,$ and $\,\widehat b>\widehat a\in\ovl{0,9}\,$ summed over the respective ranges. The first two relations express the scalar and spinorial nature of the charges $\,\cZ\,$ and $\,\cZ^{\a\a'I}$,\ respectively. The remaining ones quantify the arbitrariness of the deformation within the bounds set by the former relations, and -- once again -- we are taking into account the product structure of the symmetry group by allowing $\,(\g^\mu_0,\d^\mu_0)=(\g^\mu_1,\d^\mu_1)=(\g^\mu_2,\d^\mu_2)=(\g^\mu_3,\d^\mu_3)=(\g^\mu_4,\d^\mu_4)\,$ and $\,(\g^\mu_5,\d^\mu_5)=(\g^\mu_6,\d^\mu_6)=(\g^\mu_7,\d^\mu_7)=(\g^\mu_8,\d^\mu_8)=(\g^\mu_9,\d^\mu_9)$,\ and similarly for the $\,\vep^{\widehat a,\mu},\z^{\widehat a\widehat b,\mu},\theta^{\widehat a,K},\iota^{\widehat a\widehat b}$.\ We readily establish
\berop\label{prop:ssKRproof}
There exist no associative directional spinor-scalar Kosteleck\'y--Rabin deformations of $\,\gt{su}(2,2\,\vert\,4)$.
\eerop
\beroof
A proof, based on a systematic imposition of the super-Jacobi identities, is given in App.\,\ref{app:ssKRproof}.
\eroof

We pass to examine the spinor-vector deformation. Here, the deformation is spanned on\footnote{We may consistently set the $\,\cZ^{a'}\,$ to zero if we want to emphasise the geometric origin of the charge (there are no non-contractible 1-cycles in $\,\bS^5$). This requires that some of the parameters of the deformation be nullified. As this scenario is subsumed by the more general one considered here, we do not discuss it separately.} $\,\{Q_{\a\a'I},P_{\widehat a},$ $J_{\widehat a\widehat b},\cZ^{\a\a'I},\cZ^{\widehat a}\}$,\ and so we end up with the \emph{deformed} structure relations:
\qq\nn
&[\cZ^{\a\a'I},J_{\widehat a\widehat b}]^\sim=\tfrac{\vep_{\widehat a\widehat b}}{2}\,\bigl(\widehat\G_{\widehat a\widehat b}\ox\bd1\bigr)^{\a\a'I}_{\ \b\b'J}\,\cZ^{\b\b'J}\,,\qquad\qquad[\cZ^{\widehat a},J_{\widehat b\widehat c}]^\sim=\d^{\widehat a}_{\ \widehat b}\,\cZ_{\widehat c}-\d^{\widehat a}_{\ \widehat c}\,\cZ_{\widehat b}\,,&\cr\cr\cr
&[\cZ^{\a\a'I},P_{\widehat a}]^\sim=\widetilde\g^\mu_{\widehat a}\,\bigl(\widehat\G_{\widehat a}\,\widehat C^{-1}\ox\si_\mu\bigr)^{\a\a'I\b\b'J}\,Q_{\b\b'J}+\widetilde\d^\mu_{\widehat a}\,\bigl(\widehat\G_{\widehat a}\ox\si_\mu\bigr)\,^{\a\a'I}_{\ \b\b'J}\,\cZ^{\b\b'J}\,,&\cr\cr
&\{\cZ^{\a\a'I},Q_{\b\b'J}\}^\sim=\widetilde\vep^{\widehat a,\mu}\,\bigl(\widehat\G^{\widehat a}\ox\si_\mu\bigr)_{\ \b\b'J}^{\a\a'I}\,P_{\widehat a}+\widetilde\z^{\widehat a\widehat b,\mu}\,\bigl(\widehat\G^{\widehat a\widehat b}\ox\si_\mu\bigr)_{\ \b\b'J}^{\a\a'I}\,J_{\widehat a\widehat b}+\widetilde\eta_{\widehat a,\mu}\,\bigl(\widehat\G_{\widehat a}\ox\si_\mu\bigr)_{\ \b\b'J}^{\a\a'I}\,\cZ^{\widehat a}\,,&\cr\cr
&\{\cZ^{\a\a'I},\cZ^{\b\b'J}\}^\sim=\theta^{\widehat a,K}\,\bigl(\widehat\G^{\widehat a}\,\widehat C^{-1}\ox\si_K\bigr)^{\a\a'I\b\b'J}\,P_{\widehat a}+\widetilde\iota^{\widehat a\widehat b}\,\bigl(\widehat\G^{\widehat a\widehat b}\,\widehat C^{-1}\ox\si_2\bigr)^{\a\a'I\b\b'J}\,J_{\widehat a\widehat b}&\cr\cr
&\hspace{-2cm}+\widetilde\k_{\widehat a}^K\,\bigl(\widehat\G_{\widehat a}\,\widehat C^{-1}\ox\si_K\bigr)^{\a\a'I\b\b'J}\,\cZ^{\widehat a}\,,&\cr\cr\cr
&[\cZ^{\widehat a},P_{\widehat b}]^\sim=\eta^{\widehat a\widehat c}\,\widetilde\la_{\widehat c\widehat b}\,J_{\widehat c\widehat b}\,,\qquad\qquad[\cZ^{\widehat a},Q_{\a\a'I}]^\sim=\widetilde\mu^{\nu,\widehat a}\,\bigl(\widehat\G^{\widehat a}\ox\si_\nu\bigr)^{\b\b'J}_{\ \a\a'I}\,Q_{\b\b'J}+\widetilde\nu^{\mu,\widehat a}\,\bigl(\widehat C\,\widehat\G^{\widehat a}\ox\si_\mu\bigr)_{\a\a'I\b\b'J}\,\cZ^{\b\b'J}\,,&\cr\cr
&[\cZ^{\widehat a},\cZ^{\widehat b}]^\sim=\eta^{\widehat a\widehat c}\,\eta^{\widehat b\widehat d}\,\widetilde\rho^{\widehat c\widehat d}\,J_{\widehat c\widehat d}\,,&\cr\cr\cr
&[\cZ^{\a\a'I},\cZ^{\widehat a}]^\sim=\widetilde\varsigma^{\mu,\widehat a}\,\bigl(\widehat\G^{\widehat a}\,\widehat C^{-1}\ox\si_\mu\bigr)^{\a\a'I\b\b'J}\,Q_{\b\b'J}+\widetilde\tau^{\mu,\widehat a}\,\bigl(\widehat\G^{\widehat a}\ox\si_\mu\bigr)^{\a\a'I}_{\ \b\b'J}\,\cZ^{\b\b'J}\,,&
\qqq
in which $\,\widetilde\g^\mu_{\widehat a},\widetilde\d^\mu_{\widehat a},\widetilde\vep^{\widehat a,\mu},\widetilde\z^{\widehat a\widehat b,\mu},\widetilde\eta_{\widehat a,\mu},\widetilde\theta^{\widehat a}_K,\widetilde\iota^{\widehat a\widehat b},\widetilde\k_{a,K},\widetilde\la^{\widehat c\widehat d},\widetilde\mu^{\nu,\widehat a},\widetilde\nu^{\mu,\widehat a},\widetilde\rho^{\widehat c\widehat d},\widetilde\varsigma^{\mu,\widetilde a},\widetilde\tau^{\mu,\widetilde a}\in\bC\,$ are parameters, with the repeated indices $\,\mu,\nu\in\{0,1,2,3\},K\in\{0,1,3\}\,$ and $\,\widehat b>\widehat a\in\ovl{0,9}\,$ summed over the respective ranges, and where -- as previously -- we are taking into account the product structure of the symmetry group. Once more, the first two relations express the vectorial and spinorial nature of the charges $\,\cZ^{\widehat a}\,$ and $\,\cZ^{\a\a'I}$,\ respectively. We find
\berop\label{prop:svKRproof}
There exist no associative directional spinor-vector Kosteleck\'y--Rabin deformations of $\,\gt{su}(2,2\,\vert\,4)$.
\eerop
\beroof
A proof, once more based on a systematic imposition of the super-Jacobi identities, is given in App.\,\ref{app:svKRproof}.
\eroof

We conclude that the most natural (straightforward) deformations of the supersymmetry algebra $\,\gt{su}(2,2\,\vert\,4)\,$ of the supertarget under consideration asymptoting to the supercentral deformation \eqref{eq:ssextsMink} of the super-Minkowski superalgebra do not lead to a supersymmetric trivialisation of the Metsaev--Tseytlin super-3-cocycle as they leave the category of Lie superalgebras. This seems to point towards the absence of a trivialisation mechanism with the desired asymptotics, however\ldots

\section{Towards an  \.In\"on\"u--Wigner-contractible super-1-gerbe on $\,{\rm s}({\rm AdS}_5\x\bS^5)$}\label{sec:MTaway}

Our hitherto analysis demonstrates the supersymmetry-invariant cohomology of the supertarget $\,{\rm s}({\rm AdS}_5\x\bS^5)\,$ (in degree 2) to be rather rigid and not amenable to trivialisation through associative deformation of the underlying Lie superalgebra compatible with the \.In\"on\"u--Wigner contraction to the super-Minkowski superalgebra. Indeed, the most natural deformations of $\,\gt{su}(2,2\,\vert\,4)\,$ either fail to trivialise the Metsaev--Tseytlin super-3-cocycle in such a way as to reproduce the super-Minkowskian trivialisation of the asymptotic Green--Schwarz super-3-cocycle in the flat limit or fail to define an extended associative structure with a Lie superbracket altogether. Thinking of the asymptotic transition between the curved geometry $\,{\rm AdS}_5\x\bS^5\,$ and the flat geometry $\,{\rm Mink}^{9,1}\,$ and the tractable Lie-algebraic mechanism behind it as fundamental, we are led to invert the logic of our analysis and contemplate the possibility of deriving a Green--Schwarz super-3-cocycle, potentially different from the Metsaev--Tseytlin super-3-cocycle but with the same asymptotics, as the exterior derivative of a supersymmetric super-2-form on an extension of $\,{\rm s}({\rm AdS}_5\x\bS^5)\,$ (or even of another supermanifold with the same body and the structure of a homogeneous space of a Lie supergroup), both (the extension and the super-2-form) with the desired super-Minkowskian asymptotics. In order to give this rather non-specific and hence somewhat vague idea some flesh, we discuss a variant of a super-${\rm AdS}$ algebra conceived by Hatsuda, Kamimura and Sakaguchi in \Rcite{Hatsuda:2000mn}, with a built-in Gra\ss mann-odd deformation that makes it manifestly contractible, in the sense of \.In\"on\"u and Wigner, to the superstring-extended super-Minkowskian algebra of \Reqref{eq:ssextsMink}.

The point of departure is the so-called \textbf{${\rm AdS}$-algebra in $d+1$ dimensions}, {\it i.e.} the Lie algebra $\,\gt{so}(d,2)\,$ (of conformal transformations) which, in a suitable basis\footnote{The basis consists of simple linear combinations of the generators of translations and special conformal transformations, and those of the remaining conformal transformations: boosts, rotations and dilations.}, reads
\qq\nn
&[P_a,P_b]=J_{ab}\,,\qquad\qquad[J_{ab},J_{cd}]=\eta_{ad}\,J_{bc}-\eta_{ac}\,J_{bd}+\eta_{bc}\,J_{ad}-\eta_{bd}\,J_{ac}\,,&\cr\cr
&[P_a,J_{bc}]=\eta_{ab}\,P_c-\eta_{ac}\,P_b\,,&\cr\cr
&(\eta_{ab})=\diag(-1,\underbrace{1,1,\ldots,1}_d)\,,\qquad\qquad a,b,c,d\in\ovl{0,d}\,.&
\qqq
We consider its Gra\ss mann-odd extension $\,\widetilde{\gt{sso}(d,2)}\,$ defined in terms of Majorana-spinor generators $\,Q_\a,\cZ_\a,\ \a\in\ovl{1,2^{[\frac{d+1}{2}]}}\,$ through the relations
\qq
&[Q_\a,J_{ab}]=-\tfrac{1}{2}\,\bigl(\G_{ab}\bigr)^\b_{\ \a}\,Q_\b\,,\qquad\qquad[\cZ_\a,J_{ab}]=-\tfrac{1}{2}\,\bigl(\G_{ab}\bigr)^\b_{\ \a}\,\cZ_\b\,,&\cr\cr
&[Q_\a,P_a]=-\tfrac{\sfi}{2}\,\bigl(\G_a\bigr)^\b_{\ \a}\,\cZ_\b\,,\qquad\qquad[\cZ_\a,P_a]=\tfrac{\sfi}{2}\,\bigl(\G_a\bigr)^\b_{\ \a}\,Q_\b\,,& \label{eq:sAdSext}\\ \cr
&\{Q_\a,Q_\b\}=-2\sfi\,\bigl(C\,\G^a\bigr)_{\a\b}\,P_a\,,\qquad\qquad\{\cZ_\a,\cZ_\b\}=2\sfi\,\bigl(C\,\G^a\bigr)_{\a\b}\,P_a\,,\qquad\qquad\{Q_\a,\cZ_\b\}=\bigl(C\,\G^{ab}\bigr)_{\a\b}\,J_{ab}\,.&\nn
\qqq
Here, the $\,\G_a\equiv\eta_{ab}\,\G^b\,$ are generators of the Clifford algebra $\,\Cliff(\bR^{d,1})\,$ and
\qq\nn
\G_{ab}=\tfrac{1}{2}\,[\G_a,\G_b]\,,
\qqq
all in a Majorana representation in which together with the charge-cojugation matrix $\,C\,$ they satisfy the identities
\qq\nn
\bigl(C\,\G_a\bigr)^{\rm T}=C\,\G_a\,,\qquad\qquad\bigl(C\,\G_{ab}\bigr)^{\rm T}=C\,\G_{ab}\,.
\qqq
Such a representation and the ensuing Lie-superalgebraic extension $\,\widetilde{\gt{sso}(d,2)}\,$ of the AdS-algebra are known to exist in dimension $\,d+1=3\,$ (with the charge-conjugation matrix $\,C^{\rm T}=-C\,$ determining the symmetry properties of the generators as $\,\G_a^{\rm T}=-C\,\G_a\,C^{-1}$). Basing on this observation, we simply \emph{assume} its existence for some $d$ and study the supergeometric consequences thereof. In so doing, we draw on our super-Minkowskian intuition. The latter is well justified as upon rescaling
\qq\nn
\bigl(P_a,J_{ab},Q_\a,\cZ_\a\bigr)\longmapsto\bigl(R\,P_a,J_{ab},R^{\frac{1}{2}}\,Q_\a,R^{\frac{3}{2}}\,\cZ_\a\bigr)
\qqq
the above algebra contracts, in the limit $\,R\to\infty$,\ to the superstring extension \eqref{eq:ssextsMink} of the Lie superalgebra $\,{\rm sMink}^{d,1\,\vert\,D_{d,1}}$,\ with
\qq\label{eq:ZasZ}
Z^\a=\a\,\bigl(C^{-1}\bigr)^{\a\b}\,\cZ_\b\,,
\qqq
where $\,\a\in\bC^\x\,$ is a constant that accounts for a suitable (constant) rescaling of the supercharges and momenta.

The Lie superalgebra \eqref{eq:sAdSext}, whenever it exists, integrates to a Lie supergroup which we denote as $\,\widetilde{{\rm sSO}(d,2)}\,$ in what follows. For the sake of the present discussion, we define \textbf{the extended super-${\rm AdS}_{d+1}\,$ space} to be the homogeneous space
\qq\nn
\widetilde{{\rm sAdS}}_{d+1}:=\widetilde{{\rm sSO}(d,2)}/{\rm SO}(d,1)\,.
\qqq
The latter is patchwise embedded in the total space $\,\widetilde{{\rm sSO}(d,2)}\,$ of the principal ${\rm SO}(d,1)$-bundle $\,\widetilde{{\rm sSO}(d,2)}\too\widetilde{{\rm sSO}(d,2)}/{\rm SO}(d,1)\,$ by a collection of locally smooth sections
\qq\nn
\widehat\si_i\ :\ \cO_i\too\widetilde{{\rm sSO}(d,2)}\ :\ \widehat Z_i\equiv\bigl(x^a_i,\theta^\a_i,\z^\b_i\bigr)\longmapsto\widehat{\unl g}_i\cdot\ee^{\z^\a_i\,\cZ_\a}\cdot\ee^{x^a_i\,P_a}\cdot\ee^{\theta^\b_i\,Q_\b}\,,\qquad i\in I\,,
\qqq
defined for an open cover $\,\{\cO_i\}_{i\in I}\,$ of the base $\,\widetilde{{\rm sAdS}}_{d+1}$,\ with local coordinates $\,(x^a_i,\theta^\a_i,\z^\b_i)\,$ associated with the direct-sum complement
\qq\nn
\tgt=\bigoplus_{a=0}^d\,\corr{P_a}_\bR\oplus\bigoplus_{\a=1}^{2^{[\frac{d+1}{2}]}}\,\corr{Q_\a}_\bR\oplus\bigoplus_{\b=1}^{2^{[\frac{d+1}{2}]}}\,\corr{\cZ_\b}_\bR
\qqq
of the Lorentz algebra $\,\hgt\equiv\gt{so}(d,1)\,$ within $\,\ggt\equiv\widetilde{\gt{sso}(d,2)}$,\ and for some reference elements $\,\widehat{\unl g}_i\in\widetilde{{\rm sSO}(d,2)}$.

We begin by writing out the pullback of the Maurer--Cartan super-1-form on $\,\widetilde{{\rm sSO}(d,2)}\,$ along one of the $\,\widehat\si_i$,
\qq\nn
\widehat\si_i^*\theta_{\rm L}(\widehat Z_i)&=&\tfrac{1}{R^{\frac{1}{2}}}\,\sfd\theta^\a_i\ox Q_\a+\tfrac{1}{R}\,\bigl(\sfd x^a_i-\sfi\,\ovl\theta_i\,\G^a\,\sfd\theta_i\bigr)\ox P_a\cr\cr
&&+\tfrac{1}{R^{\frac{3}{2}}}\,\bigl(\sfd\z^\a_i+\tfrac{\sfi}{2}\,\sfd x^a_i\,\theta^\b_i\,\bigl(\G_a\bigr)^\a_{\ \b}+\tfrac{1}{3!}\,\ovl\theta_i\,\G^a\,\sfd\theta_i\,\theta^\b_i\,\bigl(\G_a\bigr)^\a_{\ \b}\bigr)\ox\cZ_\a\cr\cr
&&+\tfrac{1}{4R^2}\,\bigl(x^b_i\,\sfd x^a_i-x^a_i\,\sfd x^b_i+\sfi\,\ovl\theta_i\,\G^{ab}\,\G_c\,\theta_i\,\bigl(\sfd x^c_i+\tfrac{\sfi}{3!}\,\ovl\theta_i\,\G^c\,\sfd\theta_i\bigr)+4\ovl\theta_i\,\G^{ab}\,\sfd\z_i\bigr)\ox J_{ab}+O\bigl(R^{-\frac{5}{2}}\bigr)\,.
\qqq
From the result, we infer the asymptotics of the various component super-1-forms (in order to distinguish the one associated with the supercharge $\,Q_\a\,$ from that associated with the Gra\ss mann-odd charge $\,\cZ_\a$,\ we have put a tilde over the latter):
\qq\nn
\widehat\si_i^*\theta_{\rm L}^\a(\widehat Z_i)&=&\tfrac{1}{R^{\frac{1}{2}}}\,\sfd\theta^\a+O\bigl(R^{-\frac{5}{2}}\bigr)\,,\cr\cr
\widehat\si_i^*\widetilde\theta_{\rm L}^\a(\widehat Z_i)&=&\tfrac{1}{R^{\frac{3}{2}}}\,\bigl(\sfd\z^\a+\tfrac{\sfi}{2}\,\sfd x^a_i\,\theta^\b_i\,\bigl(\G_a\bigr)^\a_{\ \b}+\tfrac{1}{3!}\,\ovl\theta_i\,\G^a\,\sfd\theta_i\,\theta^\b_i\,\bigl(\G_a\bigr)^\a_{\ \b}\bigr)+O\bigl(R^{-\frac{5}{2}}\bigr)\,,\cr\cr
\widehat\si_i^*\theta_{\rm L}^a(\widehat Z_i)&=&\tfrac{1}{R}\,\bigl(\sfd x^a_i-\sfi\,\ovl\theta_i\,\G^a\,\sfd\theta_i\bigr)+O\bigl(R^{-3}\bigr)\,,\cr\cr
\widehat\si_i^*\theta_{\rm L}^{ab}(\widehat Z_i)&=&\tfrac{1}{4R^2}\,\bigl(x^b_i\,\sfd x^a_i-x^a_i\,\sfd x^b_i+\sfi\,\ovl\theta_i\,\G^{ab}\,\G_c\,\theta_i\,\bigl(\sfd x^c_i+\tfrac{\sfi}{3!}\,\ovl\theta_i\,\G^c\,\sfd\theta_i\bigr)+4\ovl\theta_i\,\G^{ab}\,\sfd\z_i\bigr)\ox J_{ab}+O\bigl(R^{-3}\bigr)\,,
\qqq
and so we conclude that the globally smooth left-invariant super-2-form
\qq\nn
\underset{\tx{\ciut{(2)}}}{\widehat\b}:=2\sfi\,\ovl\theta_{\rm L}\wedge\widetilde\theta_{\rm L}
\qqq
asymptotes, upon pullback\footnote{By the previous reasoning, the pullbacks glue to a globally smooth super-2-form over $\,\widetilde{{\rm sAdS}}_{d+1}\,$ which we denote as $\,\underset{\tx{\ciut{(2)}}}{\widehat\txB}$.} along $\,\widehat\si_i$,
\qq\nn
\underset{\tx{\ciut{(2)}}}{\widehat\txB}(\widehat Z_i)&\equiv&\widehat\si_i^*\underset{\tx{\ciut{(2)}}}{\widehat\b}(\widehat Z_i)=\tfrac{2\sfi}{R^2}\,\sfd\theta_i^\a\wedge C_{\a\b}\,\bigl(\sfd\z^\b_i+\tfrac{\sfi}{2}\,\sfd x^a_i\,\theta^\g_i\,\bigl(\G_a\bigr)^\b_{\ \g}+\tfrac{1}{3!}\,\ovl\theta_i\,\G^a\,\sfd\theta_i\,\theta^\g_i\,\bigl(\G_a\bigr)^\b_{\ \g}\bigr)+O\bigl(R^{-3}\bigr)\cr\cr
&=&\tfrac{1}{R^2}\,\bigl(\ovl\theta_i\,\G_a\,\sfd\theta_i\wedge\sfd x^a+2\sfi\,\sfd\ovl\theta_i\wedge\sfd\z_i\bigr)+O\bigl(R^{-3}\bigr)\,,
\qqq
an overall rescaling by $\,R^2\,$ as well as a suitable redefinition (constant rescaling) of the $\,\theta^\a_i\,$ and $\,x^a_i\,$ and the identification
\qq\nn
\xi_\a=\widetilde\a\,C_{\a\b}\,\z^\b_i
\qqq
(the source of the constant $\,\widetilde\a\in\bC^\x\,$ is similar to that of the constant $\,\a\,$ in \Reqref{eq:ZasZ}), to the supersymmetric primitive (I.5.19) of the Green--Schwarz super-3-cocycle $\,\underset{\tx{\ciut{(3)}}}{\chi}\,$ defined on the superstring-extended super-Minkowski space $\,\xcM_1^{(2)}\,$ of Prop.\,I.5.7. Thus, consistently with the postulate formulated at the beginning of the present section, we may take the left-invariant super-2-form as the point of departure of a (contractible, in the sense of \.In\"on\"u and Wigner) geometrisation of a Green--Schwarz super-3-cocycle
\qq\nn
\underset{\tx{\ciut{(3)}}}{\widehat\chi}^{\rm GS}:=\sfd\underset{\tx{\ciut{(2)}}}{\widehat\b}
\qqq
for the super-${\rm AdS}_d$ supertarget. The latter super-3-cocycle is readily derived with the help of the (super-)Maurer--Cartan equations,
\qq\nn
\underset{\tx{\ciut{(3)}}}{\widehat\chi}^{\rm GS}&=&2\sfi\,\bigl(\sfd\ovl\theta_{\rm L}\wedge\widetilde\theta_{\rm L}-\ovl\theta_{\rm L}\wedge\sfd\widetilde\theta_{\rm L}\bigr)\cr\cr
&=&2\sfi\,\bigl[\bigl(\tfrac{1}{2}\,\bigl(\G_{ab}\bigr)^\a_{\ \g}\,\theta_{\rm L}^\g\wedge\theta_{\rm L}^{ab}-\tfrac{\sfi}{2}\,\bigl(\G_a\bigr)^\a_{\ \g}\,\widetilde\theta_{\rm L}^\g\wedge\theta_{\rm L}^a\bigr)\wedge C_{\a\b}\,\widetilde\theta_{\rm L}^\b\cr\cr
&&-\theta_{\rm L}^\a\,C_{\a\b}\wedge\bigl(\tfrac{1}{2}\,\bigl(\G_{ab}\bigr)^\b_{\ \g}\,\widetilde\theta_{\rm L}^\g\wedge\theta_{\rm L}^{ab}+\tfrac{\sfi}{2}\,\bigl(\G_a\bigr)^\b_{\ \g}\,\theta_{\rm L}^\g\wedge\theta_{\rm L}^a\bigr)\bigr]\cr\cr
&=&\ovl\theta_{\rm L}\wedge\G_a\,\theta_{\rm L}\wedge\theta_{\rm L}^a+\ovl{\widetilde\theta}_{\rm L}\wedge\G_a\,\widetilde\theta_{\rm L}\wedge\theta_{\rm L}^a\,.
\qqq
While the two terms in the final expression are structurally identical, they behave differently in the flat-superspace limit $\,R\to\infty\,$ (upon pullback), to wit, the former scales as $\,R^{-2}$,\ whereas the latter scales as $\,R^{-4}$.\ Hence, after an overall rescaling by $\,R^2$,\ we wind up with the familiar Green--Schwarz super-3-cocycle \eqref{eq:sMinkGS3}. This means that the original guiding principles for the construction of the super-$\si$-model with the super-${\rm AdS}_{d+1}$ supertarget laid out in \Rcite{Metsaev:1998it} are obeyed, or -- in other words -- the super-$\si$-model with the super-3-form $\,\underset{\tx{\ciut{(3)}}}{\widehat\chi}^{\rm GS}\,$ as the Green--Schwarz super-3-cocycle is a valid\footnote{In fact, one should still verify the presence of a linearised gauged right supersymmetry, or Siegel's $\k$-symmetry. This issue was addressed in \Rcite{Hatsuda:2000mn}.} model of superloop mechanics over the homogeneous space $\,{\rm AdS}_d\,$ in the sense of Metsaev and Tseytlin.

On the other hand, at \emph{finite} $\,R$,\ the super-3-cocycle obtained above does \emph{not} descend to a sub(super)\-manifold within the extended super-${\rm AdS}_d$ space $\,\widetilde{{\rm sAdS}}_d\,$ which would correspond to a constant (zero) value of the extra coordinate $\,\z_i$.\ This implies that we should consider the super-$\si$-model with the extended supertarget $\,\widetilde{{\rm sAdS}}_d\,$ all along, and associate with it a trivial super-1-gerbe with the global curving $\,\underset{\tx{\ciut{(2)}}}{\widehat\txB}$.\ It is only in the flat-superspace limit that the supports of the (asymptotic) Green--Schwarz super-3-cocycle and that of its supersymmetric primitive split and give rise to a non-trivial Cartan--Eilenberg super-1-gerbe of Part I. We hope to return to a systematic study of this supergeometric mechanism in a future work.

\section{Conclusions \& Outlook}\label{sec:C&O}

In the present paper, we have applied the general scheme, laid out and tested in the setting of the super-Minkowski space $\,{\rm sMink}^{d,1\,\vert\,D_{d,1}}\,$ in \Rcite{Suszek:2017xlw} (Part I), of a supersymmetry-equivariant geometrisation of (the physically distinguished) Green--Schwarz super-$(p+2)$-cocycles representing classes in the Cartan--Eilenberg cohomology of supersymmetry groups $\,\txG$,\ through Lie-supergroup extensions defined by these classes, to the supergeometric data of the two-dimensional Metsaev--Tseytlin super-$\si$-model with the super-${\rm AdS}_5\x\bS^5\,$ space as the supertarget. This places the scheme in a wider context of Cartan supergeometry of homogeneous spaces $\,\xcM\cong\txG/\txH\,$ of $\,\txG\,$ corresponding to reductive decompositions of the supersymmetry algebra $\,\ggt=\tgt\oplus\hgt\,$ into a geometric (Gra\ss mann-even) Lie subalgebra $\,\hgt\,$ of the isotropy subgroup $\,\txH\,$ of a point $\,x\in\xcM\,$ and its direct-sum complement $\,\tgt$,\ with a \emph{nontrivial} topology and a non-vanishing metric curvature in the body and \emph{no} Lie-supergroup structure assumed on $\,\txG/\txH$,\ and with the latter embedded patchwise smoothly in $\,\txG\,$ by \emph{local} sections of the principal $\txH$-bundle $\,\txG\too\txG/\txH\,$ and thus endowed with a non-linear realisation of supersymmetry. The relevance of the supertarget chosen for our case study hinges upon its r\^ole, as a critical superstring background, in the formulation and analysis of the celebrated ${\rm AdS}$/CFT correspondence.

The subtlety of the superbackground under consideration stems from the apparent incompatibility of the geometrisation with the \.In\"on\"u--Wigner contraction that underlies the transition to the flat geometry $\,{\rm sMink}^{9,1\,\vert\,D_{9,1}}$,\ dual to the flat limit of an infinite common radius of the generating 1-cycle of $\,{\rm AdS}_5\,$ and of the 5-sphere in the body $\,{\rm AdS}_5\x\bS^5\,$ of the supertarget. Indeed, while the super-$\si$-model and the relevant (Metsaev--Tseytlin) super-3-cocycle do asymptote to their super-Minkowskian counterparts known from Part I, the manifestly supersymmetric primitive of the super-3-cocycle discovered in \Rcite{Roiban:2000yy} does not, and so neither does the associated trivial super-1-gerbe constructed in Section
\ref{sec:trsAdSSext}. The quest for a geometrisation (and hence also for an extension of the original Lie superalgebra $\,\gt{su}(2,2\,\vert\,4)$) consistent with the contraction has led us to analyse at some length, in Section \ref{sec:wrapanom}, the r\^ole of the (pseudo-supersymmetric) topological Wess--Zumino term in the action functional of a generic super-$\si$-model in the Polyakov formulation in the (classical) field-theoretic deformation of the supersymmetry (super)algebra furnished by the Poisson algebra of the corresponding Noether charges. The general mechanisms established in that section have been worked out explicitly, in Section \ref{sec:sMinkext}, for the one- and two-dimensional super-$\si$-models with the flat supertarget $\,{\rm sMink}^{d,1\,\vert\,D_{d,1}}\,$ and the Green--Schwarz super-$p$-cycles (for $\,p\in\{2,3\}$) studied in Part I, whereby the physical significance of the Gra\ss mann-odd Kosteleck\'y--Rabin charges, associated with a topological realisation of the Cartan--Eilenberg cohomology recalled and exploited in Part I, has been brought to light. This has prompted an asymptotic analysis, carried out in Section \ref{sec:ssextaAdSS}, of potential wrapping-charge deformations of the supersymmetry algebra $\,\gt{su}(2,2\,\vert\,4)\,$ of the super-$\si$-model with supertarget $\,{\rm s}({\rm AdS}_5\x\bS^5)\,$ induced by the de Rham cohomology class of the Roiban--Siegel supersymmetric primitive of the Metsaev--Tseytlin super-3-cocycle. Two types of (leading) contributions have been found: the Gra\ss mann-odd Kosteleck\'y--Rabin charges and purely geometric Gra\ss mann-even winding charges associated with the generating 1-cycle of $\,\bS^1\x\bR^{\x 4}\cong{\rm AdS}_5$.\ These have been used as motivation for a systematic exploration, in Sections \ref{sec:KamSakext} and \ref{sec:KostRabdef}, of the two most natural types of deformation of the supersymmetry algebra:
\ben
\item a Gra\ss mann-even central extension deforming (exclusively) the anticommutator of the supercharges in an arbitrary manner, contemplated with view to recovering the desired non-supersymmetric correction to the Roiban--Siegel primitive of the Metsaev--Tseytlin super-3-cocycle as the leading term in an asymptotic expansion of a supersymmetric super-2-form on the resultant extended supersymmetry group, and
\item a generic extension determined by a Gra\ss mann-odd deformation of the commutator $\,[Q_{\a\a'I},P_{\widehat a}]\,$ engineered so as to allow for a trivialisation, on the resultant extended supersymmetry group, of a super-2-cocycle asymptoting to the Kosteleck\'y--Rabin super-2-cocycle of Section \ref{subsect:sstringextsMink} in the limit of an infinite radius of $\,{\rm AdS}_5\x\bS^5$.
\een
Both types of deformation have been ruled out as algebraically inconsistent. This has left us with the trivial super-1-gerbe of Section \ref{sec:trsAdSSext}, manifestly incompatible with the \.In\"on\"u--Wigner contraction, as the only working geometrisation of the Metsaev--Tseytlin super-3-cocycle to date. Finally, an alternative approach to the problem of geometrisation of the Green--Schwarz super-3-cocycle has been put forward in Section \ref{sec:MTaway}, promoting the asymptotic relation between the two extended supersymmetry algebras (for the super-Minkowski space and for the super-${\rm AdS}_5\x\bS^5\,$ space, respectively), effected by an \.In\"on\"u--Wigner contraction, to the rank of the fundamental principle.

Our hitherto findings immediately suggest directions of further research. First and foremost, one should continue the search, initiated herein and motivated rather concretely in Section \ref{sec:MTaway}, for a super\-sym\-metry-equivariant geometrisation compatible with the \.In\"on\"u--Wigner contraction, of \emph{a} Green--Schwarz super-3-cocycle over \emph{a} super-${\rm AdS}_5\x\bS^5\,$ space (understood in the spirit of the remarks made in Section \ref{sec:MTaway}) with the desired super-Minkowskian asymptotics. This would be expected, from a more general perspective, to give us insights into the higher-geometric realisation of the latter mechanism, acting in the tangent of the Lie supergroup and its homogeneous space, and, potentially, to bring further evidence of the significance, indicated strongly by our considerations, of the Kosteleck\'y--Rabin charges in the geometrisation scheme developed, the latter charges being an interesting subject of study in their own right. It is tempting to look for clues in this direction in the algebraically more tractable setting of the super-$\si$-models with supertarget of the form $\,{\rm s}({\rm AdS}_p\x\bS^p)\,$ for $\,p\in\{2,3\}$,\ discussed in Refs.\,\cite{Zhou:1999sm,Rahmfeld:1998zn,Park:1998un}.

An absolutely fundamental -- from the physical point of view -- feature of any super-1-gerbe (to be constructed) over the super-${\rm AdS}_5\x\bS^5\,$ space (or any other super-${\rm AdS}_p\x\bS^p\,$ space, for that matter) is its weak $\k$-equivariance, as defined in the super-Minkowskian setting in \Rcite{Suszek:2017xlw}. This constatation determines yet another natural line of future research, understood as a continuation of the study initiated in the super-Minkowskian setting. As the symmetry is commonly regarded to be a basic building block in the construction of all the super-$\si$-models listed above, the existing literature forms a solid basis of a research thus oriented.

The relevance for our understanding of the physics of strongly coupled systems with gauge symmetry (such as, {\it e.g.}, the quark-gluon plasma) of the particular superstring background picked up for scrutiny in the present paper alone justifies pursuing the study taken up herein with view towards geometrising and thus elucidating the ${\rm AdS}$/CFT correspondence. This suggests that one ought to tackle the issue of classification and construction of supersymmetric bi-modules for any super-1-gerbe (to be constructed) over ${\rm s}({\rm AdS}_5\x\bS^5)$.\ Given the nature of the Metsaev--Tseytlin super-$\si$-model, which is that of a descendant of a (supersymmetric) Wess--Zumino--Witten model for a supersymmetry group (to a homogeneous space thereof), it is natural to expect that the much-developed cohomological and group-theoretic techniques employed in the Lie-group setting in Refs.\,\cite{Fuchs:2007fw,Runkel:2009sp,Suszek:2018maxym} might prove to be of help.

Finally, there is a host of questions independent of the specific superstring background under consideration that were raised in the paper \cite{Suszek:2017xlw} and still await an in-depth study and elucidation. These include the question as to a structural relation between the intrinsically (Lie-super)algebraic geometrisation scheme furthered herein and the theory of Lie-$n$-superalgebras and $L_\infty$-superalgebras of Baez {\it et al.} considered in Refs.\,\cite{Baez:2004hda6,Baez:2010ye,Huerta:2011ic} and anchored firmly in the present field-theoretic context in \Rcite{Fiorenza:2013nha}, as well as the issue of correspondence between the Cartan--Eilenberg-cohomological framework developed for the two-dimensional super-$\si$-models and alternative approaches to supersymmetry in the context of superstring and related models, such as, in particular, the geometrisation, originally conceived by Killingback \cite{Killingback:1986rd} and Witten \cite{Witten:1988dls}, elaborated by Freed \cite{Yau:1987}, recently revived by Freed and Moore \cite{Freed:2004yc}, and ultimately concretised in the higher-geometric language by Bunke \cite{Bunke:2009} ({\it cp.}\ also \Rcite{Waldorf:2009uf} for an explicit construction), of the Pfaffian bundle of the target-space Dirac operator, associated with fermionic contributions to the superstring path integral, in terms of a differential ${\rm String}$-structure on the target space.

We hope to return to all the above ideas in a future work.
\newpage

\appendix

\section{A proof of Proposition \ref{prop:Vielder}}\label{app:Vielder}

In our (re)derivation, we employ the notation $\,Z_{i\,t}\equiv(X_i,t\theta_i)\,$ and
\qq\nn
\si_{i\,t}^*\theta_{\rm L}(Z)=\sfd Z_{i\,t}^{\unl A}\,E_{\unl A}^{\ A}(Z_{i\,t})\ox t_A\equiv\cE_t^A(Z_i)\ox t_A\equiv E^A_t(Z_{i\,t})\ox t_A\equiv E_t(Z_{i\,t})\,.
\qqq
Upon differentiation with respect to the retraction parameter, we arrive at the initial-value problem\footnote{We are writing out the Maurer--Cartan 1-form in the matrix-group convention solely for the sake of transparency of the calculation.}
\qq\nn
\tfrac{\sfd\ }{\sfd t}\,\cE_t^A(Z_i)\ox t_A&\equiv&\tfrac{\sfd\ }{\sfd t}\,\bigl[\ee^{-t\Theta_i(Z_i)}\cdot g_i(X_i)^{-1}\,\sfd\bigl(g_i(X_i)\cdot\ee^{t\Theta_i(Z_i)}\bigr)\bigr]\cr\cr
&=&-\Theta_i(Z_i)\,\ee^{-t\Theta_i(Z_i)}\cdot g_i(X_i)^{-1}\,\sfd\bigl(g_i(X_i)\cdot\ee^{t\Theta_i(Z_i)}\bigr)\cr\cr
&&+\ee^{-t\Theta_i(Z_i)}\cdot g_i(X_i)^{-1}\,\sfd\bigl(g_i(X_i)\cdot\ee^{t\Theta_i(Z_i)}\,\Theta_i(Z_i)\bigr)\cr\cr
&=&\sfd\Theta_i(Z_i)-\bigl[\Theta_i(Z_i),\ee^{-t\Theta_i(Z_i)}\cdot g_i(X_i)^{-1}\,\sfd\bigl(g_i(X_i)\cdot\ee^{t\Theta_i(Z_i)}\bigr)\bigr]\cr\cr
&=&\bigl(\sfd\Theta_i^{\widehat\a}(Z_i)+\Theta_i^{\widehat\b}(Z_i)\,f_{a\widehat\b}^{\ \ \widehat\a}\,\cE^a_t(Z_i)\bigr)\ox F_{\widehat\a}+\Theta_i^{\widehat\a}(Z_i)\,f_{\widehat\a\widehat\b}^{\ \ a}\,\cE^{\widehat\b}(Z_i)\ox B_a
\qqq
with
\qq\nn
\bigl(\cE^{\widehat\a}_0,\cE^a_0\bigr)(Z_i)=\bigl(0,E^a(X_i,0)\equiv e^a(X_i)\bigr)\,.
\qqq
We may rewrite it succinctly as an inhomogeneous linear problem for
\qq\nn
\vec\cE_t:=\left(\begin{smallmatrix} \cE^{\widehat\a}_t \\ \cE^a_t \end{smallmatrix}\right)
\qqq
of the form
\qq\nn
\tfrac{\sfd\ }{\sfd t}\,\vec\cE_t(Z_i)=M_i(Z_i)\,\vec\cE_t(Z_i)+\vec b_i(Z_i)\,,\qquad\qquad\vec\cE_0(Z_i):=\left(\begin{smallmatrix} 0 \\ e^a(X_i) \end{smallmatrix}\right)
\qqq
with the matrix
\qq\nn
M_i(Z_i):=\left(\begin{smallmatrix} 0 & -\Theta_i^{\widehat\g}(Z_i)\,f_{\widehat\g b}^{\ \ \widehat\a} \\ \Theta_i^{\widehat\g}(Z_i)\,f_{\widehat\g\widehat\b}^{\ \ a} & 0 \end{smallmatrix}\right)
\qqq
and the inhomogeneity vector
\qq\nn
\vec b_i(Z_i):=\left(\begin{smallmatrix} \sfd\Theta_i^{\widehat\a}(Z_i) \\ 0 \end{smallmatrix}\right)\,.
\qqq
Equivalently, we may rewrite the above as a homogeneous linear problem for
\qq\nn
\widetilde\cE_t:=\vec\cE_t(Z_i)+M_i(Z_i)^{-1}\,\vec b_i(Z_i)
\qqq
of the form
\qq\nn
\tfrac{\sfd\ }{\sfd t}\,\widetilde\cE_t(Z_i)=M_i(Z_i)\,\widetilde\cE_t(Z_i)\,,\qquad\qquad\widetilde\cE_0(Z_i):=\left(\begin{smallmatrix} 0 \\ e^a(X_i) \end{smallmatrix}\right)+M_i(Z_i)^{-1}\,\left(\begin{smallmatrix} \sfd\Theta_i^{\widehat\a}(Z_i) \\ 0 \end{smallmatrix}\right)\,,
\qqq
whose solution reads
\qq\nn
\widetilde\cE_t(Z_i)=\ee^{tM_i(Z_i)}\,\widetilde\cE_0(Z_i)\,.
\qqq
Thus, we obtain the sought-after expression for the pullback super-1-forms:
\qq\nn
\left(\begin{smallmatrix} E^{\widehat\a} \\ E^a \end{smallmatrix}\right)(Z_i)\equiv\vec\cE_1(Z_i)=\ee^{M_i(Z_i)}\,\left(\left(\begin{smallmatrix} 0 \\ e^a(X_i) \end{smallmatrix}\right)+M_i(Z_i)^{-1}\,\left(\begin{smallmatrix} \sfd\Theta_i^{\widehat\a}(Z_i) \\ 0 \end{smallmatrix}\right)\right)-M_i(Z_i)^{-1}\,\left(\begin{smallmatrix} \sfd\Theta_i^{\widehat\a}(Z_i) \\ 0 \end{smallmatrix}\right)\,.
\qqq
Now, as
\qq\nn
M_i(Z_i)^2=\left(\begin{smallmatrix} f_{\widehat\b\widehat\g}^{\ \ c}\,f_{c\widehat\d}^{\ \ \widehat\a} & 0 \\ 0 & f_{b\widehat\g}^{\ \ \widehat\vep}\,f_{\widehat\vep\widehat\d}^{\ \ a} \end{smallmatrix}\right)\,\Theta_i^{\widehat\g}(Z_i)\,\Theta_i^{\widehat\d}(Z_i)
\qqq
is block-diagonal, whereas $\,M_i(Z_i)\,$ is block-off-diagonal, we have
\qq\nn
\left(\begin{smallmatrix} E^{\widehat\a}(Z_i) \\ 0 \end{smallmatrix}\right)&=&{\rm sh}M_i(Z_i)\,\left(\begin{smallmatrix} 0 \\ e^a(X_i) \end{smallmatrix}\right)+\tfrac{{\rm sh}M_i(Z_i)}{M_i(Z_i)}\,\left(\begin{smallmatrix} \sfd\Theta_i^{\widehat\a}(Z_i) \\ 0 \end{smallmatrix}\right)=\tfrac{{\rm sh}M_i(Z_i)}{M_i(Z_i)}\,\left(\begin{smallmatrix} \sfd\Theta_i^{\widehat\a}(Z_i)+f_{a\widehat\b}^{\ \ \widehat\a}\,e^a(X_i)\,\Theta_i^\b(Z_i) \\ 0 \end{smallmatrix}\right)\cr\cr
&\equiv&\tfrac{{\rm sh}M_i(Z_i)}{M_i(Z_i)}\,\left(\begin{smallmatrix} \sfD\Theta_i^{\widehat\a}(Z_i) \\ 0 \end{smallmatrix}\right)\,,\cr\cr\cr
\left(\begin{smallmatrix} 0 \\ E^a(Z_i) \end{smallmatrix}\right)&=&{\rm ch}M_i(Z_i)\,\left(\begin{smallmatrix} 0 \\ e^a(X_i) \end{smallmatrix}\right)+\tfrac{{\rm ch}M_i(Z_i)-\bd1}{M_i(Z_i)}\,\left(\begin{smallmatrix} \sfd\Theta_i^{\widehat\a}(Z_i) \\ 0 \end{smallmatrix}\right)\cr\cr
&=&\vec\cE_0(Z_i)+\bigl({\rm ch}M_i(Z_i)-\bd1\bigr)\,\left(\begin{smallmatrix} 0 \\ e^a(X_i) \end{smallmatrix}\right)+\tfrac{{\rm ch}M_i(Z_i)-\bd1}{M_i(Z_i)}\,\left(\begin{smallmatrix} \sfd\Theta_i^{\widehat\a}(Z_i) \\ 0 \end{smallmatrix}\right)\cr\cr
&\equiv&\vec\cE_0(Z_i)+\tfrac{2\,{\rm sh}^2\tfrac{M_i(Z_i)}{2}}{M_i(Z_i)^2}\,M_i(Z_i)\,\left(M_i(Z_i)\,\left(\begin{smallmatrix} 0 \\ e^a(X_i) \end{smallmatrix}\right)+\left(\begin{smallmatrix} \sfd\Theta_i^{\widehat\a}(Z_i) \\ 0 \end{smallmatrix}\right)\right)\cr\cr
&\equiv&\vec\cE_0(Z_i)+\tfrac{2\,{\rm sh}^2\tfrac{M_i(Z_i)}{2}}{M_i(Z_i)^2}\,M_i(Z_i)\,\left(\begin{smallmatrix} \sfD\Theta_i^{\widehat\a}(Z_i) \\ 0 \end{smallmatrix}\right)=\vec\cE_0(Z_i)+\tfrac{2\,{\rm sh}^2\tfrac{M_i(Z_i)}{2}}{M_i(Z_i)^2}\,\left(\begin{smallmatrix} 0 \\ [\sfD\Theta_i(Z_i),\Theta_i(Z_i)]^a \end{smallmatrix}\right)\,.
\qqq
This gives us the solution stated in the proposition. $\qed$

\section{A proof of Proposition \ref{prop:canlifunK}}\label{app:canlifunK}

The defining equation \eqref{eq:canlifunK} of the canonical lift rewrites as
\qq\nn
\sum_{\t\in\Pgt}\,\int_{\xcC_p\cap\t}\,\Vol(\xcC_p)\,\bigl[\D_{\t\,\unl A}\bigl(X;\si_{i_\t}\circ\unl\xi_\t(\cdot)\bigr)\,\theta_{\rm L}^{\unl A}\bigl(\si_{i_\t}\circ\unl\xi_\t(\cdot)\bigr)+\pi_{\t\,\unl A}(\cdot)\,\pLie{\cK_X}\theta_{\rm L}^{\unl A}\bigl(\si_{i_\t}\circ\xi_\t(\cdot)\bigr)\bigr]=0\,,
\qqq
and so it makes sense only if
\qq\nn
\ceL_{J_\k}\con\pLie{\cK_X}\theta_{\rm L}^{\unl A}\rstr_{\si(\txG/\txH)}=0\,,\qquad\k\in\ovl{1,\dim\,\hgt}\,,
\qqq
in which case a solution reads
\qq\nn
\D_{\t\,\unl A}\bigl(X;\si_{i_\t}\circ\unl\xi_\t(\cdot)\bigr)=-\pi_{\t\,\unl B}(\cdot)\,\ceL_{t_{\unl A}}\con\pLie{\cK_X}\theta_{\rm L}^{\unl B}\bigl(\si_{i_\t}\circ\xi_\t(\cdot)\bigr)\,.
\qqq
That the former condition is satisfied readily follows from the transformation properties of the components of the Maurer--Cartan super-1-form along $\,\tgt\,$ discussed previously. Indeed, in the light of \Reqref{eq:thetgtens}, we obtain (for $\,x\in\cO_i$)
\qq\nn
\pLie{\cK_X}\theta_{\rm L}^{\unl A}\bigl(\si_i(x)\bigr)&\equiv&\tfrac{\sfd\ }{\sfd t}\rstr_{t=0}\,\theta_{\rm L}^{\unl A}\bigl(\ee^{tX}\cdot\si_i(x)\cdot\ee^{t Y_i(X;x)}\bigr)=\tfrac{\sfd\ }{\sfd t}\rstr_{t=0}\,\bigl(\sfT_e\Ad_{\ee^{-t Y_i(X;x)}}\bigr)^{\unl A}_{\ \unl B}\,\theta_{\rm L}^{\unl B}\bigl(\si_i(x)\bigr)\cr\cr
&=&-\bigl(\ad_{Y_i(X;x)}\bigr)^{\unl A}_{\ \unl B}\,\theta_{\rm L}^{\unl B}\bigl(\si_i(x)\bigr)\,.
\qqq
We conclude that the constraints \eqref{eq:canlifunK} are soluble, and a solution has the form indicated in the statement of the proposition.
$\qed$

\section{A proof of Proposition \ref{prop:PoiNoe}}\label{app:PoiNoe}

We compute
\qq\nn
&&\{h_{X_1},h_{X_2}\}_{\Om^{({\rm NG})}_{{\rm GS},p}}[\unl\xi,\pi]\equiv\widetilde\cK_{X_2}\con\widetilde\cK_{X_1}\con\Om^{({\rm NG})}_{{\rm GS},p}[\unl\xi,\pi]\cr\cr
&=&-\sum_{\t\in\Pgt}\,\int_{\xcC_p\cap\t}\,\Vol(\xcC_p)\,\pi_{\t\,\unl A}(\cdot)\,\pLie{\cK_{X_2}}\bigl(\cK_{X_1}\con\theta^{\unl A}_{\rm L}\bigr)\bigl(\si_{i_\t}\circ\unl\xi_\t(\cdot)\bigr)\cr\cr
&&-\widetilde\cK_{X_2}[\unl\xi,\pi]\con\sum_{\t\in\Pgt}\,\int_{\xcC_p\cap\t}\,\Vol(\xcC_p)\,\d\pi_{\t\,\unl A}(\cdot)\,\bigl(\cK_{X_1}\con\theta_{\rm L}^{\unl A}\bigr)\bigl(\si_{i_\t}\circ\unl\xi_\t(\cdot)\bigr)\cr\cr
&&-\sum_{\t\in\Pgt}\,\int_{\xcC_p\cap\t}\,\ev^*\bigl[\pLie{\cK_{X_2}}\bigl(\cK_{X_1}\con\underset{\tx{\ciut{(p+1)}}}{\b}-\underset{\tx{\ciut{(p)}}}{\G_{X_1}}\bigr)-\sfd\bigl(\cK_{X_2}\con\bigl(\cK_{X_1}\con\underset{\tx{\ciut{(p+1)}}}{\b}-\underset{\tx{\ciut{(p)}}}{\G_{X_1}}\bigr)\bigr)\bigr]\bigl(\si_{i_\t}\circ\unl\xi_\t(\cdot)\bigr)\cr\cr
&=&-\sum_{\t\in\Pgt}\,\bigg[\int_{\xcC_p\cap\t}\,\Vol(\xcC_p)\,\pi_{\t\,\unl A}(\cdot)\,\bigl([\cK_{X_2},\cK_{X_1}]\con\theta^I_{\rm L}\bigr)\bigl(\si_{i_\t}\circ\unl\xi_\t(\cdot)\bigr)+\int_{\xcC_p\cap\t}\,\ev^*[\cK_{X_2},\cK_{X_1}]\con\underset{\tx{\ciut{(p+1)}}}{\b}\bigl(\si_{i_\t}\circ\unl\xi_\t(\cdot)\bigr)\cr\cr
&&-\int_{\xcC_p\cap\t}\,\ev^*\bigl(-\cK_{X_1}\con\sfd\underset{\tx{\ciut{(p)}}}{\G_{X_2}}+\pLie{\cK_{X_2}}\underset{\tx{\ciut{(p)}}}{\G_{X_1}}+\sfd\bigl(\cK_{X_2}\con\cK_{X_1}\con\underset{\tx{\ciut{(p+1)}}}{\b}\bigr)-\sfd\bigl(\cK_{X_2}\con\underset{\tx{\ciut{(p)}}}{\G_{X_1}}\bigr)\bigr)\bigl(\si_{i_\t}\circ\unl\xi_\t(\cdot)\bigr)\bigg]\,,
\qqq
and so, upon invoking relation \eqref{eq:KKKh}, we obtain
\qq\nn
&&\{h_{X_1},h_{X_2}\}_{\Om^{({\rm NG})}_{{\rm GS},p}}[\unl\xi,\pi]\cr\cr
&=&h_{-[X_1,X_2]}[\unl\xi,\pi]+\sum_{\t\in\Pgt}\,\bigg[\int_{\xcC_p\cap\t}\,\ev^*\bigl(\pLie{\cK_{X_2}}\underset{\tx{\ciut{(p)}}}{\G_{X_1}}-\pLie{\cK_{X_1}}\underset{\tx{\ciut{(p)}}}{\G_{X_2}}-\underset{\tx{\ciut{(p)}}}{\G_{[X_1,X_2]}}\bigr)\bigl(\si_{i_\t}\circ\unl\xi_\t(\cdot)\bigr)\cr\cr
&&+\int_{\xcC_p\cap\t}\,\ev^*\sfd\bigl(\cK_{X_1}\con\underset{\tx{\ciut{(p)}}}{\G_{X_2}}-\cK_{X_2}\con\underset{\tx{\ciut{(p)}}}{\G_{X_1}}+\cK_{X_2}\con\cK_{X_1}\con\underset{\tx{\ciut{(p+1)}}}{\b}\bigr)\bigl(\si_{i_\t}\circ\unl\xi_\t(\cdot)\bigr)\bigg]\cr\cr
&\equiv&h_{-[X_1,X_2]}[\unl\xi,\pi]+\sum_{\t\in\Pgt}\,\bigg[\int_{\xcC_p\cap\t}\,\ev^*\sfd\bigl(\cK_{X_1}\con\underset{\tx{\ciut{(p)}}}{\G_{X_2}}-\cK_{X_2}\con\underset{\tx{\ciut{(p)}}}{\G_{X_1}}+\cK_{X_2}\con\cK_{X_1}\con\underset{\tx{\ciut{(p+1)}}}{\b}\bigr)\bigl(\si_{i_\t}\circ\unl\xi_\t(\cdot)\bigr)\cr\cr
&&-\int_{\xcC_p\cap\t}\,\ev^*\underset{\tx{\ciut{(p)}}}{\a_{X_1,X_2}}\bigl(\si_{i_\t}\circ\unl\xi_\t(\cdot)\bigr)\bigg]\,,
\qqq
as claimed. $\qed$

\section{Conventions for \& facts about the $\,{\rm AdS}_5\x\bS^5\,$ Clifford algebras}\label{app:CliffAdSS}

In the present paper, we are dealing with a ditignuished, geometrically/physically motivated realisation of the Clifford algebra $\,\Cliff(\bR^{9,1})\,$ in terms of the generators of the Clifford algebras $\,\Cliff(\bR^{4,1})$, $\Cliff(\bR^{5,0})\,$ and $\,\Cliff(\bR^{2,1})$.\ Let us denote the generators of $\,\Cliff(\bR^{4,1})\,$ (in the 4-dimensional spinor representation, in which they are traceless) as
\qq\nn
\{\G^a\}_{a\in\ovl{0,4}}\,,
\qqq
those of $\,\Cliff(\bR^{5,0})\,$ (also in the 4-dimensional spinor representation, in which they are traceless) as
\qq\nn
\{\G^{a'}\}_{a'\in\ovl{5,9}}\,,
\qqq
and -- finally -- those of $\,\Cliff(\bR^{2,1})\,$ (in the 2-dimensional spinor representation, in which they are traceless) as
\qq\nn
\{\g^0\equiv\sfi\,\si_2,\g^1\equiv\si_1,\g^2\equiv-\si_3\}\,,
\qqq
where
\qq\nn
\si_1=\bigl(\begin{smallmatrix} 0 & 1 \\ 1 & 0 \end{smallmatrix}\bigr)\,,\qquad\qquad\si_2=\bigl(\begin{smallmatrix} 0 & -\sfi \\ \sfi & 0 \end{smallmatrix}\bigr)\,,\qquad\qquad\si_3=\bigl(\begin{smallmatrix} 1 & 0 \\ 0 & -1 \end{smallmatrix}\bigr)
\qqq
are the standard Pauli matrices. The latter generators satisfy the algebra
\qq\nn
\g^I\cdot\g^J=\eta^{IJ}\,\bd1+\eta^{II}\,\eta^{JJ}\,\vep^{IJ}_{\ \ K}\,\g^K\,,\qquad\qquad(\eta^{IJ})=\diag(-1,1,1)\,.
\qqq
Denote
\qq\label{eq:Gext1}
\widehat\G^{\widehat b}=\left\{ \barr{cl} \G^b\ox\bd1 & \tx{ if }\ \widehat b=b\in\ovl{0,4} \cr\cr \bd1\ox\sfi\,\G^{b'} & \tx{ if }\ \widehat b=b'\in\ovl{5,9} \earr \right.\,.
\qqq
We readily prove the identity
\qq\label{eq:GamasandGama}
\widehat\G_{\widehat a}\,\widehat\G_{\widehat b}\,\widehat\G^{\widehat a}=-8\vep_{\widehat b}\,\widehat\G_{\widehat b}\,,
\qqq
alongside
\qq\label{eq:GambsandGamaa}\qquad\qquad\qquad
\widehat\G_a\,\widehat\G_{\widehat b}\,\widehat\G^a=\left\{ \barr{cl} -3\widehat\G_b & \tx{ if }\ \widehat b=b\in\ovl{0,4} \cr\cr
5\widehat\G_{b'} & \tx{ if }\ \widehat b=b'\in\ovl{5,9} \earr \right.\,,\qquad\qquad\widehat\G_{a'}\,\widehat\G_{\widehat b}\,\widehat\G^{a'}=\left\{ \barr{cl} -5\widehat\G_b & \tx{ if }\ \widehat b=b\in\ovl{0,4} \cr\cr
3\widehat\G_{b'} & \tx{ if }\ \widehat b=b'\in\ovl{5,9} \earr \right.\,.
\qqq
The standard generators of $\,\Cliff(\bR^{9,1})\,$ are now given by
\qq\nn
\{\unl\g^a\equiv\widehat\G^a\ox\g^1,\unl\g^{a'}\equiv-\widehat\G^{a'}\ox\g^0\}_{a\in\ovl{0,4},a'\in\ovl{5,9}}\,,
\qqq
{\it cp} \Rcite{Metsaev:1998it}. Out of these, we form bi-vectorial objects
\qq\nn
\G^{ab}:=\tfrac{1}{2}\,[\G^a,\G^b]\,,\qquad\qquad\G^{a'b'}:=\tfrac{1}{2}\,[\G^{a'},\G^{b'}]\,,
\qqq
and
\qq\nn
\widehat\G^{\widehat a\widehat b}:=\tfrac{1}{2}\,[\widehat\G^{\widehat a},\widehat\G^{\widehat b}]\,.
\qqq
linearly independent from the generators. These satisfy the identities
\qq\label{eq:GamacommGambc}
[\widehat\G_{\widehat a},\widehat\G^{\widehat b\widehat c}]=2\vep_{\widehat a}\,\bigl(\d_{\widehat a}^{\ \widehat b}\,\widehat\G^{\widehat c}-\d_{\widehat a}^{\ \widehat c}\,\widehat\G^{\widehat b}\bigr)\,.
\qqq
The antisymmetrised products of generators of order $\,n\geq 3\,$ are related to those of order $\,5-n\,$ by the (Hodge-type) duality relations:
\qq\nn
&\G^{a_1 a_2\ldots a_n}=\tfrac{c(4,1;n)}{(5-n)!}\,\vep^{a_1 a_2\ldots a_n}_{\qquad\quad a_{n+1} a_{n+2}\ldots a_5}\,\G^{a_{n+1} a_{n+2}\ldots a_5}\,,\quad n\in\ovl{1,5}\,,&\cr\cr
&\G^{a_1' a_2'\ldots a_n'}=\tfrac{c(5,0;n)}{(5-n)!}\,\vep^{a_1' a_2'\ldots a_n'}_{\qquad\quad a_{n+1}' a_{n+2}'\ldots a_5'}\,\G^{a_{n+1}' a_{n+2}'\ldots a_5'}\,,\quad n\in\ovl{1,5}\,,&
\qqq
in which $\,c(p,q;n)\in\{-1,1\}\,$ is a sign depending upon the signature $\,(p,q)\,$ of the metric tensor and upon $\,n$,\ and $\,\vep^{a_1 a_2\ldots a_n}_{\qquad\quad a_{n+1} a_{n+2}\ldots a_5}\,$ (resp.\ $\,\vep^{a_1' a_2'\ldots a_n'}_{\qquad\quad a_{n+1}' a_{n+2}'\ldots a_5'}$) is obtained from the Levi--Civita symbol
\qq\nn
\vep_{a_1 a_2\ldots a_n\ldots a_5}=\left\{ \barr{cl} \sign(\si) & \tx{if }\ (a_1,a_2,\ldots,a_5)=\bigl(\si(1),\si(2),\si(3),\si(4),\si(5)\bigr) \tx{ for } \si\in\Sgt_{\ovl{1,5}} \\ 0 & \tx{otherwise} \earr\right.\,.
\qqq
by contraction of the first $n$ indices with the inverse metric $\,\eta^{ab}\,$ (resp.\ $\,\d^{a'b'}$).

The set of generators of the Clifford algebra is augmented with the charge-conjugation matrices -- for $\,\Cliff(\bR^{4,1})$:
\qq\nn
C=-C^{\rm T}\,,
\qqq
for $\,\Cliff(\bR^{5,0})$:
\qq\nn
C'=-C'{}^{\rm T}\,,
\qqq
for $\,\Cliff(\bR^{2,1})$:
\qq\nn
c\equiv\g_0=-c^{\rm T}\,,
\qqq
and -- in the end -- also for $\,\Cliff(\bR^{9,1})$:
\qq\label{eq:CCc}
\cC=C\ox C'\ox c\equiv\widehat C\ox c=-\cC^{\rm T}\,.
\qqq
These enable us to descibe the basic symmetry properties of the said generators,
\qq\nn
\bigl(\G^a\bigr)^{\rm T}=C\,\G^a\,C^{-1}\,,\qquad\qquad\bigl(\G^{a'}\bigr)^{\rm T}=C'\,\G^{a'}\,C'{}^{-1}\,,\qquad\qquad\bigl(\g^I\bigr)^{\rm T}=-c\,\g^I\,c^{-1}\,,
\qqq
which we may rewrite equivalently as
\qq\nn
\bigl(C\,\G^a\bigr)^{\rm T}=-C\,\G^a\,,\qquad\qquad\bigl(C'\,\G^{a'}\bigr)^{\rm T}=-C'\,\G^{a'}\,,\qquad\qquad(c\,\g^I)^{\rm T}=c\,\g^I\,.
\qqq
Taking these identities into account, we may readily construct a Clifford basis of the full matrix algebra
\qq\nn
\bC(4)=\corr{\bd1,\G^a,\G^{a_1 a_2}\ \vert\ a,a_1,a_2\in\ovl{0,4}}_\bC\equiv\corr{\bd1,\G^{a'},\G^{a_1' a_2'}\ \vert\ a',a_1',a_2'\in\ovl{5,9}}_\bC\,,
\qqq
and of its subalgebra composed of the symmetric matrices
\qq\nn
\bC(4)^{\rm sym}=\corr{C\,\G^{a_1 a_2}\ \vert\ a_1,a_2\in\ovl{0,4}}_\bC\equiv\corr{C'\,\G^{a_1' a_2'}\ \vert\ a',a_1',a_2'\in\ovl{5,9}}_\bC\,,
\qqq
as well as the Clifford basis of the analogous subalgebra in dimension $\,4\x 4\x 2=32$,
\qq
\bC(32)^{\rm sym}=\langle\,\widetilde\G^I,\widetilde\G^J\cdot\bigl(\sfi\,\vep_{\widehat a}\,\widehat\G^{\widehat a}\ox\bd1\bigr)\equiv\widetilde\G^{J\widehat a},\vep_{\widehat b\widehat c}\,\widehat C\,\widehat\G^{\widehat b\widehat c}\ox\g^0\equiv\widetilde\G^{\widehat b\widehat c},C\,\G^{de}\ox C'\,\G^{f'}\ox\g^0\equiv\widetilde\G^{def'},\cr\cr
\sfi\,C\,\G^g\ox C'\,\G^{h'i'}\ox\g^0\equiv\widetilde\G^{gh'i'},\widetilde\G^K\,\widehat\G^j\,\widehat\G^{k'}\equiv\widetilde\G^{Kjk'},\widetilde\G^L\cdot(\G^{lm}\ox\G^{n'o'}\ox\bd1)\equiv\widetilde\G^{Llmn'o'}\ \vert\cr\cr
I,J,K,L\in\{0,1,2\},\ \widehat a,\widehat b,\widehat c\in\ovl{0,9},\ d,e,g,j,l,m\in\ovl{0,4},\ f',h',i',k',n',o'\in\ovl{5,9}\,\rangle_\bC\,,\label{eq:C32symbas}
\qqq
written out above, in conformity with the previously announced factorisation of the ${\rm Spin}$-group, in terms of the matrices
\qq\nn
\widetilde\G^0:=-\widehat C\ox\bd1\,,\qquad\qquad\widetilde\G^1:=\widehat C\ox\si_3\,,\qquad\qquad\widetilde\G^2:=\widehat C\ox\si_1\,,\qquad\qquad\qquad\cC:=\widehat C\ox\sfi\,\si_2\,.
\qqq
For later convenience, we represent the basis elements as
\qq\nn
\cC\,\unl\G^\la=\bigl(\cC\,\unl\G^\la\bigr)^{\rm T}\,,
\qqq
where
\qq\nn
\unl\G^\la&\in&\{\bd1\ox\g^I\equiv\unl\G^I,\sfi\,\vep_{\widehat a}\,\widehat\G^{\widehat a}\ox\g^J\equiv\unl\G^{J\widehat a},\vep_{\widehat b\widehat c}\,\widehat\G^{\widehat b\widehat c}\ox\bd1\equiv\unl\G^{\widehat b\widehat c},\G^{de}\ox\G^{f'}\ox\bd1\equiv\unl\G^{def'},\sfi\,\G^g\ox\G^{h'i'}\ox\bd1\equiv\unl\G^{gh'i'},\cr\cr
&&\hspace{0.25cm}\widehat\G^j\,\widehat\G^{k'}\ox\g^K\equiv\unl\G^{Kjk'},\G^{lm}\ox\G^{n'o'}\ox\g^L\equiv\unl\G^{Llmn'o'}\}\,.
\qqq
These satisfy the orthonormality relations:
\qq\label{eq:ortoGamma}
\tfrac{1}{32}\,\tr_{\bC^{\x 32}}\,\bigl(\bigl(\cC\,\unl\G^\la\bigr)\cdot\bigl(\cC\,\unl\G^\mu\bigr)^{-1}\bigr)=\d^{\la\mu}
\qqq
and the completeness relations:
\qq\nn
\tfrac{1}{32}\,\bigl(\cC\,\unl\G^\la\bigr)_{\a\a'I\b\b'J}\,\bigl(\bigl(\cC\,\unl\G^\la\bigr)^{-1}\bigr)^{\g\g'K\d\d'L}=\tfrac{1}{2}\,\d_{\a\a'I}^{\ (\g\g'K}\,\d_{\b\b'J}^{\ \d\d'L)}\,.
\qqq

\section{A proof of Proposition \ref{prop:KS}}\label{app:KSproof}

The point of departure is the computation of the various structure constants $\,c^{\la\,0\widehat a}_{\ \ \ \ \mu}$.\ We obtain
\qq\nn
c^{I\,0\widehat a}_{\ \ \ \ \mu}\ &:&\ [\unl\G^I,\unl\G^{0\widehat a}]=2\eta^{II}\,\vep^{0I}_{\ \ J}\,\unl\G^{J\widehat a}\,,\cr\cr\cr
c^{I\widehat b\,0\widehat a}_{\ \ \ \ \ \mu}\ &:&\ [\unl\G^{I\widehat b},\unl\G^{0\widehat a}]=\left\{ \barr{cl} 2\bigl(\eta^{0I}\,\vep_{\widehat a\widehat b}\,\unl\G^{\widehat a\widehat b}-\vep_{\widehat a}\,\eta^{\widehat a\widehat b}\,\eta^{II}\,\vep^{0I}_{\ \ J}\,\unl\G^J\bigr) & \tx{ if }\ (\widehat a,\widehat b)\in\ovl{0,4}^{\x 2}\cup\ovl{5,9}^{\x 2} \cr\cr 2\eta^{II}\,\vep^{0I}_{\ \ J}\,\unl\G^{Jab'} & \tx{ if }\ (\widehat a,\widehat b)=(a,b')\in\ovl{0,4}\x\ovl{5,9} \cr\cr -2\eta^{II}\,\vep^{0I}_{\ \ J}\,\unl\G^{Jba'} & \tx{ if }\ (\widehat a,\widehat b)=(a',b)\in\ovl{5,9}\x\ovl{0,4} \earr \right.\,,\cr\cr\cr
c^{\widehat b\widehat c\,0\widehat a}_{\ \ \ \ \ \mu}\ &:&\ [\unl\G^{\widehat b\widehat c},\unl\G^{0\widehat a}]=\left\{ \barr{cl} 2\eta^{\widehat a\widehat a}\,\bigl(\d^{\widehat c\widehat a}\,\d^{\widehat b}_{\ \widehat d}-\d^{\widehat b\widehat a}\,\d^{\widehat c}_{\ \widehat d}\bigr)\,\unl\G^{0\widehat d} & \tx{ if }\ (\widehat a,\widehat b,\widehat c)\in\ovl{0,4}^{\x 3}\cup\ovl{5,9}^{\x 3} \cr\cr 0 & \tx{ otherwise } \earr \right.\,,\cr\cr\cr
c^{bcd'\,0\widehat a}_{\qquad\ \mu}\ &:&\ [\unl\G^{bcd'},\unl\G^{0\widehat a}]=\left\{ \barr{cl} 2\eta^{\widehat a\widehat a}\,\bigl(\d^{c\widehat a}\,\d^b_{\ e}-\d^{b\widehat a}\,\d^c_{\ e}\bigr)\,\unl\G^{0ed'} & \tx{ if }\ \widehat a\in\ovl{0,4} \cr\cr 2\sfi\,\unl\G^{0bca'd'} & \tx{ if }\ \widehat a=a'\in\ovl{5,9} \earr \right.\,,\cr\cr\cr
c^{bc'd'\,0\widehat a}_{\qquad\ \, \mu}\ &:&\ [\unl\G^{bc'd'},\unl\G^{0\widehat a}]=\left\{ \barr{cl} 2\unl\G^{0abc'd'}  & \tx{ if }\ \widehat a=a\in\ovl{0,4} \cr\cr 2\sfi\,\eta^{\widehat a\widehat a}\,\bigl(\d^{c'\widehat a}\,\d^{d'}_{\ e'}-\d^{d'\widehat a}\,\d^{c'}_{\ e'}\bigr)\,\unl\G^{0be'} & \tx{ if }\ \widehat a\in\ovl{5,9} \earr \right.\,,\cr\cr\cr
c^{Ibc'\,0\widehat a}_{\qquad\ \mu}\ &:&\ [\unl\G^{Ibc'},\unl\G^{0\widehat a}]=\left\{ \barr{cl} 2\,\bigl(\eta^{0I}\,\bigl(1-\d^{ba}\bigr)\,\unl\G^{abc'}-\d^{ba}\,\eta^{aa}\,\eta^{II}\,\vep^{0I}_{\ \ J}\,\unl\G^{Jc'}\bigr)\  & \tx{ if }\ \widehat a=a\in\ovl{0,4} \cr\cr 2\,\bigl(\d^{c'a'}\,\eta^{II}\,\vep^{0I}_{\ \ J}\,\unl\G^{Jb}-\eta^{0I}\,\bigl(1-\d^{c'a'}\bigr)\,\unl\G^{ba'c'}\bigr)\ & \tx{ if }\ \widehat a=a'\in\ovl{5,9} \earr \right.\,,\cr\cr\cr
c^{Ibcd'e'\,0\widehat a}_{\qquad\quad\ \mu}\ &:&\ [\unl\G^{Ibcd'e'},\unl\G^{0\widehat a}]=\left\{ \barr{cl} 2\eta^{aa}\,\eta^{0I}\,\bigl(\d^{ca}\,\d^b_{\ f}-\d^{ba}\,\d^c_{\ f}\bigr)\,\unl\G^{fd'e'}+\sfi\,c(4,1;3)\,\eta^{II}\,\vep^{0I}_{\ \ J}\,\vep^{abc}_{\ \ \ fg}\,\unl\G^{Jfgd'e'}\cr\cr \tx{ if }\ \widehat a=a\in\ovl{0,4} \cr\cr\cr 2\sfi\,\eta^{0I}\,\bigl(\d^{d'a'}\,\d^{e'}_{\ f'}-\d^{e'a'}\,\d^{d'}_{\ f'}\bigr)\,\unl\G^{bcf'}+c(5,0;3)\,\eta^{II}\,\vep^{0I}_{\ \ J}\,\vep^{a'd'e'}_{\ \ \ \ f'g'}\,\unl\G^{Jbcf'g'}\cr\cr \tx{ if }\ \widehat a=a'\in\ovl{5,9} \earr \right.\,.
\qqq
Given these, we compute, for $\,\widehat a=a\in\ovl{0,4}$,
\qq\nn
0&=&\cZ_\la\,c^{\la\,0a}_{\quad\ \mu}\,\cC\,\unl\G^\mu=2\cZ_I\,\eta^{II}\,\vep^{0I}_{\ \ J}\,\unl\G^{Ja}+2\cZ_{Ib}\,\bigl(\eta^{0I}\,\unl\G^{ab}-\eta^{ab}\,\eta^{II}\,\vep^{0I}_{\ \ J}\,\unl\G^J\bigr)+2\cZ_{Ib'}\,\eta^{II}\,\vep^{0I}_{\ \ J}\,\unl\G^{Jab'}-4\cZ_{ab}\,\eta^{aa}\,\unl\G^{0b}\cr\cr
&&-4\cZ_{abc'}\,\eta^{aa}\,\unl\G^{0bc'}+2\cZ_{bc'd'}\,\unl\G^{0abc'd'}+2\cZ_{Ibc'}\,\bigl(\eta^{0I}\,\bigl(1-\d^{ba}\bigr)\,\unl\G^{abc'}-\d^{ba}\,\eta^{aa}\,\eta^{II}\,\vep^{0I}_{\ \ J}\,\unl\G^{Jc'}\bigr)\cr\cr
&&+\cZ_{Ibcd'e'}\,\bigl(-4\eta^{aa}\,\eta^{0I}\,\d^{ba}\,\unl\G^{cd'e'}+\sfi\,c(4,1;3)\,\eta^{II}\,\vep^{0I}_{\ \ J}\,\vep^{abc}_{\ \ \ fg}\,\unl\G^{Jfgd'e'}\bigr)
\qqq
and, for $\,\widehat a=a'\in\ovl{5,9}$,
\qq\nn
0&=&\cZ_\la\,c^{\la\,0a'}_{\quad\ \mu}\,\cC\,\unl\G^\mu=2\cZ_I\,\eta^{II}\,\vep^{0I}_{\ \ J}\,\unl\G^{Ja'}+2\cZ_{Ib'}\,\bigl(\d^{a'b'}\,\eta^{II}\,\vep^{0I}_{\ \ J}\,\unl\G^J-\eta^{0I}\,\unl\G^{a'b'}\bigr)-2\cZ_{Ib}\,\eta^{II}\,\vep^{0I}_{\ \ J}\,\unl\G^{Jba'}-4\cZ_{a'b'}\,\unl\G^{0b'}\cr\cr
&&+2\sfi\,\cZ_{bcd'}\,\unl\G^{0bca'd'}+4\sfi\,\cZ_{ba'c'}\,\unl\G^{0bc'}+2\cZ_{Ibc'}\,\bigl(\d^{c'a'}\,\eta^{II}\,\vep^{0I}_{\ \ J}\,\unl\G^{Jb}-\eta^{0I}\,\bigl(1-\d^{c'a'}\bigr)\,\unl\G^{ba'c'}\bigr)\cr\cr
&&+\cZ_{Ibcd'e'}\,\bigl(2\sfi\,\eta^{0I}\,\bigl(\d^{d'a'}\,\d^{e'}_{\ f'}-\d^{e'a'}\,\d^{d'}_{\ f'}\bigr)\,\unl\G^{bcf'}+c(5,0;3)\,\eta^{II}\,\vep^{0I}_{\ \ J}\,\vep^{a'd'e'}_{\ \ \ \ f'g'}\,\unl\G^{Jbcf'g'}\bigr)\,.
\qqq
On the basis of the orthonormality relations \eqref{eq:ortoGamma}, we now infer the vanishing of the following components of the admissible central charge:
\qq\nn
\cZ_1,\cZ_2,\cZ_{I\widehat a},\cZ_{\widehat a\widehat b},\cZ_{abc'},\cZ_{ab'c'},\cZ_{Iab'},\cZ_{Iabc'd'}=0\,.
\qqq
In other words, we are left, by the end of the day, with $\,\cZ_0\,$ as the only non-zero (and otherwise unconstrained) parameter of the deformation, in conformity with the claim of the Theorem. $\qed$

\section{A proof of Proposition \ref{prop:ssKRproof}}\label{app:ssKRproof}

Impose the super-Jacobi identity
\qq\nn
{\rm sJac}\bigl(Q_{\a\a'I},P_{\widehat a},P_{\widehat b}\bigr)=0
\qqq
for $\,(\widehat a,\widehat b)\in\ovl{0,4}^{\x 2}\cup\ovl{5,9}^{\x 2}\,$ to derive the constraints
\qq\nn
0&=&-\tfrac{\sfi\a}{2}\,\bigl(\sfi\,\bigl(\widehat C\,\widehat\G_{\widehat a\widehat b}\ox\si_1\bigr)_{\a\a'I\b\b'J}\,\cZ^{\b\b'J}+2\g^\mu_{\widehat a}\,\bigl(\widehat C\,\widehat\G_{\widehat a\widehat b}\,\widehat C^{-1}\ox\si_3\,\si_\mu\bigr)_{\a\a'I}^{\ \b\b'J}\,Q_{\b\b'J}\cr\cr
&&\hspace{.75cm}+2\d^\mu_{\widehat a}\,\bigl(\widehat C\,\widehat\G_{\widehat a\widehat b}\ox\si_3\,\si_\mu\bigr)_{\a\a'I\b\b'J}\,\cZ^{\b\b'J}\bigr)\,,
\qqq
and hence
\qq\nn
\g^\mu_{\widehat a}=0\,,\qquad\mu\in\{0,1,2,3\}\,;\qquad\qquad\d^\nu_{\widehat a}=0\,,\qquad\nu\in\{0,1,3\}\,;\qquad\qquad\d^2_{\widehat a}=\tfrac{1}{2}\,,
\qqq
so that we necessarily have
\qq\nn
[\cZ^{\a\a'I},P_{\widehat a}]^\sim=\tfrac{1}{2}\,\bigl(\widehat\G_{\widehat a}\ox\si_2\bigr)\,^{\a\a'I}_{\ \b\b'J}\,\cZ^{\b\b'J}\,.
\qqq
The super-Jacobi identity of the same type for $\,(\widehat a,\widehat b)\in\ovl{0,4}\x\ovl{5,9}\,$ is now automatically satisfied.

Next, we require
\qq\nn
{\rm sJac}\bigl(Q_{\a\a'I},Q_{\b\b'J},Q_{\g\g'K}\bigr)=0\,,
\qqq
which yields
\qq\nn
0&=&\a\,\bigl(\widehat C\,\widehat\G^{\widehat a}\ox\bd1\bigr)_{\a\a'I\b\b'J}\,\bigl(\widehat C\,\widehat\G_{\widehat a}\ox\si_3\bigr)_{\g\g'K\d\d'L}\,\cZ^{\d\d'L}\cr\cr
&&+\b^{\unl\mu}_{\rm s}\,\bigl(\widehat C\ox\si_{\unl\mu}\bigr)_{\a\a'I\b\b'J}\,\bigl(\mu^\nu\,\bigl(\bd1\ox\si_\nu\bigr)^{\d\d'L}_{\ \g\g'K}\,Q_{\d\d'L}+\nu^\rho\,\bigl(\widehat C\ox\si_\rho\bigr)_{\g\g'K\d\d'L}\,\cZ^{\d\d'L}\bigr)\cr\cr
&&+\a\,\bigl(\widehat C\,\widehat\G^{\widehat a}\ox\bd1\bigr)_{\g\g'K\a\a'I}\,\bigl(\widehat C\,\widehat\G_{\widehat a}\ox\si_3\bigr)_{\b\b'J\d\d'L}\,\cZ^{\d\d'L}\cr\cr
&&+\b^{\unl\mu}_{\rm s}\,\bigl(\widehat C\ox\si_{\unl\mu}\bigr)_{\g\g'K\a\a'I}\,\bigl(\mu^\nu\,\bigl(\bd1\ox\si_\nu\bigr)^{\d\d'L}_{\ \b\b'J}\,Q_{\d\d'L}+\nu^\rho\,\bigl(\widehat C\ox\si_\rho\bigr)_{\b\b'J\d\d'L}\,\cZ^{\d\d'L}\bigr)\cr\cr
&&+\a\,\bigl(\widehat C\,\widehat\G^{\widehat a}\ox\bd1\bigr)_{\b\b'J\g\g'K}\,\bigl(\widehat C\,\widehat\G_{\widehat a}\ox\si_3\bigr)_{\a\a'I\d\d'L}\,\cZ^{\d\d'L}\cr\cr
&&+\b^{\unl\mu}_{\rm s}\,\bigl(\widehat C\ox\si_{\unl\mu}\bigr)_{\b\b'J\g\g'K}\,\bigl(\mu^\nu\,\bigl(\bd1\ox\si_\nu\bigr)^{\d\d'L}_{\ \a\a'I}\,Q_{\d\d'L}+\nu^\rho\,\bigl(\widehat C\ox\si_\rho\bigr)_{\a\a'I\d\d'L}\,\cZ^{\d\d'L}\bigr)\,,
\qqq
or, equivalently,
\qq\nn
\b^{\unl\mu}_{\rm s}\,\mu^\nu\,\bigl(\bigl(\widehat C\ox\si_{\unl\mu}\bigr)_{\a\a'I\b\b'J}\,\bigl(\bd1\ox\si_\nu\bigr)^{\d\d'L}_{\ \g\g'K}+\bigl(\widehat C\ox\si_{\unl\mu}\bigr)_{\g\g'K\a\a'I}\,\bigl(\bd1\ox\si_\nu\bigr)^{\d\d'L}_{\ \b\b'J}\cr\cr
+\bigl(\widehat C\ox\si_{\unl\mu}\bigr)_{\b\b'J\g\g'K}\,\bigl(\bd1\ox\si_\nu\bigr)^{\d\d'L}_{\ \a\a'I}\bigr)=0\,,\cr\cr\cr
\bigl(\widehat\G^{\widehat a}\ox\bd1\bigr)^{\a\a'I}_{\ \b\b'J}\,\bigl(\widehat C\,\widehat\G_{\widehat a}\ox\si_3\bigr)_{\d\d'L\g\g'K}+\tfrac{\b^{\unl\mu}_{\rm s}\,\nu^\rho}{\a}\,\bigl(\bd1\ox\si_{\unl\mu}\bigr)^{\a\a'I}_{\ \b\b'J}\,\bigl(\widehat C\ox\si_\rho^{\rm T}\bigr)_{\d\d'L\g\g'K}\cr\cr
+\bigl(\widehat\G^{\widehat a}\ox\bd1\bigr)^{\a\a'I}_{\ \g\g'K}\,\bigl(\widehat C\,\widehat\G_{\widehat a}\ox\si_3\bigr)_{\d\d'L\b\b'J}+\tfrac{\b^{\unl\mu}_{\rm s}\,\nu^\rho}{\a}\,\bigl(\bd1\ox\si_{\unl\mu}^{\rm T}\bigr)^{\a\a'I}_{\ \g\g'K}\,\bigl(\widehat C\ox\si_\rho^{\rm T}\bigr)_{\d\d'L\b\b'J}\cr\cr
\bigl(\widehat C\,\widehat\G^{\widehat a}\ox\bd1\bigr)_{\g\g'K\b\b'J}\,\bigl(\widehat\G_{\widehat a}\ox\si_3\bigr)^{\a\a'I}_{\ \d\d'L}+\tfrac{\b^{\unl\mu}_{\rm s}\,\nu^\rho}{\a}\,\bigl(\widehat C\ox\si_{\unl\mu}^{\rm T}\bigr)_{\g\g'K\b\b'J}\,\bigl(\bd1\ox\si_\rho\bigr)^{\a\a'I}_{\ \d\d'L}=0
\qqq
Upon contracting the latter equation with $\,(\widehat\G_{\widehat b}\ox\bd1)^{\b\b'J}_{\ \ \a\a'I}\,$ and invoking identity \eqref{eq:GamasandGama} as well as the tracelessness of the $\,\widehat\G_{\widehat b}$,\ we arrive at the contraints
\qq\nn
0=8\vep_{\widehat b}\,\si_3+\tfrac{\b^{\unl\mu}_{\rm s}\,\nu^\rho}{2\a}\,\bigl(\si_{\unl\mu}+\si_{\unl\mu}^{\rm T}\bigr)\,\si_\rho\equiv 8\vep_{\widehat b}\,\si_3+\tfrac{\b^{\unl\mu}_{\rm s}\,\nu^\rho}{\a}\,\si_{\unl\mu}\,\si_\rho\,.
\qqq
Clearly, these cannot be satisfied as the sign in front of the (non-zero) first term flips as we pass from $\,\widehat b\in\ovl{0,4}\,$ to $\,\widehat b\in\ovl{5,9}$,\ whereas the second term is independent of $\,\widehat b$.\ $\qed$

\section{A proof of Proposition \ref{prop:svKRproof}}\label{app:svKRproof}

The super-Jacobi identity
\qq\nn
{\rm sJac}\bigl(Q_{\a\a'I},P_{\widehat a},P_{\widehat b}\bigr)=0
\qqq
yields, as previously, the constraints
\qq\nn
\widetilde\g^\mu_{\widehat a}=0\,,\qquad\mu\in\{0,1,2,3\}\,;\qquad\qquad\widetilde\d^\nu_{\widehat a}=0\,,\qquad\nu\in\{0,1,3\}\,;\qquad\qquad\widetilde\d^2_{\widehat a}=\tfrac{1}{2}\,,
\qqq
so that, again,
\qq\nn
[\cZ^{\a\a'I},P_{\widehat a}]^\sim=\tfrac{1}{2}\,\bigl(\widehat\G_{\widehat a}\ox\si_2\bigr)\,^{\a\a'I}_{\ \b\b'J}\,\cZ^{\b\b'J}\,.
\qqq

Considering, next, the super-Jacobi identity
\qq\nn
{\rm sJac}\bigl(Q_{\a\a'I},Q_{\b\b'J},Q_{\g\g'K}\bigr)=0\,,
\qqq
we obtain, through an analysis fully analogous to the one conducted for the spinor-scalar deformation, the constraints
\qq\nn
\b^{\unl\mu}_{{\rm v}\,\widehat a}\,\widetilde\mu^{\nu,\widehat a}\,\bigl(\bigl(\widehat C\,\widehat\G_{\widehat a}\ox\si_{\unl\mu}\bigr)_{\a\a'I\b\b'J}\,\bigl(\widehat\G^{\widehat a}\ox\si_\nu\bigr)^{\d\d'L}_{\ \g\g'K}+\bigl(\widehat C\,\widehat\G_{\widehat a}\ox\si_{\unl\mu}\bigr)_{\g\g'K\a\a'I}\,\bigl(\widehat\G^{\widehat a}\ox\si_\nu\bigr)^{\d\d'L}_{\ \b\b'J}\cr\cr
+\bigl(\widehat C\,\widehat\G_{\widehat a}\ox\si_{\unl\mu}\bigr)_{\b\b'J\g\g'K}\,\bigl(\widehat\G^{\widehat a}\ox\si_\nu\bigr)^{\d\d'L}_{\ \a\a'I}\bigr)=0\,,\cr\cr\cr
\bigl(\widehat\G^{\widehat a}\ox\bd1\bigr)^{\a\a'I}_{\ \b\b'J}\,\bigl(\widehat C\,\widehat\G_{\widehat a}\ox\si_3\bigr)_{\d\d'L\g\g'K}+\tfrac{\b^{\unl\mu}_{{\rm v}\,\widehat a}\,\widetilde\nu^{\rho,\widehat a}}{\a}\,\bigl(\widehat\G_{\widehat a}\ox\si_{\unl\mu}\bigr)^{\a\a'I}_{\ \b\b'J}\,\bigl(\widehat C\,\widehat\G^{\widehat a}\ox\si_\rho^{\rm T}\bigr)_{\d\d'L\g\g'K}\cr\cr
+\bigl(\widehat\G^{\widehat a}\ox\bd1\bigr)^{\a\a'I}_{\ \g\g'K}\,\bigl(\widehat C\,\widehat\G_{\widehat a}\ox\si_3\bigr)_{\d\d'L\b\b'J}+\tfrac{\b^{\unl\mu}_{{\rm v}\,\widehat a}\,\widetilde\nu^{\rho,\widehat a}}{\a}\,\bigl(\widehat\G_{\widehat a}\ox\si_{\unl\mu}\bigr)^{\a\a'I}_{\ \g\g'K}\,\bigl(\widehat C\,\widehat\G^{\widehat a}\ox\si_\rho^{\rm T}\bigr)_{\d\d'L\b\b'J}\cr\cr
+\bigl(\widehat C\,\widehat\G^{\widehat a}\ox\bd1\bigr)_{\g\g'K\b\b'J}\,\bigl(\widehat\G_{\widehat a}\ox\si_3\bigr)^{\a\a'I}_{\ \d\d'L}+\tfrac{\b^{\unl\mu}_{{\rm v}\,\widehat a}\,\widetilde\nu^{\rho,\widehat a}}{\a}\,\bigl(\widehat C\,\widehat\G_{\widehat a}\ox\si_{\unl\mu}\bigr)_{\g\g'K\b\b'J}\,\bigl(\widehat\G^{\widehat a}\ox\si_\rho\bigr)^{\a\a'I}_{\ \d\d'L}=0\,.
\qqq
Upon contracting the latter one with $\,(\widehat\G_{\widehat b}\ox\bd1)^{\b\b'J}_{\ \a\a'I}\,$ and employing \Reqref{eq:GamasandGama}, alongside the refined identities \eqref{eq:GambsandGamaa}, we infer from it a pair of independent constraints:
\qq\nn
8\si_3+\tfrac{16\b^0_{{\rm v}\,b}\,\widetilde\nu^{\rho,b}}{\a}\,\d_{\unl\mu 0}\,\si_\rho^{\rm T}-\tfrac{1}{\a}\,\bigl(3\b^{\unl\mu}_{{\rm v}\,b}\,\widetilde\nu^{\rho,b}+5\b^{\unl\mu}_{{\rm v}\,b'}\,\widetilde\nu^{\rho,b'}\bigr)\,\si_\rho^{\rm T}\,\si_{\unl\mu}&=&0\,,\cr\cr
8\si_3+\tfrac{16\b^0_{{\rm v}\,b'}\,\widetilde\nu^{\rho,b'}}{\a}\,\d_{\unl\mu 0}\,\si_\rho^{\rm T}-\tfrac{1}{\a}\,\bigl(5\b^{\unl\mu}_{{\rm v}\,b}\,\widetilde\nu^{\rho,b}+3\b^{\unl\mu}_{{\rm v}\,b'}\,\widetilde\nu^{\rho,b'}\bigr)\,\si_\rho^{\rm T}\,\si_{\unl\mu}&=&0\,.
\qqq
with the unique solution
\qq\nn
\widetilde\nu^{\rho,\widehat b}=-\left(\d_{\unl\mu 0}\,\d^{\rho 3}+\sfi\,\d_{\unl\mu 1}\,\d^{\rho 2}-\d_{\unl\mu 3}\,\d^{\rho 0}\right)\,\tfrac{\a}{\b^{\unl\mu}_{{\rm v}\,\widehat b}}\,.
\qqq
Note, in particular, that we necessarily have
\qq\label{eq:dirnontriv}
\b^{\unl\mu}_{{\rm v}\,\widehat b}\neq 0\,.
\qqq
It is straightforward to verify that the deformation thus determined satisfies the second of the two constraints obtained above. However, as our goal is to show its inconsistency (as an associative deformation), we leave the proof out and proceed with the other constraints. These we contract with $\,(\widehat\G_{\widehat b}\ox\si_\rho)^{\g\g'K}_{\ \d\d'L}\,$ and with the help of identities \eqref{eq:GambsandGamaa} transform into a pair of independent constraints
\qq\nn
16\b^{\unl\mu}_{{\rm v}\,b}\,\widetilde\mu^{\rho,b}\,\si_{\unl\mu}-\bigl(3\b^{\unl\mu}_{{\rm v}\,b}\,\widetilde\mu^{\nu,b}+5\b^{\unl\mu}_{{\rm v}\,b'}\,\widetilde\mu^{\nu,b'}\bigr)\,\tfrac{\si_{\unl\mu}\,\si_\rho\,\si_\nu+\si_\nu^{\rm T}\,\si_\rho^{\rm T}\,\si_{\unl\mu}}{2}&=&0\,,\cr\cr
16\b^{\unl\mu}_{{\rm v}\,b'}\,\widetilde\mu^{\rho,b'}\,\si_{\unl\mu}-\bigl(5\b^{\unl\mu}_{{\rm v}\,b}\,\widetilde\mu^{\nu,b}+3\b^{\unl\mu}_{{\rm v}\,b'}\,\widetilde\mu^{\nu,b'}\bigr)\,\tfrac{\si_{\unl\mu}\,\si_\rho\,\si_\nu+\si_\nu^{\rm T}\,\si_\rho^{\rm T}\,\si_{\unl\mu}}{2}&=&0\,.
\qqq
with the unique solution (readily extracted by considering first $\,\rho=\unl\mu\,$ and subsequently $\,\rho=2\,$ in the above)
\qq\nn
\widetilde\mu^{\rho,\widehat b}=0\,.
\qqq
Hence, at this stage, the result
\qq\nn
[\cZ^{\widehat a},Q_{\a\a'I}]^\sim=\widetilde\nu^{\mu,\widehat a}\,\bigl(\widehat C\,\widehat\G^{\widehat a}\ox\si_\mu\bigr)_{\a\a'I\b\b'J}\,\cZ^{\b\b'J}\,.
\qqq

We now impose the super-Jacobi identity
\qq\nn
{\rm sJac}\bigl(Q_{\a\a'I},Q_{\b\b'J},P_{\widehat a}\bigr)=0
\qqq
to obtain the constraints
\qq\nn
0&=&\tfrac{\sfi\,\a\,\widetilde\vep^{\widehat b,\nu}}{2}\,\bigl(\widehat C\,\widehat\G_{\widehat a}\,\widehat\G^{\widehat b}\ox\si_3\,\si_\nu+\widehat C\,\widehat\G^{\widehat b}\,\widehat\G_{\widehat a}\ox\si_\nu^{\rm T}\,\si_3\bigr)_{\a\a'I\b\b'J}\,P_{\widehat b}\cr\cr
&&+\tfrac{1}{2}\,\bigl(\b^{\unl\mu}_{{\rm v}\,\widehat b}\,\bigl(\widehat C\,\widehat\G_{\widehat b}\,\widehat\G_{\widehat a}\ox\si_{\unl\mu}\,\si_2-\widehat C\,\widehat\G_{\widehat a}\,\widehat\G_{\widehat b}\ox\si_2\,\si_{\unl\mu}\bigr)+\sfi\,\a\,\widetilde\eta_{\widehat b,\nu}\,\bigl(\widehat C\,\widehat\G_{\widehat a}\,\widehat\G_{\widehat b}\ox\si_3\,\si_\nu+\widehat C\,\widehat\G_{\widehat b}\,\widehat\G_{\widehat a}\ox\si_\nu^{\rm T}\,\si_3\bigr)\bigr)_{\a\a'I\b\b'J}\,\cZ^{\widehat b}\cr\cr
&&+\bigl(\b^{\unl\mu}_{{\rm v}\,\widehat b}\,\widetilde\la_{\widehat b\widehat a}\,\d^{\widehat c}_{\ \widehat a}\,\bigl(\widehat C\,\widehat\G^{\widehat b}\ox\si_{\unl\mu}\bigr)+\tfrac{\sfi\,\a\,\widetilde\z^{\widehat b\widehat c,\nu}}{2}\,\bigl(\widehat C\,\widehat\G_{\widehat a}\,\widehat\G^{\widehat b\widehat c}\ox\si_3\,\si_\nu-\widehat C\,\widehat\G^{\widehat b\widehat c}\,\widehat\G_{\widehat a}\ox\si_\nu^{\rm T}\,\si_3\bigr)\bigr)_{\a\a'I\b\b'J}\,J_{\widehat b\widehat c}\,,
\qqq
or, equivalently,
\qq\nn
\widetilde\vep^{\widehat b,\nu}\,\bigl(\widehat C\,\widehat\G_{\widehat a}\,\widehat\G^{\widehat b}\ox\si_3\,\si_\nu+\widehat C\,\widehat\G^{\widehat b}\,\widehat\G_{\widehat a}\ox\si_\nu^{\rm T}\,\si_3\bigr)=0\,,\cr\cr
\b^{\unl\mu}_{{\rm v}\,\widehat b}\,\bigl(\widehat C\,\widehat\G_{\widehat b}\,\widehat\G_{\widehat a}\ox\si_{\unl\mu}\,\si_2-\widehat C\,\widehat\G_{\widehat a}\,\widehat\G_{\widehat b}\ox\si_2\,\si_{\unl\mu}\bigr)+\sfi\,\a\,\widetilde\eta_{\widehat b,\nu}\,\bigl(\widehat C\,\widehat\G_{\widehat a}\,\widehat\G_{\widehat b}\ox\si_3\,\si_\nu+\widehat C\,\widehat\G_{\widehat b}\,\widehat\G_{\widehat a}\ox\si_\nu^{\rm T}\,\si_3\bigr)=0\,,\cr\cr
2\b^{\unl\mu}_{{\rm v}\,\widehat b}\,\widetilde\la_{\widehat b\widehat a}\,\d^{\widehat c}_{\ \widehat a}\,\bigl(\widehat C\,\widehat\G^{\widehat b}\ox\si_{\unl\mu}\bigr)+\sfi\,\a\,\widetilde\z^{\widehat b\widehat c,\nu}\,\bigl(\widehat C\,\widehat\G_{\widehat a}\,\widehat\G^{\widehat b\widehat c}\ox\si_3\,\si_\nu-\widehat C\,\widehat\G^{\widehat b\widehat c}\,\widehat\G_{\widehat a}\ox\si_\nu^{\rm T}\,\si_3\bigr)=0\,.
\qqq
For $\,(\widehat a,\widehat b)\in\ovl{0,4}^{\x 2}\cup\ovl{5,9}^{\x 2}$,\ the first of them rewrites as
\qq\nn
\widetilde\vep^{\widehat b,0}\,\vep_{\widehat a}\,\d^{\widehat b}_{\ \widehat a}\,\bigl(\widehat C\ox\si_3\bigr)+\sfi\,\widetilde\vep^{\widehat b,1}\,\eta^{\widehat b\widehat c}\,\bigl(\widehat C\,\widehat\G_{\widehat a\widehat c}\ox\si_2\bigr)-\sfi\,\widetilde\vep^{\widehat b,2}\,\vep_{\widehat a}\,\d^{\widehat b}_{\ \widehat a}\,\bigl(\widehat C\ox\si_1\bigr)+\widetilde\vep^{\widehat b,3}\,\vep_{\widehat a}\,\d^{\widehat b}_{\ \widehat a}\,\bigl(\widehat C\ox\bd1\bigr)=0\,,
\qqq
and so yields
\qq\nn
\widetilde\vep^{\widehat b,\nu}=0\,.
\qqq
Writing out the second one in a similar fashion, we arrive at the result
\qq\nn
\widetilde\eta_{\widehat b,\nu}=-\bigl(\d_{\unl\mu 0}\,\d_{\nu 1}+\d_{\unl\mu 1}\,\d_{\nu 0}-\sfi\,\d_{\unl\mu 3}\,\d_{\nu 2}\bigr)\,\tfrac{\b^{\unl\mu}_{{\rm v}\,\widehat b}}{\a}\,.
\qqq
The last constraints can readily be solved in a similar fashion, to the effect
\qq\nn
\widetilde\z^{\widehat b\widehat c,\nu}=0\,,\qquad\qquad\widetilde\la_{\widehat b\widehat c}=0\,,
\qqq
but the result is immaterial to the rest of our reasoning, and so we do not spend any time proving it.

Thus equipped, we examine the super-Jacobi identity
\qq\nn
{\rm sJac}\bigl(\cZ^{\a\a'I},Q_{\b\b'J},P_{\widehat a}\bigr)=0\,.
\qqq
Its projection on $\,\cZ^{\widehat b}\,$ along the remaining generators of the deformation (marked as $\,[\cdot]_{\widehat b}$) reads
\qq\nn
0&=&-\tfrac{1}{2}\,\bigl(\widehat\G_{\widehat a}\ox\si_2\bigr)^{\a\a'I}_{\ \g\g'K}\,\bigl[\bigl\{\cZ^{\g\g'K},Q_{\b\b'J}\bigr\}^\sim\bigr]_{\widehat b}+\tfrac{1}{2}\,(\widehat\G_{\widehat a}\ox\si_2)^{\g\g'K}_{\ \ \b\b'J}\,\bigl[\bigl\{\cZ^{\a\a'I},Q_{\g\g'K}\bigr\}^\sim\bigr]_{\widehat b}\cr\cr
&&+\tfrac{\sfi\,\a}{2}\,\bigl(\widehat C\,\widehat\G_{\widehat a}\ox\si_3\bigr)_{\b\b'J\g\g'K}\,\bigl[\bigl\{\cZ^{\g\g'K},\cZ^{\a\a'I}\bigr\}^\sim\bigr]_{\widehat b}\cr\cr
&=&\bigl(\d_{\unl\mu 0}\,\d_{\nu 1}+\d_{\unl\mu 1}\,\d_{\nu 0}-\sfi\,\d_{\unl\mu 3}\,\d_{\nu 2}\bigr)\,\tfrac{\b^{\unl\mu}_{{\rm v}\,\widehat b}}{2\a}\,\bigl(\widehat\G_{\widehat a}\,\widehat\G_{\widehat b}\ox\si_2\,\si_\nu-\widehat\G_{\widehat b}\,\widehat\G_{\widehat a}\ox\si_\nu\,\si_2\bigr)^{\a\a'I}_{\ \b\b'J}\cr\cr
&&+\tfrac{\sfi\,\a\,\widetilde\k_{\widehat b}^K}{2}\,\bigl(\widehat\G_{\widehat b}\,\widehat\G_{\widehat a}\ox\si_K\,\si_3\bigr)^{\a\a'I}_{\ \b\b'J}
\qqq
and gives us, once again for $\,(\widehat a,\widehat b)\in\ovl{0,4}^{\x 2}\cup\ovl{5,9}^{\x 2}$,\ the constraints
\qq\nn
-\d_{\unl\mu 0}\,\tfrac{\sfi\,\vep_{\widehat a}\,\b^0_{{\rm v}\,\widehat b}\,\eta_{\widehat a\widehat b}}{\a}\,\bigl(\bd1\ox\si_3\bigr)+\d_{\unl\mu 1}\,\tfrac{\b^1_{{\rm v}\,\widehat b}}{\a}\,\bigl(\widehat\G_{\widehat a\widehat b}\ox\si_2\bigr)-\d_{\unl\mu 3}\,\tfrac{\sfi\,\b^3_{{\rm v}\,\widehat b}}{\a}\,\bigl(\widehat\G_{\widehat a\widehat b}\ox\bd1\bigr)+\tfrac{\sfi\,\a\,\widetilde\k_{\widehat b}^K}{2}\,\bigl(\bigl(\vep_{\widehat a}\,\eta_{\widehat a\widehat b}\,\bd1-\widehat\G_{\widehat a\widehat b}\bigr)\ox\si_K\,\si_3\bigr)=0\,,
\qqq
or, equivalently,
\qq\nn
\d_{\unl\mu 0}\,\tfrac{2\b^0_{{\rm v}\,\widehat b}}{\a^2}\,\bd1-\widetilde\k_{\widehat b}^K\,\si_K&=&0\,,\cr\cr
\d_{\unl\mu 1}\,\tfrac{2\b^1_{{\rm v}\,\widehat b}}{\a^2}\,\si_2-\d_{\unl\mu 3}\,\tfrac{2\sfi\,\b^3_{{\rm v}\,\widehat b}}{\a^2}\,\bd1-\sfi\,\widetilde\k_{\widehat b}^K\,\si_K\,\si_3&=&0\,.
\qqq
These cannot be imposed consistently with our previous result \eqref{eq:dirnontriv}. Indeed, {\it e.g.}, for $\,\unl\mu=0$,\ the latter equation yields $\,\widetilde\k_{\widehat b,K}=0$,\ but then the former requires $\,\b^0_{{\rm v}\,\widehat b}=0$,\ and similarly for $\,\unl\mu\in\{1,3\}$. $\qed$

\end{document}

%% file: 1armagedef.tex
\newcommand{\alxydim}[2]{\begin{aligned}\xymatrix#1{#2}\end{aligned}}

\newcommand{\brem}{\begin{Rem}}
\newcommand{\erem}{\end{Rem}\medskip}
\newcommand{\beg}{\begin{Eg}}
\newcommand{\eeg}{\end{Eg}}
\newcommand{\bedef}{\begin{Def}}
\newcommand{\exdef}{\begin{flushright}$\diamond$\end{flushright}
\end{Def}\vskip0.1cm}
\newcommand{\berop}{\begin{Prop}}
\newcommand{\eerop}{\end{Prop}}
\newcommand{\belem}{\begin{Lem}}
\newcommand{\elem}{\end{Lem}}
\newcommand{\bethe}{\begin{Thm}}
\newcommand{\ethe}{\end{Thm}}
\newcommand{\becor}{\begin{Cor}}
\newcommand{\ecor}{\end{Cor}}
\newcommand{\beroof}{\noindent\begin{proof}}
\newcommand{\eroof}{\end{proof}}
\newcommand{\becon}{\begin{Conv}}
\newcommand{\econ}{\begin{flushright}$\checkmark$\end{flushright}\end{Conv}}
\newcommand{\befact}{\begin{Fact}}
\newcommand{\efact}{\begin{flushright}$\checkmark$\end{flushright}\end{Fact}}
\newcommand{\bequest}{\begin{Quest}}
\newcommand{\equest}{\end{Quest}}
\newcommand{\brob}{\begin{Prob}}
\newcommand{\erob}{\end{Prob}}

\newcommand{\barr}{\begin{array}}
\newcommand{\earr}{\end{array}}
\newcommand{\ben}{\begin{enumerate}}
\newcommand{\een}{\end{enumerate}}
\newcommand{\bit}{\begin{itemize}}
\newcommand{\eit}{\end{itemize}}

\newcommand{\qq}{\begin{eqnarray}}
\newcommand{\qqq}{\end{eqnarray}}

\newcommand{\nn}{\nonumber}

\newcommand{\ovl}[1]{\overline{#1}}
\newcommand{\unl}[1]{\underline{#1}}

\newcommand{\Reqref}[1]{Eq.\,\eqref{#1}}
\newcommand{\Rcite}[1]{Ref.\,\cite{#1}}

\newcommand\void[1]{}

\newcommand{\tx}[1]{\textrm{#1}} 
\newcommand{\ciut}[1]{\tiny$#1$}

\newcommand{\gt}[1]{\mathfrak{#1}}

\def\cA{\mathcal{A}}

\def\cC{\mathcal{C}}

\def\cE{\mathcal{E}}

\def\cG{\mathcal{G}}
\def\ceH{\mathcal{H}}
\def\cI{\mathcal{I}}

\def\cK{\mathcal{K}}
\def\ceL{\mathcal{L}}

\def\cO{\mathcal{O}}

\def\cR{\mathcal{R}}

\def\cZ{\mathcal{Z}}

\def\xcC{\mathscr{C}}

\def\xcL{\mathscr{L}}
\def\xcM{\mathscr{M}}

\def\xcP{\mathscr{P}}

\def\xcR{\mathscr{R}}

\def\xcT{\mathscr{T}}

\def\xcW{\mathscr{W}}

\def\t{\mathbf{t}}

\def\bC{{\mathbb{C}}}

\def\bR{{\mathbb{R}}}
\def\bS{{\mathbb{S}}}

\def\bX{{\mathbb{X}}}

\def\a{\alpha}
\def\b{\beta}
\def\g{\gamma}
\def\G{\Gamma}
\def\d{\delta}
\def\D{\Delta}
\def\ep{\epsilon}
\def\vep{\varepsilon}

\def\k{\kappa}

\def\la{\lambda}
\def\La{\Lambda}

\def\Om{\Omega}

\def\si{\sigma}
\def\Si{\Sigma}

\def\t{\tau}

\def\z{\zeta}

\def\ggt{\gt{g}}

\def\hgt{\gt{h}}

\def\Pgt{\gt{P}}

\def\rgt{\gt{r}}

\def\Sgt{\gt{S}}
\def\tgt{\gt{t}}

\def\zgt{\gt{z}}

\newcommand{\sfd}{{\mathsf d}}
\newcommand{\sfD}{{\mathsf D}}

\newcommand{\sfi}{{\mathsf i}}

\newcommand{\sfp}{{\mathsf p}}
\newcommand{\sfP}{{\mathsf P}}

\newcommand{\sfT}{{\mathsf T}}

\newcommand{\sfY}{{\mathsf Y}}

\newcommand{\txA}{{\rm A}}
\newcommand{\txb}{{\rm b}}
\newcommand{\txB}{{\rm B}}

\newcommand{\txD}{{\rm D}}
\newcommand{\ee}{{\rm e}}

\newcommand{\txg}{{\rm g}}
\newcommand{\txG}{{\rm G}}

\newcommand{\txH}{{\rm H}}

\newcommand{\txK}{{\rm K}}

\def\Cv{\v{C}}

\def\vH{\check{H}}

\def\exp{{\rm exp}}
\def\id{{\rm id}}
\newcommand{\pr}{{\rm pr}}
\def\sign{{\rm sign}}

\def\too{\longrightarrow}
\def\ev{{\rm ev}}

\def\1morf{1{\rm -Mor}}
\def\2morf{2{\rm -Mor}}
\def\dim{{\rm dim}}

\def\ker{{\rm ker}}

\def\Vol{{\rm Vol}}

\newcommand{\pLie}[1]{\,{-\hspace{-8pt}\xcL}_{#1}}
\def\p{\partial}

\def\con{\righthalfcup}
\newcommand{\Diff}{{\rm Diff}}

\newcommand{\sG}{\mathcal{sG}}

\def\bd1{{\boldsymbol{1}}}
\def\brd0{{\boldsymbol{0}}}

\def\det{{\rm det}}
\def\tr{{\rm tr}}
\def\diag{\textrm{diag}}

\def\ad{{\rm ad}}
\def\Ad{{\rm Ad}}

\def\Cliff{{\rm Cliff}}

\newcommand{\uj}{{\rm U}(1)}

\def\x{\times}
\def\ox{\otimes}

\def\rx{\rtimes}

\def\lact{\vartriangleright}
\def\must{\stackrel{!}{=}}

\def\rstr{\mathord{\restriction}}

\newcommand{\corr}[1]{\left\langle #1 \right\rangle}
